\documentclass[11pt, reqno,preprint]{article}
\usepackage{jheppub}
\usepackage{epsfig}
\usepackage{caption}
\usepackage{amssymb}
\usepackage{amsmath}
\usepackage{mathrsfs}
\usepackage{hyperref}
\usepackage{multirow}
\usepackage{graphicx}
\usepackage{textcomp}
\usepackage{subcaption}
\usepackage{tikz}
\usepackage{float}
\usetikzlibrary{decorations.pathmorphing}
\usetikzlibrary{decorations.markings}
\usetikzlibrary{decorations.markings}
\usepackage{nicematrix}
\allowdisplaybreaks

\newcommand{\beq}{\begin{equation}}
\newcommand{\eeq}{\end{equation}}
\newcommand{\bba}{\begin{align}}
\newcommand{\eea}{\end{align}}
\def\avgbosonic#1{\langle\!\langle\ {#1} \ \rangle\!\rangle}
\def\SDYM{{\scriptscriptstyle\text{SDYM}}}
\def\NMHV{{\scriptscriptstyle\text{NMHV}}}
\def\basicR#1#2#3{R_{#1#2\mu}^{#3}}
\def\<{\langle}
\def\>{\rangle}
\newcommand{\dd}{\mathrm{d}}
\newcommand{\tr}{\mathrm{tr}}

\title{Determinants in self-dual $\mathcal{N}=4$ SYM and twistor space}
\author{Simon Caron-Huot, Frank Coronado and Beatrix M\"uhlmann}
\affiliation{
Department of Physics, McGill University, 3600 Rue University, Montr\'eal, H3A 2T8, QC Canada
}
\emailAdd{schuot@physics.mcgill.ca}
\emailAdd{fcidrogo@gmail.com}
\emailAdd{beatrix.muehlmann@mcgill.ca}

\abstract{We consider correlation functions of supersymmetrized determinant operators in self-dual super Yang-Mills (SYM).  These provide a generating function for correlators of arbitrary single-trace half-BPS operators, including, for appropriate Grassmann components, the so-called loop
integrand of the non-self-dual theory.
We introduce a novel twistor space representation for determinant operators which makes contact with the recently studied $m=2$ amplituhedron.
By using matrix duality we rewrite the $n$-point determinant correlator  as a $n\times n$ matrix integral where the gauge group rank $N_c$ is turned into a coupling.
The correlators are rational functions whose denominators,
in the planar limit, contain only ten-dimensional distances.
Using this formulation, we verify a recent conjecture regarding the ten-dimensional symmetry of
the components with maximal Grassmann degree and we
obtain new formulas for correlators
of Grassmann degree four.}

\begin{document}

\maketitle

\section{Introduction}

Correlation functions of local operators are central observables in a conformal field theory.  
In $\mathcal{N}=4$ super Yang-Mills (SYM) the most studied ones involve protected operators.
Their four- and higher-point correlation functions depend nontrivially on the gauge coupling and carry a wealth of dynamical information about the theory.
This includes correlators of the stress tensor, which captures physics as diverse
as energy flux in conformal collider physics, planar scattering amplitudes of gluons
through null polygonal Wilson loops,
as well as graviton scattering amplitudes in a dual AdS${}_5\times $S$_5$ spacetime.

The generic half-BPS operators in this theory are spanned by traces of the form
\beq
\tr[\left(y{\cdot}\phi(x)\right)^k] +\mbox{susy-descendants}
\eeq
where $x$ is a spacetime coordinate and $y$ is a null six vector conjugate to the ${\rm Spin}(6)$ R-symmetry group.
There are many motivations for considering these operators for general $k\neq 2$, beyond the $k=2$ stress-tensor multiplet.
At strong coupling, these are related to Kaluza-Klein
harmonics of the graviton which reveal the geometry of the S$^5$ internal manifold \cite{Lee:1998bxa}.
The large-$k$ limit has also been particularly fruitful from the integrability perspective,
since large-$k$ operators define a natural ``vacuum'' state of an infinite-length
spin chain \cite{Berenstein:2002jq,Beisert:2010jr}.  Correlation functions also dramatically simplify in that limit, see for example
\cite{Jiang:2016ulr, Basso:2017khq, Basso:2019diw, Coronado:2018ypq, Coronado:2018cxj, Kostov:2019stn, Belitsky:2020qrm, Fleury:2020ykw,Aprile:2020luw}.
Finite-$k$ or higher-point correlators \cite{Chicherin:2015edu, Chicherin:2018avq, Fleury:2019ydf, Bargheer:2022sfd} are currently an important source of data in the program of computing correlation functions with integrability \cite{Basso:2015zoa, Fleury:2016ykk, Eden:2016xvg, Cavaglia:2021mft, Bercini:2022jxo}.

Recently, it has been observed that single-trace operators with different weight $k$ often naturally
combine into a generating function,
\begin{equation}\begin{aligned}\label{eq:introsingletrace}
\mathbb{O}(x,y,\theta)&\,\equiv\,-\log\det\left(1-y{\cdot}\phi(x)\right) + \mbox{susy-descendants}
\\ &= \tr\left[y{\cdot}\phi+ \frac{1}{2}\,\left(y{\cdot}\phi\right)^2\,+\,\frac{1}{3}\,\left(y{\cdot} \phi\right)^3\, +\cdots\right]\,.
\end{aligned}\end{equation}
We have retained the first term since it contributes in the $U(N_c)$ version of the theory, which we will consider.
The usefulness of such a repackaging was first observed at strong 't Hooft coupling,
where it unites tree-level correlation functions of arbitrary Kaluza-Klein modes
into a single object that enjoys a ten-dimensional conformal
symmetry \cite{Caron-Huot:2018kta}.
This symmetry means that correlators of (the top component of) $\mathbb{O}$ depend only on conformal cross-ratios built out of ten-dimensional distances, which combine spacetime and R-charge kinematics:
\beq\label{eq:introX10d}
X \equiv (x,y) \qquad \text{and} \qquad X_{ij}^{2} \equiv (x_{i}-x_{j})^2 + (y_{i}-y_{j})^2\,.
\eeq
The origin of this symmetry remains mysterious.
It implies concise formulas for these correlators and degeneracies in the double-trace spectrum \cite{Rastelli:2016nze, Aprile:2018efk}. It has also been observed in other geometries of the form AdS${}_k\times $S$^k$
\cite{Rastelli:2019gtj,Giusto:2020neo,Aprile:2021mvq,Abl:2021mxo,Drummond:2022dxd}, but it is known to be violated by stringy corrections away from infinite 't Hooft coupling \cite{Abl:2020dbx,Aprile:2020mus}. Yet, the symmetry has been observed at weak 't Hooft coupling in the so-called integrand that controls loop corrections to correlation functions \cite{Caron-Huot:2021usw}.

The integrand appears naturally when expanding
full Yang-Mills theory as a perturbation around the self-dual theory.
As will be reviewed below, the $\ell$-loop correction to some observables can be expressed as an integral over correlators featuring $\ell$ additional insertions of the chiral Lagrangian $L_{\rm int}$. 
The ``integrand'' is then a \emph{bona fide} observable in self-dual (super) Yang-Mills.
In fact, the chiral Lagrangian is nothing but the top component (of Grassmann degree four) of \eqref{eq:introsingletrace},
so the integrand is really a particular component
of the supersymmetrized correlator:
\beq\label{eq:introgeneralization}
\left\langle \prod_{i=1}^{n}\mathbb{O}(x_i,y_i,\theta_i)\prod_{i=n+1}^{n+\ell}L_{\text{int}}(x_i) \right\rangle_{\SDYM} \overset{\text{SUSY}}{\longrightarrow}~\left\langle \prod_{i=1}^{n+\ell}\mathbb{O}(x_i,y_i,\theta_i) \right\rangle_{\SDYM}~.
\eeq
A limit of this correlation function includes the integrand for planar scattering amplitudes \cite{Arkani-Hamed:2010zjl,Mason:2010yk,Eden:2010ce,Caron-Huot:2010ryg}.
This is due to the duality between amplitudes and null polygonal Wilson loops in the SYM theory (understood as T-duality of the AdS${}_5\times $S$^5$ superstring \cite{Berkovits:2008ic,Beisert:2008iq}); at the integrand level, the latter are a null limit of stress-tensor multiplet correlators, see for example \cite{Eden:2011yp,Eden:2011ku,Adamo:2011dq}.
The $y$ dependence of the above correlators is expected to map to the mass dependence
of scattering amplitudes of massive gluons along the Coulomb branch \cite{Caron-Huot:2021usw}.
The fact that the right-hand-side enjoys a full permutation symmetry in $(n+\ell)$ coordinates
simplifies many calculations, as was first noticed for stress tensor correlators in \cite{Eden:2011we}.

The full $y$ and $\theta$ dependence of \eqref{eq:introgeneralization} has to our knowledege never been studied.
For example, the previous work \cite{Caron-Huot:2021usw} by two of the present authors
considered only the bottom components ($\theta_i=0$ for $i\leq n$) and treated all Lagrangians as functions of $x$ only; it was then observed that the result was the restriction of a ten-dimensional-invariant object.
The main goal of this paper is to understand how to calculate the super-correlator on the right-hand-side in a way that preserves its full dependence on $y$ and $\theta$.

Our main tool will be the twistor space formulation of self-dual SYM.
This is a natural starting point since solving self-dual theories was
one of the first main applications of the Penrose(-Ward)
twistor transform \cite{doi:10.1063/1.1705200}.
We will benefit from a vast body of literature on the twistor approach to $\mathcal{N}=4$ SYM,
starting from Witten's formulation which accounts for both the self-dual theory and
for perturbations around it.
In the original works \cite{Witten:2003nn, Boels:2006ir} the chiral Lagrangian $L_{\text{int}}(x)$ was constructed as a non-local operator in twistor space with support over an entire $\mathbb{CP}^{1}$ line, 
interpreted from a topological string theory viewpoint
in terms of D1 branes. Half-BPS operators with $k\geq 2$ were later constructed \cite{Adamo:2011dq,Koster:2016ebi,Chicherin:2016soh}.

One of the main results of this paper is a novel construction of the
generating function \eqref{eq:introsingletrace} in twistor space.
In fact, we find a simple formula  for determinant operators
\beq\label{eq:introSuperD}
\mathbb{D}(x,y,\theta) \,=\, \exp[-\mathbb{O}(x,y,\theta)] =\det\left(1-y{\cdot}\phi(x)\right) + \text{susy-descendants}
\eeq
as the partition function of a Gaussian model on a $\mathbb{CP}^{1|2}$ superline.
Interestingly, the amplitudes of this model coincide with those of the $m=2$ amplituhedron \cite{Lukowski:2020dpn}.
Furthermore, the twistor-space propagators connecting different superlines naturally lead to a geometric series that sums up to ten-dimensional propagators $1/X_{ij}^2$. We will use this formulation to confirm
the validity of formulas obtained in \cite{Caron-Huot:2021usw} for four-point integrands
in situations in which all $y_i\neq 0$, which were previously beyond reach.

Our second main result is a reformulation of $n$-point determinant correlation functions
in the $U(N_c)$ theory as an integral over matrices of dimension $n\times n$ using matrix duality \cite{GopakumarTalk, Jiang:2019xdz}.
The rank $N_c$ of the gauge group then becomes a coupling of the matrix integral. 
The bosonic limit of our matrix integral was discussed previously in refs.~\cite{Jiang:2019xdz,Budzik:2021fyh,Chen:2019gsb}
and generates scalar correlators in the free theory.
Our extension covers the full $\theta$ dependence, thus crucially opening up access to correlation functions involving the chiral Lagrangian.
Using this approach, we will obtain new results for 
NMHV --- we present explicit results for up to $n=6$ ---
as well as an avenue to obtain non-planar or higher order N$^{k}$MHV correlators.   The ten-dimensional distance \eqref{eq:introX10d} automatically appears in denominators in the large-$N_c$ expansion.
\newline\newline
This article is organised as follows. In section \ref{sec:twistor} we review twistor space, its geometry and the formulation of $\mathcal{N}=4$ SYM in twistor space. We also discuss the subset of half-BPS operators. The supersymmetrized determinant operator we introduce in section \ref{section3}. We use this section to compare our result to related work. In section \ref{section4} we discuss Feynman rules and obtain two- to five-point planar correlators manifesting ten-dimensional denominators.
Matrix duality is carried out in section \ref{section5}. In this section we discuss properties and examples of the resulting model and explain the large $N_c$ limit. We uncover the ten-dimensional structure from the matrix integral perspective and in section \ref{section6} we systematically calculate higher point NMHV correlators using matrix model manipulations. We conclude with a summary of our results and future directions in section \ref{section8}.

\section{Self-dual super Yang-Mills and twistor space}\label{sec:twistor}

In this section we describe the formulation of $\mathcal{N}=4$ SYM as an expansion around its self-dual part (SDYM) in both spacetime and twistor space. In this formulation the integrands can be regarded as Born-level correlators in SDYM of the operators in question and the non self-dual part of the Lagrangian. Furthermore in the twistor formulation the self-dual part can be made manifestly Gaussian thanks to the larger gauge freedom in twistor space (CSW gauge). This constitutes our main motivation to uplift to twistor space. For once it allows to simplify the computation of integrands or Born correlators of determinants as shown in section \ref{section4}. Besides, being a Gaussian in CSW gauge, it allows to apply the matrix duality as shown in section \ref{section5}.

\subsection{Self-dual super Yang-Mills in spacetime}

The Chalmers-Siegel procedure \cite{PhysRevD.54.7628} gives $\mathcal{N}=4$ SYM as an expansion around its (anti-)self-dual part. Following the notations in  \cite{Caron-Huot:2010ryg} we have 
\beq\label{eq:N4aroundSDYM}
S_{\mathcal{N}=4} = S_{\text{self-dual}} \,+\, g^2\int \frac{\dd^{4}x}{\pi^2}\,L_{\text{int}}(x)~,
\eeq
where $g^2=\frac{g^2_{\rm YM}N_c}{(4\pi)^2}$
is related to the 't Hooft coupling and the action of (anti-)self-dual Yang Mills is:
\beq\label{eq:selfdualaction}
S_{\text{self-dual}} = \frac{N_c}{4\pi^2} \int \dd^4 x \,\tr \left(G_{\alpha\beta}F^{\alpha\beta} - \tilde{\psi}_A^{\dot{\alpha}} D_{\alpha\dot{\alpha}} \psi^{\alpha A} + \frac{1}{4}\phi_{AB}D^2\phi^{AB} + \frac{1}{2}\phi^{AB}[\tilde{\psi}_{\dot{\alpha}A}, \tilde{\psi}^{\dot{\alpha}}_B]\right)~.
\eeq
The field $G_{\alpha\beta}$ is a Lagrange multiplier which serves to impose on-shell the self-dual condition of the field strength $F_{\alpha\beta}=0$.  
In order to complete the action of full SYM we need to add the non-self dual part which includes the remaining interaction terms including the complex conjugate of the Yukawa interaction in \eqref{eq:selfdualaction}. This is given by the chiral Lagrangian:
\begin{eqnarray}\label{eq:LintSpacetime}
L_{\text{int}}(x) &=& -N_c~\tr\left( \frac{1}{2}G_{\alpha\beta}G^{\alpha\beta}- \frac{1}{2}\phi_{AB}[\psi_\alpha^A,\psi^{\alpha B}]  - \frac{1}{32}[\phi_{AB},\phi_{CD}][\phi^{AB}, \phi^{CD}]\right)~.
\end{eqnarray}
This extra term now gives the on-shell condition $G_{\alpha\beta}\propto F_{\alpha\beta}$ which transforms the action \eqref{eq:N4aroundSDYM} into the more standard form of SYM.
As we explain below, $L_{\text{int}}$ is a chiral operator and belongs to the $1/2$-BPS multiplet
of the $20'$ operator or stress-tensor multiplet.\footnote{
One could incorporate a $\theta$-term by adding a term
$\sim \tau F_{\dot\alpha\dot\beta}F^{\dot\alpha\dot\beta}$ to the (anti-)self-dual part of the action,
where $\tau=\frac{\theta}{2\pi} + \frac{4\pi i}{g_{\rm YM}^2}$ is the complexified gauge coupling.
The coefficient of $L_{\rm int}$ would then involve $\bar\tau^{-1}$.
Formally, \eqref{eq:Lagrange_insertion} is then obtained by expanding at large $\bar\tau$
with fixed $\tau$. The nonperturative status of this method is unclear to the authors, but it certainly makes sense in perturbation theory which will be the focus of this paper.
}

In this formulation the Lagrangian insertion method gives the loop-integrands of determinant operators as:
\beq\label{eq:Lagrange_insertion}
\hspace{-2mm}
\left\langle  \prod_{i=1}^{n} \mathbb{D}(x_{i},y_{i},\theta_i)\right\rangle_{\!\!\!\scriptscriptstyle\text{SYM}}\!\!  = \sum_{\ell=0}^{\infty} \frac{(-g^{2})^\ell}{\ell!}\int \frac{\dd^4 x_{n+1}}{\pi^2}\cdots 
\frac{\dd^4 x_{n+\ell}}{\pi^2} \left\langle  \prod_{i=1}^{n}\mathbb{D}(x_{i},y_{i},\theta_i)\,\prod_{k=1}^{\ell }L_{\text{int}}(x_{n+k}) \right\rangle_{\!\!\!\scriptscriptstyle \SDYM}\!\!.
\eeq 
Notice that the ``integrand'' in this formulation is an exact correlation function in the self-dual theory.  

Let us briefly comment on our normalization of the self-dual action.  It is motivated by a desire to remove factors of $N_c$ from the planar two-point functions discussed below, as well as factors of $1/(4\pi^2)$ from propagators, so as to manifest the symmetry between $x$ and $y$ variables as much as possible:
\begin{equation} \label{normalization}
    \langle y_1{\cdot}\phi(x_1)\, y_2{\cdot}\phi(x_2)\rangle_\SDYM
    = \frac{d_{12}}{N_c} \quad\mbox{with}\quad
    d_{ij} \equiv \frac{2y_i{\cdot}y_j}{(x_i{-}x_j)^2} \equiv \frac{-y_{ij}^2}{x_{ij}^2}~.
\end{equation}

With foresight on the application to matrix duality on the loop integrands, we notice two challenges in this formulation. The first difficulty is that SDYM does not have a Gaussian action, but contains
cubic interactions. The second problem is that the chiral Lagrangian  $L_{\text{int}}$ is not a determinant operator. The first obstacle will be overcome by using the twistor formulation of the self-dual action \eqref{eq:selfdualaction}, which admits a Gaussian form.
The second challenge will be overcome by insisting to account for the $\theta$-dependence of all determinants, so that we can extract $L_{\rm int}(x)$ as a component of $\mathbb{D}(x,y,\theta)$.
This will make the right-hand-side of \eqref{eq:Lagrange_insertion} more symmetrical.

\subsection{Self-dual super Yang-Mills in twistor space}\label{sec:SDYMtwistor}

\begin{figure}[t]
\begin{center}
\begin{tikzpicture}[scale=2,thick]   
\draw [fill,red] (-1.43,.6) circle [radius=0.03];
\draw [fill,blue] (-1.15,1.2) circle [radius=0.03];
\node[scale=1,red] at (-1.6,.6)   {$x_i$ ~};
\node[scale=1,blue] at (-1.28,1.22)   {$x_j$ ~};

\draw[red,thick] (1.,0)--(3,1.);
\node[scale=1,red] at (3.1,1.1)   {$i$};
\draw[blue,thick] (1,1)--(1.8,.5);
\draw[blue,thick] (2,.4)--(3,-.22);
\node[scale=1,blue] at (3.1,-.22)   {$j$};

\draw [fill,black] (1.4,0.2) circle [radius=0.02];
\draw [fill,black] (2.5,0.75) circle [radius=0.02];
\draw [fill,black] (1.48,0.7) circle [radius=0.02];
\draw [fill,black] (2.75,-.065) circle [radius=0.02];
\node[scale=1] at (1.4,0.35)   {$Z_{i,1}$ ~};
\node[scale=1] at (2.5,0.9)   {$Z_{i,2}$ ~};
\node[scale=1] at (1.52,0.88)   {$Z_{j,1}$ ~};
\node[scale=1] at (2.8,.1)   {$Z_{j,2}$ ~};

\end{tikzpicture}
\end{center}
\caption{Map between points in Minkowski space
and lines in twistor space. The spacetime points $x_i$ and $x_j$ correspond to the $i^{\text{th}}$ and $j^{\text{th}}$ $\mathbb{CP}^1$ line respectively. Each line is fully determined by two twistors $Z_{a,1}, Z_{a,2}$, $a\in \{i,j\}$.}
\label{geometry_twistor}
\end{figure}
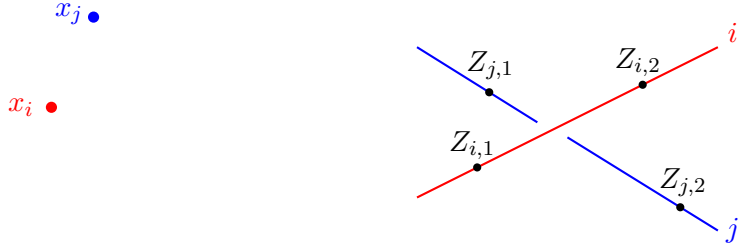

Twistor space is a complex 3-dimensional space in which spacetime points are represented by (complex) lines.
For the maximally supersymmetric theory, the 
super-twistor space $\mathbb{CP}^{3|4}$
is parameterized by homogeneous coordinates:
\beq
\mathcal{Z} = (Z^{I},\eta^{A})\quad\text{with}\quad I,A =1,2,3,4~.
\eeq
These are defined only projectively  so $\mathcal{Z}$ and  $c\,\mathcal{Z}$ are to be identified, for any nonzero complex
number $c$. The index on the fermions $\eta^A$ transforms in the fundamental of the $SU(4)$ R-charge symmetry, and the Lorentz indices in the bosonic part transform as Weyl spinors:
\beq\label{bosonic_twistor}
Z^{I} = \left(\lambda_{\alpha},\mu^{\dot{\alpha}}\right).
\eeq
The relationship with a point $(x,\theta)$ in spacetime is given by the incidence relations:
 \beq\label{incidence}
 \mu^{\dot{\alpha}} \,=\, x^{\dot{\alpha}\beta}\lambda_{\beta} \quad \text{and}\quad \eta^{A} = \theta^{A\alpha} \lambda_{\alpha}\,.
\eeq
For fixed $x$ and $\theta$, this defines a $\mathbb{CP}^1\in \mathbb{CP}^{3|4}$.
Conversely, given a line in twistor space, characterized by any two points $\mathcal{Z}_1$ and
$\mathcal{Z}_2$ on it, 
the incidence relation can be reverted. For example the bosonic $x$ can be extracted by forming the $4\times 4$ matrix
\beq
 Z^{I}_{1} Z^{J}_{2} - Z^{I}_{2} Z^{J}_{1}  \propto \begin{pmatrix}
 \epsilon_{\alpha\beta} & - i x_{\alpha}^{\dot{\beta}} \\ 
 i x_{\beta}^{\dot{\alpha}} & - \frac{1}{2}x^2\epsilon^{\dot{\alpha}\dot{\beta}}
\end{pmatrix}~.
\eeq
The bosonic distance $x_{ij}$ is proportional to a $4\times 4$ determinant,
\begin{equation}\label{bosonic_distance}
     x_{ij}^2 \sim \frac{1}{4}\epsilon_{IJKL} \epsilon^{ab} Z_{i,a}^I Z_{i,b}^J \epsilon^{cd}Z_{j,c}^K Z_{j,d}^L = \langle Z_{i,1}Z_{i,2} Z_{j,1}Z_{j,2} \rangle ~.
 \end{equation}

The Penrose \cite{doi:10.1063/1.1705200} transform relates fields
defined on spacetime and on twistor space.
All SYM fields fit within a single $(0,1)$-form connection $\mathcal{A}$ on twistor space, with a Grassman expansion of the form
\begin{align}\label{expansionA}
\mathcal{A}(Z,\bar{Z},\psi) &=  A (Z,\bar{Z}) + \eta^A \psi_{A}(Z,\bar{Z}) + \frac{1}{2!}\eta^A \eta^B \phi_{A B} (Z,\bar{Z})
\cr
&+ \frac{1}{3!} \epsilon_{ABCD} \eta^A \eta^B \eta^C \tilde{\psi}^D(Z,\bar{Z}) + \frac{1}{4!} \epsilon_{ABCD}\eta^A \eta^B\eta^C\eta^D G(Z,\bar{Z})
~.
\end{align}
As reviewed in 
\cite{Boels:2006ir,Mason:2010yk} (see also \cite{Witten:2003nn}),
the transformation can be defined off-shell in such a way that the self-dual action \eqref{eq:selfdualaction} maps to the holomorphic Chern-Simons action:
\beq \label{SDYM}
S_{\text{self-dual}} = N_c\!\int \Omega^{3|4} \,\text{tr}\left(\mathcal{A}\bar{\partial}\mathcal{A}+\frac{2}{3}\mathcal{A}^{3}\right) \,,
\eeq
where the measure is proportional to the unique holomorphic form on $\mathbb{CP}^{3|4}$:
\begin{equation}\label{measureCP34}
\Omega^{3|4} \equiv 
\frac{1}{(2\pi i)^3}
\frac{1}{4!}\epsilon_{IJKL} Z^I \dd Z^J \dd Z^K \dd Z^L \dd^4 \eta~.
\end{equation}
Notice that this part of the action is local in twistor space. On the contrary, the interaction part takes the form of an integral $\dd^4x$ over the moduli space of an embedded $\mathbb{CP}^1$, as in \eqref{eq:N4aroundSDYM}, with:
\bba\label{Lint twistor}
L_{\text{int}}(x) &=  N_c\int \dd^{8}\theta\log\int\,\mathcal{D}\alpha\,\mathcal{D}\beta\, \exp \left(\int_{\mathbb{CP}^{1}} \frac{\langle\lambda\,\dd\lambda\rangle}{2\pi i} \alpha(\lambda)\,(\bar{\partial}+\mathcal{A})\big|_{\mathbb{CP}^{1}}\,\beta(\lambda)\right) \\
&=N_c\int \dd^{8}\theta\log \det \left(\bar{\partial} + \mathcal{A}\right)\big{|}_{\mathbb{CP}^{1}}~,
\end{align}
where $\langle\lambda \, \dd \lambda\rangle \equiv \epsilon_{\alpha\beta}\lambda^{\alpha}\dd\lambda^\beta$.
The fields $\alpha$ and $\beta$ are respectively in the antifundamental and fundamental of the gauge group U($N_c)$.
In practice, this formula is usually formally expanded as
\begin{align}\label{expansion_intTerm}
\, (\log \det \left(\bar{\partial} + \mathcal{A}\right) - \log \det \bar{\partial} )\big|_{\mathbb{CP}^1}=  \sum_{m=1}^\infty\frac{(-1)^{m+1}}{m} \tr\left(\bar{\partial}^{-1}\mathcal{A}_1\ldots \bar{\partial}^{-1}\mathcal{A}_m\right)~.
\end{align}
In \cite{Witten:2003nn}, it was proposed that the twistor action
\eqref{SDYM} describes strings ending on a stack of $N_c$ ``space-filling" D5 branes in topological string theory, while $L_{\rm int}$ originates from D1 instantons. The fields $\alpha$ and $\beta$ were then interpreted as D1-D5 strings.
A version for non-supersymmetric Yang-Mills theories, where
$L_{\rm int}\propto {\rm tr}(G^2)$, was discussed recently in \cite{Costello:2022wso}.

A key feature of twistor space is its enlarged gauge redundancies: in one gauge, the relation to spacetime fields becomes particularly simple; in another gauge, self-dual interactions are linearized.  Both will play an important role below.

\paragraph{Spacetime gauge}
The spacetime or harmonic gauge \cite{Boels:2006ir,Boels:2007qn}
makes the connection to spacetime fields particularly simple.
It amounts to imposing that, for any line corresponding to a \emph{real} Euclidean point, the connection starts at order $\eta^2$:
\begin{equation} \label{A spacetime gauge}
\mathcal{A}\big|_{{\rm real\, }\mathbb{CP}^1} = \mathcal{O}(\eta^2)\,.
\end{equation}
The quadratic term is then related to the spacetime scalar field
through the conventional (bosonic) Penrose transform:
\begin{equation}\label{scalarTT}
\phi_{AB}(x)=\int_{\mathbb{C}\mathbb{P}^1}  \frac{\langle \lambda \,\dd \lambda\rangle}{2\pi i} h_{(x,\theta)}^{-1}(\lambda)\frac{\partial^2 \mathcal{A}(\mathcal{Z}_{(x,\theta)}(\lambda))}{\partial \eta^A \partial\eta^B}h_{(x,\theta)}(\lambda)\big|_{\theta=0}~
\qquad\mbox{(spacetime gauge)}~,
\end{equation}
where $h$ is a suitable flat section, see \cite{doi:10.1063/1.1705200,Boels:2006ir}.
Since the condition \eqref{A spacetime gauge} ensures that the expansion
\eqref{expansion_intTerm} terminates, this gauge also makes it relatively straightforward to demonstrate the equality between \eqref{eq:N4aroundSDYM} and \eqref{Lint twistor}, see \cite{Boels:2006ir,Boels:2007qn}.

We stress that the simple map \eqref{scalarTT} is only
valid in this special gauge.  A general $\mathbb{CP}^{3|4}$ gauge transformations will add fermions to it.
For example, by consideration of the half-BPS condition of certain operators or by using harmonic superspace, it was shown in
\cite{Koster:2016ebi,Chicherin:2016soh} that the more general form of the (super) Penrose transform is 
\begin{align}\label{penrosemap_phi}
&\hspace{-8mm}\phi_{AB}(x)=\int  \frac{\langle \lambda\, \dd \lambda\rangle}{2\pi i} h_{(x,\theta)}^{-1}(\lambda)\frac{\partial^2 \mathcal{A}(\lambda)}{\partial \eta^A \partial\eta^B}h_{(x,\theta)}(\lambda)\big|_{\theta=0}\cr
&\hspace{-8mm}+\int \frac{\langle \lambda\, \dd \lambda\rangle\langle \lambda' \,\dd \lambda'\rangle}{(2\pi i)^2\langle\lambda\,\lambda'\rangle}h_{(x,\theta)}^{-1}(\lambda)\frac{\partial \mathcal{A}( \lambda)}{\partial\eta^A}h_{(x,\theta)}^{-1}(\lambda)h_{(x,\theta)}(\lambda')\frac{\partial \mathcal{A}( \lambda')}{\partial\eta^{'B}}h_{(x,\theta)}(\lambda')\big|_{\theta=0} - (A\leftrightarrow B)~.
\end{align}

\paragraph{Axial or CSW gauge}
In axial or CSW gauge one chooses a particular but fixed reference twistor $\mathcal{Z}_*$ along which the gauge field $\mathcal{A}$ vanishes \cite{Cachazo:2004kj}. This choice is particularly useful since it removes the cubic interaction in the Chern-Simons action \eqref{SDYM}: self-dual Yang-Mills becomes a free theory. The propagator
in this gauge is \cite{Mason:2010yk}
\beq\label{twistor propagator}
\langle \mathcal{A}^{a}(\mathcal{Z}_{1})\,\mathcal{A}^{b}(\mathcal{Z}_{2})\rangle \, = \, 
\frac{1}{N_c}
\Delta_{*}(\mathcal{Z}_1,\mathcal{Z}_2)\,\delta^{ab}
\eeq
where
\beq\label{eq:delta44}
\Delta_{*}(\mathcal{Z}_1,\mathcal{Z}_2) \,=\,\bar{\delta}^{2|4}(\mathcal{Z}_1,\mathcal{Z}_2,\mathcal{Z}_{*})\,\equiv\, \frac{1}{(2\pi i)^2}
\int_{\mathbb{C}^2} \frac{\dd s}{s}\frac{\dd t}{t}\,\bar{\delta}^{4|4}(s\,\mathcal{Z}_1+t\,\mathcal{Z}_2+\,\mathcal{Z}_{*})~.
\eeq
Here $\bar\delta(z)$ is a $(0,1)$-form distribution which can be defined by either of the following properties (for smooth $f(z)$):\footnote{
Our definitions differ by $2\pi i$ factors from those in \cite{Mason:2010yk,Chicherin:2014uca},
but it appears to us that various factors of $2\pi i$ were also discarded there. We have tried to be internally consistent.
}
\beq\label{eq:defbardelta}
\bar\delta(z)=-\bar\partial \frac{1}{z}, \qquad
\int_{\mathbb{C}} \frac{dz \wedge \bar\delta(z)}{2\pi i} f(z) =f(0)~.
\eeq
The propagator \eqref{eq:delta44} inverts the $\bar{\partial}$-operator including appropriate gauge-fixing terms \cite{Mason:2010yk}:
\beq\label{GreensFunction}
\bar{\partial}\,\Delta_{*}(\mathcal{Z}_{1},\mathcal{Z}_2)+
\bar\delta^{3|4}(\mathcal{Z}_{1},\mathcal{Z}_{2})
+\bar\delta^{3|4}(\mathcal{Z}_{1},\mathcal{Z}_{*})
+\bar\delta^{3|4}(\mathcal{Z}_{2},\mathcal{Z}_{*}) =0\,.
\eeq
The fact that the propagator is supported on a $\delta$-function of complex codimension two greatly simplifies calculations.

\subsection{Half-BPS operators and their geometry}

We now introduce the half-BPS operators that we will study in this paper, as well as the kinematic restrictions that the BPS condition imposes in both spacetime and twistor space. 

\paragraph{A family of half-BPS operators}
We consider supersymmetric generalizations of the scalar field
determinant $\mathsf{D}(x,y)$ and its logarithm, which we denote as $\mathsf{O}(x,y)$:
\begin{alignat}{3}\label{eq:OpDefinitions}
    \mathsf{D}(x,y) &\equiv \det (1- y{\cdot}\phi(x)) ~  &&\mapsto ~ \mathbb{D}(x,y,\theta) &&\equiv \det(1-y{\cdot}\phi(x)) + \text{susy\, descendants}\cr 
    \mathsf{O}(x,y)&\equiv -\log \mathsf{D}(x,y) ~ &&\mapsto~ \mathbb{O}(x,y,\theta) &&\equiv -\log\mathbb{D}(x,y,\theta) ~\cr 
    \mathsf{O}^{\text{sp}}(x,y) &\equiv \mathsf{O}(x,y)+ \text{multi-traces} ~~ &&\mapsto ~ \mathbb{O}^{\text{sp}}(x,y,\theta) &&\equiv \mathbb{O}(x,y,\theta)+\text{multi-traces}~.
\end{alignat}
The supersymmetrized determinant $\mathbb{D}(x,y,\theta)$ and its single-trace version $\mathbb{O}(x,y,\theta)$ contain its bosonic counterparts $\mathsf{D}(x,y)$ and $\mathsf{O}(x,y)$ as their bottom components. Besides, they include an expansion in (chiral) superspace Grassmann coordinates $\theta$ which truncates at order $\theta^4$ and contains all (chiral) superdescendants.  The single-particle operators $\mathbb{O}^{\text{sp}}(x,y,\theta)$ will be discussed in detail in section \ref{sec:singleParticle}. 

Furthermore, our operators are parametrized by the null vector $y$ in R-charge space. They admit an infinite series expansion in powers of $y$. In $\mathbb{O}(x,y,\theta)$, the terms proportional to  $y^k$ form a supermultiplet dual to the $k$-th Kaluza-Klein mode of the supergraviton multiplet in AdS$_{5}\times S^5$. The (chiral) stress-tensor multiplet, dual to the supergraviton itself, can be obtained as the quadratic term in this series:
\beq \label{L from O}
\mathbb{O}(x,y,\theta)\big{|}_{y^2\text{ projection}} \,=\, \tr[(y {\cdot}\phi(x))^2]+ \cdots -\frac12(\theta{\cdot}y{\cdot}\theta)^2\frac{1}{N_c} \,{L}_{\text{int}}(x)~.
\eeq
The bottom scalar operator lives in the $20'$ representation of the SU$(4)$ R-charge
(ie. the symmetric traceless 2-tensor of ${\rm so}(6)$).
Here we also point out that its top component is the chiral Lagrangian used in the Lagrangian insertion method in \eqref{eq:Lagrange_insertion} to compute integrands.
This observation suffices to recognize that correlators of $\mathbb{O}(x,y,\theta)$'s contain all the integrands  as a special components of their $y$- and $\theta$-series.
This motivates our goal to study the correlators of $\mathbb{O}(x,y,\theta)$'s, or of determinants,
as a generalization of the integrands of \cite{Caron-Huot:2021usw}.

\paragraph{The half-BPS condition in chiral superspace}
Supermultiplets in $\mathcal{N}=4$ SYM can be described in a superspace that has eight chiral $\theta^{A}_{\alpha}$ and eight antichiral $\bar{\theta}_{A\bar{\alpha}}$ variables. Half-BPS multiplets, by definition, depend only on half of these variables (four chiral and four anti-chiral). In this paper we specialize to the chiral half by setting $\bar{\theta}=0$. (This will be sufficient to recover $L_{\rm int}$.)
The BPS condition  is parametrized by the R-charge polarization $y$, which in chiral superspace is:
\begin{equation}\label{halfBPSMink}
y^{AB}\frac{\partial}{ \partial \theta_\alpha^B}\,\mathbb{O}(x,y,\theta)=0~,\quad \forall ~\alpha=1,2,\quad A=1,2,3,4~.
\end{equation}
The above expression holds for the operator $\mathbb{O}(x,y,\theta)$ (\ref{eq:OpDefinitions}) or any other BPS operator. 
This gives seemingly eight conditions, however these are only four constraints due to the null condition of the polarization vector $\left(\epsilon_{ABCD}y^{AB}y^{CD}=0\right)$. In order to write these conditions without redundancy we use the twistor-like decomposition in R-charge space:
\beq\label{BPS_chiralsuperspace}
y^{AB} = \epsilon^{a'b'}\,Y^{A}_{a'} Y^{B}_{b'}~. 
\eeq
These four-component vectors serve to make a $SU(2)\times SU(2)'$ decomposition of the $SU(4)$ R-charge space. Using them we can split the chiral superspace in two halves:\footnote{In \cite{Chicherin:2014uca} this is called harmonic space decomposition, there $\theta$ and $\Psi$ in this equation are denoted by $\theta^{\pm}$.}
\begin{equation}\label{theta_decomp}
\theta_\alpha^A = W_a^A \theta^a_\alpha + Y^A_{a'}\Psi^{a'}_\alpha~,
\end{equation}
where the vectors $W$ are introduced to parametrize the unprimed $SU(2)$ copy. We assume the normalization $\epsilon_{ABCD}W^A W^B y^{CD}=1$.
Now the BPS condition can be understood as the independence on the four Grassmann variables $\Psi$:
\beq\label{BPSTwistor}
\frac{\partial}{\partial \Psi^{a'}_{\alpha}} \mathbb{O}(x,y,\theta)\,\equiv\,Y^{A}_{a'}\frac{\partial}{\partial \theta^{A}_{\alpha}} \mathbb{O}(x,y,\theta) = 0\qquad\text{with}\quad \alpha=1,2,\;a'=1,2\,.
\eeq
Hence the half-BPS operators live in the superspace spanned by the four Grassmann variables $\theta^{a}_{\alpha}$. Note that we use the same symbol $\theta$ with different indices to denote the superspace coordinate and its projection orthogonally to $Y$.

The half-BPS condition is readily uplifted to twistor space. We simply multiply both sides of \eqref{theta_decomp} by $\lambda$ leading to 
\begin{equation}\label{etaConstraint}
 \eta^A=
 \theta_\alpha^A \lambda^\alpha = W_a^A \theta^a_\alpha \lambda^\alpha + Y^A_{a'}\psi^{a'}~,
\end{equation}
where we used the incidence relation (\ref{incidence}) and defined $\psi^{a'} \equiv \Psi^{a'}_\alpha \lambda^\alpha$. Therefore, in twistor space, the condition (\ref{halfBPSMink}) that an operator is annihilated by four supercharges can be written as
\begin{equation}\label{eq:BPSconstraint}
\frac{\partial}{\partial \psi^{a'}}\mathbb{O}(x,y,\theta)
\,\equiv\,Y^{A}_{a'} \frac{\partial}{\partial\eta^A}\mathbb{O}(x,y,\theta)=0\qquad\text{with}\quad a'=1,2~.
\end{equation}
This admits a natural geometrical interpretation:  a half-BPS operator can be labelled by a $\mathbb{CP}^{1|2}\subset \mathbb{CP}^{3|4}$ subspace spanned by superline coordinates $(\lambda, \psi)$.  This observation will form the basis of our construction below.

We finish by providing an example of a supersymmetrized operator constructed in twistor space.  According to \cite{Chicherin:2014uca} the chiral stress-tensor multiplet can be obtained from the same $\alpha$-$\beta$ path integral that gives $L_{\rm int}$, see \eqref{Lint twistor},  but integrating over only half of the super-space coordinates instead. We integrate only over the four $\Psi^{a'}_{\alpha}$ in \eqref{theta_decomp}
\bba
\mathbb{O}(x,y,\theta)\big{|}_{y^2\text{ proj.}} &= \int \dd^{4}\Psi\log\int\,\mathcal{D}\alpha\,\mathcal{D}\beta\, \exp \left(\int_{\mathbb{CP}^{1}}\langle\lambda\,\dd\lambda\rangle \alpha(\lambda)\,(\bar{\partial}+\mathcal{A})\big|_{\mathbb{CP}^{1}}\,\beta(\lambda)\right)\\
&= \tr[(y {\cdot}\phi(x))^2]+  \cdots -\frac12\left(\theta^A_\alpha\  y_{AB}\  \theta^{B\alpha} \right)^2 \,\frac{1}{N_c}{L}_{\text{int}}~,\nonumber
\end{align}
where on the second line we also made explicit the indices in \eqref{L from O}\footnote{
The Grassmann product can also be written as $-\frac12\left(\theta^A_\alpha\  y_{AB}\  \theta^{B\alpha} \right)^2= (\theta^a_\alpha)^4$, see footnote on page 3 of \cite{Chicherin:2014uca}.
}.

\section{Supersymmetrized determinants in twistor space}\label{section3}

In this section we present a twistor-space description of determinant operators, which extends the half-BPS operators described
in \cite{Caron-Huot:2021usw} by including their supersymmetry descendants:
\beq\label{eq:bosonDet}
\mathbb{D}(x, y,\theta)\,\equiv\,\det(1- y{\cdot}\phi(x)) + \text{susy descendants}~.
\eeq
Here $y$ is the six-dimensional R-symmetry vector and $\phi$ is a six vector combining the six scalar fields of $\mathcal{N}=4$ SYM; $x$ is a spacetime point that translates to twistor space via (\ref{incidence}) and $\theta$ are four chiral Grassmann variables.

Our main proposal is that this  determinant is computed by a simple Gaussian model supported on the $\mathbb{CP}^{1|2}$ line associated with the data $(x, y,\theta)$:
\beq\label{mainConjecture}
\hspace{-1mm}\mathbb{D}(x,y,\theta)=\;\int \mathcal{D}\alpha \mathcal{D}\beta\, \exp\left({\int_{\mathbb{CP}^{1|2}}
\frac{\langle\lambda\, \dd \lambda \rangle{\dd^2 \psi}}{2\pi i} \,\alpha(\lambda,\psi)\left(\bar\partial+\mathcal{A} +\bar\delta^{1|2}_{\mu,\lambda}\right)\Big{|}_{\mathbb{CP}^{1|2}}\,
\beta(\lambda,\psi)}\right)~,
\eeq
where $\alpha$ and $\beta$ are
fermionic superfields on $\mathbb{CP}^{1|2}$, transforming respectively in the antifundamental and fundamental of $U(N_c)$, and homogeneous of degree zero in $\lambda, \psi$. Here $(\lambda,\psi)$ are any set of homogeneous coordinates on $\mathbb{CP}^{1|2}$; $\lambda$ could be chosen to be the top two components of the ambient twistor coordinates \eqref{bosonic_twistor}, but this is not needed.

Eq.~\eqref{mainConjecture} is a natural generalization of the $\mathbb{CP}^1$ model reviewed in \eqref{Lint twistor}.  A main new feature is the presence of the $\mathbb{CP}^{1|2}$ delta function $\bar\delta^{1|2}_{\mu,\lambda}$ which localises the $\alpha(\lambda,\psi)$-$\beta(\lambda,\psi)$ superfields onto an arbitrary reference point $\mu\in \mathbb{CP}^{1|2}$, however we will see that the result is actually independent of $\mu$.\footnote{The reference point $\mu$ should not be confused with the bottom component of the bosonic twistor $Z$ \eqref{bosonic_twistor}.}  Therefore, the formula \eqref{mainConjecture}
is manifestly covariant under supertranslations in $\theta$
and automatically satisfies the half-BPS condition \eqref{BPSTwistor}, which is simply translation in $\psi$. 


In the following subsections we explain the different ingredients of $\mathbb{D}(x,y,\theta)$ and derive the $\alpha$-$\beta$ propagators on a single $\mathbb{CP}^{1|2}$ line, which we will use to justify the proposal and compare with the existing literature.
We then connect the theory in \eqref{mainConjecture} with the so-called $m=2$ amplituhedron \cite{Lukowski:2020dpn}, and give the Feynman rules for the case where multiple $\mathbb{CP}^{1|2}$ lines are connected via gauge field propagators. 
The techniques and relations discussed in this section will be used in subsequent sections to calculate correlation functions of determinant $\mathbb{D}(x,y,\theta)$ operators.

\subsection{The Gaussian model on $\mathbb{CP}^{1|2}$}

A superfield on $\mathbb{CP}^{1|2}$ can be characterized by its
expansion
\begin{align} \label{expansion of alpha}
\alpha(\lambda,\psi) &= \alpha^{(0)}(\lambda) +
\alpha^{(1)}_{1}(\lambda)\psi^{1}+\alpha^{(1)}_{2}(\lambda)\psi^{2} + \alpha^{(2)}(\lambda)\,\psi^1\psi^2~,
\end{align}
and similarly for $\beta$ and $\mathcal{A}$.
We parenthesize the subscript on the different components for future convenience.

For the next few steps we will assume for concreteness that $\mu$ has no Grassmann components.
Then the $\bar{\delta}^{1|2}_{\lambda,\mu}$ term in the action \eqref{mainConjecture} integrates simply to $\alpha^{(0)}(\mu)\beta^{(0)}(\mu)$.
The equation of motion from varying $\alpha^{(2)}$ thus gives simply
\begin{equation} \label{EOM b0}
    (\bar\partial+\mathcal{A}^{(0)})\beta^{(0)}(\lambda) = 0~.
\end{equation}
Since $\beta^{(0)}$ has homogeneity degree zero, this equation always admits a solution.  Were it not for the $\mu$-dependent term in \eqref{mainConjecture}, this ``zero mode'' would make the path integral ill-defined (zero). Instead, we will find a finite but $\mu$-dependent propagator.

Let us discuss the perturbative evaluation of the path integral in powers of $\mathcal{A}$.
We start with the propagator with $\mathcal{A}=0$ and study various component equations. Varying $\alpha^{(1)}_{2}(\lambda)$ for example gives (the $\bar\delta$ distribution is defined in \eqref{eq:defbardelta}):
\begin{equation}
 \bar\partial_\lambda   \langle  \beta^{(1)}_{1}(\lambda)\ \alpha^{(1)}_{2}(w)\rangle + \bar\delta(\langle \lambda\, w\rangle)=0 \quad\Rightarrow\quad
    \langle \beta^{(1)}_{1}(\lambda)\ \alpha^{(1)}_{2}(w)\rangle = \frac{1}{\langle \lambda\, w\rangle}\,.
\end{equation}
Since $\alpha^{(1)}_{1}$ has homogeneity minus one in $\lambda$, this solution is unique.  Varying $\alpha^{(0)}$ yields a similar equation which now features the $\mu$ term:
\begin{equation} \label{EOM a0}
   \bar\partial_\lambda \langle \beta^{(2)}(\lambda)\  \alpha^{(0)}(w)\rangle  =
    \bar\delta(\langle \lambda\, w\rangle) -\bar\delta(\langle \lambda\, \mu\rangle) \langle \beta^{(0)}(\mu)\ \alpha^{(0)}(w) \rangle \,.
\end{equation}
To solve this we must find a function
$\langle \beta^{(2)}(\lambda)\  \alpha^{(0)}(w)\rangle$ which is homogeneous of degree minus two in $\lambda$ and has simple poles at most at $w$ and $\mu$.  Such a function is unique, and exists only when its two poles have equal and opposite residues. Eq.~\eqref{EOM a0} thus simultaneously determines two propagators:
\begin{equation}
    \langle \beta^{(2)}(\lambda)\ \alpha^{(0)}(w)\rangle
    = \frac{\langle w\, \mu\rangle}{\langle \lambda\, w\rangle\langle\mu\, \lambda\rangle}~,
    \qquad \langle \beta^{(0)}(\lambda)\,  \alpha^{(0)}(w)\rangle=1\,.
\end{equation}
Here we have also used that the second average is constant thanks to the $\beta^{(0)}$ equation of motion.   Other components of the propagator are similar; we see that all are finite but some are $\mu$-dependent.
Summing up all components into \eqref{expansion of alpha},
we find that the superpropagator takes a simple form:
\begin{equation}\label{CP12_prop}
\langle \beta(\lambda_i,\psi_i)\ \alpha(\lambda_j,\psi_j)\rangle  = 1+ R(\lambda_i, \lambda_j, \mu) \equiv {\Delta}(\lambda_i,\lambda_j,\mu)~,
\end{equation}
where using the notation $\langle \lambda_i\, \lambda_j\rangle = \epsilon_{ab} \lambda_i^a\lambda_j^b$ the $R$-invariant is defined by 
\begin{equation}\label{r123}
R(\lambda_i, \lambda_j, \lambda_k) \equiv \frac{\delta^{0|2}\left(\langle \lambda_{i} \,\lambda_j\rangle \psi_{k} + \langle \lambda_{j} \,\lambda_{k}\rangle \psi_{i} + \langle \lambda_k\,\lambda_{i}\rangle \psi_{j} \right)}{\langle \lambda_{i}\, \lambda_j\rangle\langle \lambda_j\,\lambda_{k}\rangle\langle\lambda_{k}\,\lambda_{i}\rangle}~.
\end{equation}
This is a natural superconformal invariant which imposes the
(fermionic) constraint that three points in $\mathbb{CP}^{1|2}$ lie on the same $\mathbb{CP}^1$.
This object was used in earlier twistor-space studies (see for example eq.~3.45 of \cite{Chicherin:2014uca}); it generalizes a similar five-index object in $\mathbb{CP}^{3|4}$ that plays an important role for planar scattering amplitudes \cite{Arkani-Hamed:2009nll, Mason:2009qx}.
The $R$-invariant is antisymmetric in its three arguments and vanishes if $\lambda_i = \lambda_j$.  It is easy to check
that \eqref{CP12_prop} reduces to all the cases quoted above, and solves all the equations of motion from \eqref{mainConjecture}.  Note that we have
suppressed color indices and the propagator \eqref{CP12_prop} is an $N_c\times N_c$ matrix proportional to the identity.

Let us now briefly explain why the path integral \eqref{mainConjecture}
is independent of $\mu$.  This can be understood from the zero-mode of \eqref{EOM b0}, which is associated with a shift symmetry of the kinetic term in the action. Since this shift symmetry is explicitly broken by the $\mu$-dependent term, it can be used to relate theories with different $\mu$'s. Namely, consider the following change of variables:
\begin{equation}
    \beta(\lambda) \mapsto \beta(\lambda) +
    H(\lambda,\nu)\beta(\nu)-H(\lambda,\mu)\beta(\mu)~,
\end{equation}
where $H$ is a $N_c\times N_c$ gauge link satisfying $(\bar\partial_\lambda + \mathcal{A})H=0$ and $H(\lambda,\lambda)=1$.
The only effect of this shift is to replace $\beta(\mu)$ by $H(\mu,\nu)\beta(\nu)$ in \eqref{mainConjecture}. Applying a similar shift to $\alpha$ completes the replacement of $\mu$ by $\nu$.

Using the propagator $\Delta$ we can integrate out the $\mathbb{CP}^{1|2}$ fields $\alpha$ and $\beta$ and obtain the (logarithm of) our determinant as a sum over single-trace vertices: 
\beq\label{eq:logDetoutAlphaBeta}
-\log\mathbb{D}(x,y,\theta)=
\sum_{n=1}^{\infty}\frac{1}{n}\int \prod_{a=1}^{n}\Omega_a^{1|2}\Delta(\lambda_{a},\lambda_{a+1},\mu)\,\text{tr}\left(\prod_{a=1}^{n}\mathcal{A}\left(\mathcal{Z}(\lambda_{a},\psi_{a},\theta)\right)\right)\bigg{|}_{\mathbb{CP}^{1|2}}
\eeq
with the measure $\Omega^{1|2}_{a} \,\equiv\frac{\langle\lambda_{a}\,\dd\lambda_{a} \rangle\,\dd^{2}\psi_{a}}{2\pi i}$
and cyclic identification $\lambda_{n+1}=\lambda_1$.
The overall minus sign is from the fermion loop. Note that the left-hand-side is precisely the single-trace generating function $\mathbb{O}(x,y,\theta)$ from \eqref{eq:OpDefinitions}, so this formula can also be used to calculate correlators of single-trace half-BPS operators.

From this expression, we can easily confirm
that the model computes the claimed determinant.
Because \eqref{mainConjecture} is manifestly supersymmetric, it suffices to demonstrate this in the $\theta=0$ case, ie. for the leading component in \eqref{eq:bosonDet}.  Furthermore, because the model is manifestly invariant under $\mathbb{CP}^{3|4}$ gauge transformations,
it suffices to consider the spacetime gauge \eqref{A spacetime gauge}. In this gauge, $\mathcal{A}\big|_{\mathbb{CP}^{1|2}}\sim \psi^2$, so that we can ignore all the $R$ terms in the propagator \eqref{CP12_prop} (which contain too many fermions) and set $\Delta\mapsto 1$:
\begin{align}\label{eq:logDetoutAlphaBeta sg}
\log\mathbb{D}(x,y,0) &= -\sum_{n=1}^\infty \frac{1}{n} \text{tr}
\prod_{a=1}^n \int_{\mathbb{CP}^{1|2}} \Omega_a^{1|2}
\mathcal{A}\left(\mathcal{Z}(\lambda_{a},\psi_{a},0)\right)
\qquad \mbox{(spacetime gauge)}
\nonumber\\ &= -\sum_{n=1}^\infty \frac{1}{n} \text{tr}
\left[ (y{\cdot}\phi(x))^n\right] = \log\det(1-y{\cdot}\phi(x))~,
\end{align}
where the integral was performed using
the simple instance of the Penrose transform in \eqref{scalarTT}.

We stress that the steps in \eqref{eq:logDetoutAlphaBeta sg} are only valid in a special gauge.
They suffice to establish our proposal thanks to its invariance under $\mathbb{CP}^{3|4}$ gauge transformations.  But in general, the $R$ terms in the propagator cannot be discarded and the expansion of \eqref{eq:logDetoutAlphaBeta} contains for example twistor-space fermions even when $\theta=0$.
This is consistent with the findings of \cite{Koster:2016ebi,Chicherin:2016soh}, who arrived at the supersymmetrized Penrose transform in \eqref{penrosemap_phi} by imposing the half-BPS condition, or by considering SYM theory in harmonic chiral superspace, respectively.
In our approach, the BPS condition is manifest and all terms are automatically generated by expanding the general formula \eqref{eq:logDetoutAlphaBeta}.

The $\mathbb{CP}^{1|2}$ model \eqref{mainConjecture} seems rather unique and unaffected by many of the ambiguities which were noted in Witten's original $\alpha\beta$ model on $\mathbb{CP}^1$ \cite{Witten:2003nn}.  For example, one could try assigning other homogeneity degrees to $\alpha$ and $\beta$ such that the product $\alpha\beta$ has degree zero (as needed for the action to make sense), but that would create more zero-modes and we would no longer be able to find a sensible propagator.
We could also choose $\alpha^{(0)}$ and $\beta^{(0)}$ to be bosons instead of fermions, but that would simply replace $\mathbb{D}$ by $1/\mathbb{D}$.

\subsection{Amplitudes on a single $\mathbb{CP}^{1|2}$ and the $m=2$ amplituhedron}

The $R$-invariant defined above satisfies many interesting identities, which will be useful in calculations below.  In fact, many of these identities have been discussed previously in the context of the so-called $m=2$ amplituhedron \cite{Arkani-Hamed:2013jha,Lukowski:2020dpn}. As we now elaborate on, single-trace correlation functions (of the current conjugate to $\mathcal{A}$) in the model \eqref{mainConjecture} are precisely the $m=2$ scattering amplitudes.

The main properties of the $R$-invariants are the four-term identity,
where the index in $\lambda_{a}$ distinguishes points on the $\mathbb{CP}^{1|2}$ line:\footnote{It can be proven component-by-component using the Schouten identity
\begin{equation}
\langle \lambda_4\, \lambda_1\rangle\langle \lambda_2\,\lambda_3\rangle + \langle \lambda_4\, \lambda_3\rangle\langle \lambda_1\,\lambda_2\rangle + \langle \lambda_4\, \lambda_2\rangle\langle \lambda_3\,\lambda_1\rangle =0~.
\end{equation}}
\begin{equation}
\label{four term identity}
    R(\lambda_1, \lambda_2, \lambda_3) - R(\lambda_1, \lambda_2, \lambda_4) +
    R(\lambda_1, \lambda_3, \lambda_4) - R(\lambda_2, \lambda_3, \lambda_4) =0~,
\end{equation}
together with fusion rules like:
\begin{equation}\label{eq:fusionRule}
    R(\lambda_1,\lambda_2,\lambda_3)R(\lambda_1,\lambda_3,\lambda_4)
\equiv R(\lambda_1,\lambda_2,\lambda_3,\lambda_4) =
    R(\lambda_2,\lambda_3,\lambda_4)R(\lambda_1,\lambda_2,\lambda_4)~,
\end{equation}
which state that the middle object (defined by that relation) is cyclically invariant.
The latter also implies $R(\lambda_1,\lambda_2,\lambda_3)R(\lambda_1,\lambda_2,\lambda_4)R(\lambda_1,\lambda_3,\lambda_4)=0$.  Together, these imply the following identities satisfied by the propagator:
\bba\label{eq:DeltaIdentities}
{\Delta}(\lambda_i,\lambda_j,\mu){\Delta}(\lambda_j,\lambda_i,\mu)= 1~,\quad 
{\Delta}(\lambda_i,\lambda_j,\mu){\Delta}(\lambda_j,\lambda_k,\mu){\Delta}(\lambda_k,\lambda_i,\mu) = \Delta(\lambda_i,\lambda_j,\lambda_k)~.
\end{align}
These simplify respectively the amplitudes that appear in the $n=2$ and $n=3$ cases of 
\eqref{eq:logDetoutAlphaBeta}.
More generally,
these identities are useful to manifest the $\mu$-independence of $\mathbb{CP}^{1|2}$ amplitudes,
for example the four-point amplitude is:
\begin{align}\label{n4}
&\Delta(\lambda_1,\lambda_2,\mu)\Delta(\lambda_2,\lambda_3,\mu)\Delta(\lambda_3,\lambda_4,\mu)\Delta(\lambda_4,\lambda_1,\mu) \cr 
&= \Delta(\lambda_1,\lambda_2,\mu)\Delta(\lambda_2,\lambda_3,\mu)\times\Delta(\lambda_3,\lambda_1,\mu)\Delta(\lambda_1,\lambda_3,\mu)\times\Delta(\lambda_3,\lambda_4,\mu)\Delta(\lambda_4,\lambda_1,\mu)\cr 
&= \Delta(\lambda_1,\lambda_2,\lambda_3)\Delta(\lambda_1,\lambda_3,\lambda_4)\cr 
&= 1+\big[R(\lambda_1,\lambda_2,\lambda_3)+ R(\lambda_1,\lambda_3,\lambda_4)\big] + R(\lambda_1,\lambda_2,\lambda_3,\lambda_4)~.
\end{align}
In the second line we used the first identity in \eqref{eq:DeltaIdentities} to introduce a unit factor and then used the second identity \eqref{eq:DeltaIdentities} twice to obtain the third line. In the last line we simply grouped terms by their Grassmann degree.

For general $n$, by using the identities \eqref{eq:DeltaIdentities} recursively we obtain a manifestly $\mu$-independent expression for the $n$-point amplitude in the model \eqref{mainConjecture}:
\beq\label{generaln}
\prod_{a=1}^{n}\Delta(\lambda_{a},\lambda_{a+1},\mu)= \prod_{a=2}^{n-1}\Delta(\lambda_{1},\lambda_{a},\lambda_{a+1})~.
\eeq
Notice we made the arbitrary choice of taking $\lambda_1$ as a special reference when using the identities. We could make a different choice and obtain different but equivalent expressions that are also $\mu$-independent.
As a final example, this gives the five-point amplitude as
\begin{align}\label{5points}
&\hspace{-5mm}{\Delta}(\lambda_1,\lambda_2,\mu){\Delta}(\lambda_2,\lambda_3,\mu){\Delta}(\lambda_3,\lambda_4,\mu){\Delta}(\lambda_4,\lambda_5,\mu){\Delta}(\lambda_5,\lambda_1,\mu) \cr
=&\  1 + \big[R(\lambda_1,\lambda_2,\lambda_3) + R(\lambda_1,\lambda_3,\lambda_4) + R(\lambda_1,\lambda_4,\lambda_5)\big]\cr
&+ \big[R(\lambda_1,\lambda_2,\lambda_3,\lambda_4) + R(\lambda_1,\lambda_3,\lambda_4,\lambda_5) + R(\lambda_1,\lambda_2,\lambda_3)R(\lambda_1,\lambda_4,\lambda_5)\big] \cr 
&+  R(\lambda_1,\lambda_2,\lambda_3,\lambda_4,\lambda_5)~.
\end{align}
The $n$-point amplitude enjoys the following properties:
it is superconformal-invariant as a function of $(\lambda_i,\psi_i)\in \mathbb{CP}^{1|2}$;
it is cyclically invariant;
all its singularities are single poles of the form $1/\< \lambda_i\ \lambda_{i+1}\>$.
The $n$-point amplitude reduces to the $(n{-}1)$-point one in either of two ways:
by taking residues on adjacent poles we get the residue of $R(\lambda_i\,\lambda_{i+1},\mu)$
times the lower-point amplitude, or by taking the supersymmetric coincidence limit $(\lambda_j,\psi_j)\to (\lambda_i,\psi_i)$ we get smoothly to the lower-point amplitude.
These properties, which could be used to uniquely determine the amplitudes in a recursive fashion, are manifested by \eqref{generaln}. 

The term of Grassmann degree $2k$ is called the $m=2$ amplitude $\mathcal{A}_{n,k,2}$. (For instance, $\mathcal{A}_{5,2,2}$ in example 4.14 of \cite{Lukowski:2020dpn} appears to coincide with the $\sim R^2$ terms in \eqref{5points}. More generally, we believe that \eqref{5points} solves the recursion in \cite{Bao:2019bfe}). The $m=2$ case was studied as a simplified model for the original amplituhedron, which involves $(m{=}4)$-dimensional supertwistors.

\subsection{Multiple interacting $\mathbb{CP}^{1|2}$ superlines}

In this section we show how the computation of correlation functions of our determinants simplifies in the CSW gauge. We summarize the simplified Feynmann rules in section \ref{sec:Feynman}. 

In the CSW gauge the twistor action is Gaussian and the correlation functions are computed by the partition function:
\begin{multline}\label{eq:correlatorAction}
    \left\langle \prod_{i=1}^{n}\mathbb{D}(x_i,y_i,\theta_i) \right\rangle_{\scriptscriptstyle\text{SDYM}}
    = \int \mathcal{D}\mathcal{A} \exp\bigg[\int\Omega^{3|4}\left( \frac{1}{2}\mathcal{A}\,\bar{\partial}\mathcal{A}\right)\cr 
    -\sum_{i=1}^{n}\sum_{m=1}^{\infty}\frac{1}{m}\int \prod_{a=1}^{m}\Omega_{ia}^{1|2}\Delta(\lambda_{ia},\lambda_{i\,a+1},\mu_i)\,\text{Tr}\left(\prod_{a=1}^{m}\mathcal{A}\left(\mathcal{Z}_{ia}\right)\right)\bigg{|}_{\mathbb{CP}^{1|2}}\bigg]~.
\end{multline}
Since we are now dealing with various operators we introduce an extra label for each correspondent $\mathbb{CP}^{1|2}$ superline. In \eqref{eq:correlatorAction}, the first label in the twistor $\mathcal{Z}_{ia}$ distinguishes the superline and the second label corresponds to a point in that line with coordinates $(\lambda_{ia},\psi_{ia})$.

Each superline is specified by two supertwistors $\mathcal{Z}_{i,1},\,\mathcal{Z}_{i,2}$, which we can use to parametrize any other supertwistor $\mathcal{Z}=\left(Z,\eta\right)$  in the same $\mathbb{CP}^{1|2}$
line using the coordinates $(\lambda,\psi)$ as:
\bba \label{line parametrization}
Z_{i}\left(\lambda\right)&= \lambda^{1}\,Z_{i,1}+\lambda^{2}\,Z_{i,2}\,, \\
\eta_{i}\left(\lambda,\psi\right)\,&=\,\lambda^{\alpha} \left(\theta_{i}\right)^{a}_{\alpha}\,W_{i,a}+ \, \psi^{a'}\,Y_{i,a'}\,.
\end{align}
The relationship between $Z$, $W$ and $Y$ and the spacetime coordinates are the incidence relations reviewed in (\ref{incidence}).

\begin{figure}[t]
\begin{center}
\begin{tikzpicture}[thick]
\draw[blue] (-7,1)-- (-6,2.5);
\draw[blue] (-4,1)-- (-5,2.5);
\draw[decorate,decoration=snake] (-6.57,1.65) -- (-4.44,1.65);
\draw [fill,black] (-6.57,1.65) circle [radius=0.06];
\draw [fill,black] (-4.44,1.65) circle [radius=0.06];
\node[scale=.6,blue] at (-7,.9)   {$\mathbb{CP}^{1|2}$ ~};
\node[scale=.6,blue] at (-3.8,.9)   {$\mathbb{CP}^{1|2}$ ~};
\end{tikzpicture}
\caption{Two operators, with support on $\mathbb{CP}^{1|2}$ lines, exchange gauge field propagators in the $\mathbb{CP}^{3|4}$ bulk.
}
\label{connectedCP1}
\end{center}
\end{figure}
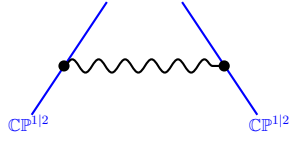

In order to compute correlation functions at Born level we have to connect the vertices $\text{Tr}\left(\mathcal{A}\cdots \mathcal{A}\right)$ from the operator $i$ with the ones in operator $j$ using the $\mathbb{CP}^{3|4}$ propagator for the superfield $\mathcal{A}$, see figure \ref{connectedCP1}. As presented in section \ref{sec:SDYMtwistor}, this propagator is a delta function which will allow us to easily perform the $\mathbb{CP}^{1|2}$ integrals. After this integration we will obtain some effective Feynman rules which only include an on-shell version of the $\mathbb{CP}^{1|2}$ propagator $\Delta$ and the spacetime propagator $d_{ij}$. Later we show that this simplification allows us to partially resum the propagators between a pair of operators as a geometric series and obtain an effective ten-dimensional propagator $D_{ij}$.

In order to perform the $\mathbb{CP}^{1|2}$ integrals, we first 
pull back the bulk propagator \eqref{twistor propagator}
to the lines parametrized by \eqref{line parametrization}. In these new variables, the bulk propagator neatly factorizes:
\begin{multline}\label{eq:deltaBosonicCP34}
\delta^{4}(\eta_i(\psi_a)+\eta_j(\psi_b)+\eta_{*})\, = \,\langle Y_{i1}Y_{i2} Y_{j1} Y_{j2}\rangle\,\delta\left(\psi_a^1+\psi_{ij}^{1}\right)\, \delta\left(\psi_a^{2}+\psi_{ij}^{2}\right)\delta\left(\psi^{1}_{b}+\psi_{ji}^{1}\right)\, \delta\left(\psi^{2}_{b}+\psi_{ji}^{2}\right)~,\cr \delta^{4}(Z_{i}(\lambda)+Z_{j}(w)+Z_{*})\, = \, \frac{\delta\left(\lambda^{1}+\lambda_{ij}^{1}\right)\, \delta\left(\lambda^{2}+\lambda_{ij}^{2}\right)\delta\left(w^{1}+\lambda_{ji}^{1}\right)\, \delta\left(w^{2}+\lambda_{ji}^{2}\right)}{\langle Z_{i1}Z_{i2} Z_{j1} Z_{j2}\rangle}~,
\end{multline}
where the four-bracket factors, defined in (\ref{bosonic_distance}), are Jacobians due to the change of variables. For instance, the on-shell  bosonic coordinates $\lambda_{ij}$ are obtained after solving the four-component equation:
\beq
 \underbrace{\lambda^{1}\,Z_{i,1}+\lambda^{2}\,Z_{i,2}}_{Z_{i}(\lambda)}+ \underbrace{w^{1}\,Z_{j,1}+w^{2}\,Z_{j,2}}_{Z_j(w)}+Z_{*} =0~.
\eeq
By solving for $\lambda$ and $w$ we find the solutions $\lambda_{ij}$ and $\lambda_{ji}$, which we refer to as the on-shell $\mathbb{CP}^{1|2}$ coordinates. The components of these spinors can be written in terms of four-brackets: 
\beq\label{tilde_lambda}
\lambda_{ij}^{1}\,=\, -\frac{\langle Z_{*} Z_{i,2} Z_{j,1} Z_{j,2}\rangle}{\langle Z_{i,1} Z_{i,2} Z_{j,1} Z_{j,2} \rangle}\qquad\text{and}\qquad \lambda_{ij}^{2}\,=\, -\frac{\langle Z_{i,1} Z_{*} Z_{j,1} Z_{j,2}\rangle}{\langle Z_{i,1} Z_{i,2} Z_{j,1} Z_{j,2} \rangle}~,
\eeq
and similarly for the fermionic coordinates
\beq\label{onshellpsi}
\psi_{ij}^{1}\,=\, -\frac{\langle E_{ij}\,Y_{i,2} Y_{j,1} Y_{j,2}\rangle}{\langle Y_{i,1} Y_{i,2} Y_{j,1} Y_{j,2} \rangle}\qquad\text{and}\qquad \psi_{ij}^{2}\,=\, -\frac{\langle Y_{i,1} E_{ij}\,Y_{j,1} Y_{j,2}\rangle}{\langle Y_{i,1} Y_{i,2} Y_{j,1} Y_{j,2} \rangle}~,
\eeq
where
\beq
E_{ij}= \,\lambda^{\alpha}_{ij}\,\left(\theta_{i}\right)^{a}_{\alpha}\,W_{i,a}+\lambda^{\alpha}_{ji}\,\left(\theta_{j}\right)^{a}_{\alpha}\,W_{j,a} ~.
\eeq
The factorized form  \eqref{eq:deltaBosonicCP34},
schematically $\bar\delta^{2|4}\propto \bar\delta^{1|2}\times \bar\delta^{1|2}$,
allows to easily perform the integrals in \eqref{eq:delta44}. The integration variables are fully localized to the on-shell values in (\ref{tilde_lambda}, \ref{onshellpsi}). In this way we now replace the $\mathbb{CP}^{1|2}$ propagator $\Delta$ in \eqref{CP12_prop} by its ``on-shell" version:
\begin{equation}
\Delta^{i}_{jk\mu} \equiv \Delta(\lambda_{ij},\lambda_{ik},\mu_{i})~.
\end{equation}

\begin{figure}[t]
 \begin{minipage}[h]{1\textwidth}
 \centering 
  \includegraphics[width=1\textwidth]{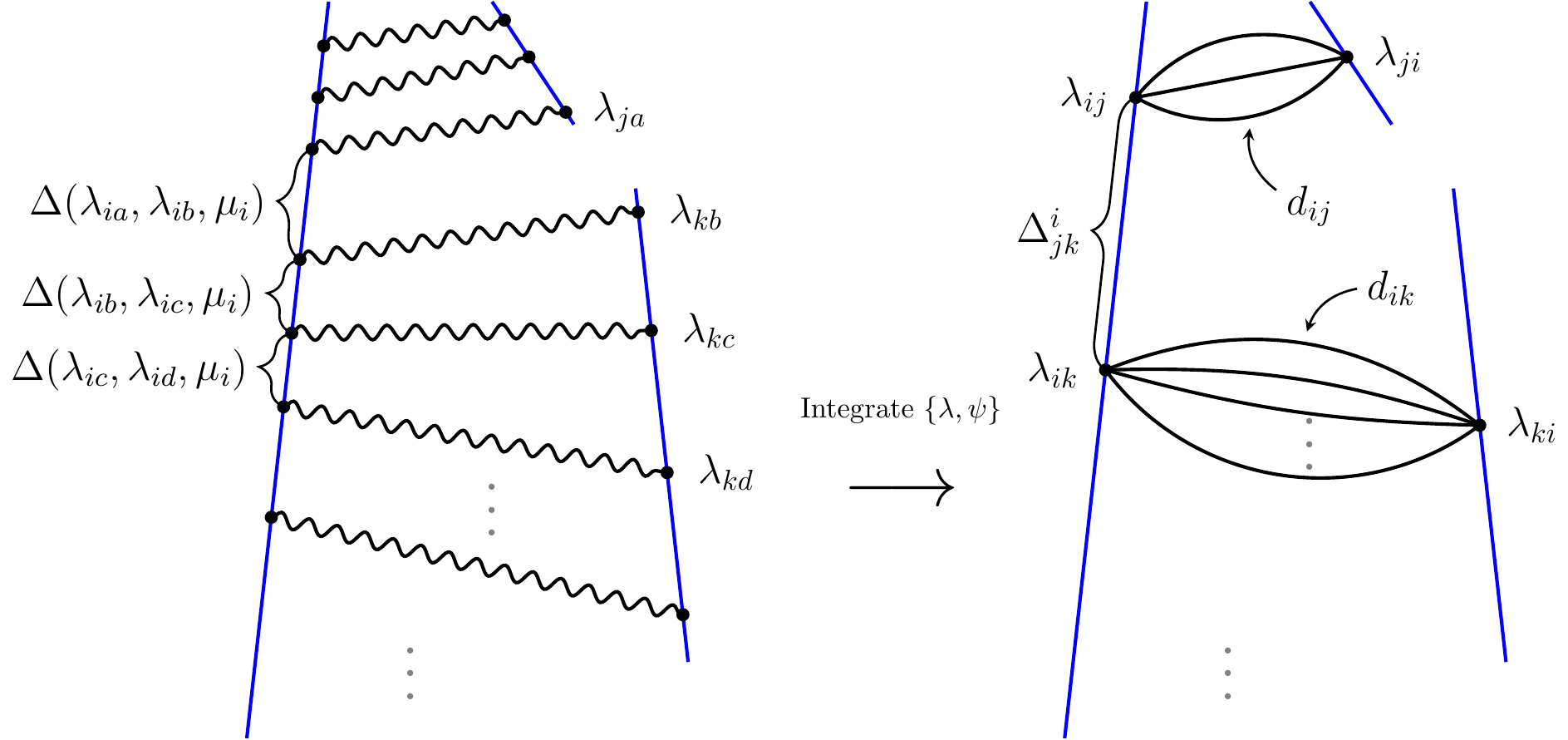}
 \end{minipage}
 \caption{Feynman rules in twistor space. The blue lines are different $\mathbb{CP}^{1|2}$ superlines connected via $\mathbb{CP}^{3|4}$ gauge field propagators. Left: On each $\mathbb{CP}^{1|2}$ line with coordinates $(\lambda_i,\psi_i)$ we have a series of $\alpha$-$\beta$ propagators (\ref{CP12_prop}) stretching between points $a,b,c\ldots$; $\mu_i$ is the reference position. The black wavy propagators connecting two superlines are the $\mathbb{CP}^{3|4}$ gauge field propagators which simplify to (\ref{eq:delta44}) in CSW gauge. Right: The Delta-function in the  $\mathbb{CP}^{3|4}$ gauge field propagator localises the $\mathbb{CP}^{1|2}$ $\alpha$-$\beta$ propagators $\Delta$ onto on-shell coordinates (\ref{tilde_lambda}-\ref{onshellpsi}). Note that we are evaluating the propagators counterclockwise. After stripping out these Delta-functions the effective propagator that connects the $i^{\text{th}}$ and $j^{\text{th}}$ superlines is given by $d_{ij}$.}
 \label{fig:onshellprops}
 \end{figure}

After stripping out the localizing delta functions from the $\mathbb{CP}^{3|4}$ propagator, we are left with an effective propagator formed by the residual Jacobian factors (four-brackets) in \eqref{eq:deltaBosonicCP34}:
\beq\label{dij_propagator}
\frac{\langle Y_{i1}Y_{i2} Y_{j1} Y_{j2}\rangle}{\langle Z_{i1}Z_{i2} Z_{j1} Z_{j2}\rangle} \sim -\frac{y_{ij}^2}{x_{ij}^2} \equiv d_{ij}\,.
\eeq
This coincides with the standard four-dimensional scalar propagator. It is represented by solid black lines on the right panel of figure \ref{fig:onshellprops}. 
 
As shown in figure \ref{fig:onshellprops}, we can have multiple bulk propagators stretching between different points of the same pair of superlines. We distinguish these points at superline $i$ with labels $a,b,c,\cdots$ and $\mathbb{CP}^{1|2}$ coordinates $(\lambda_{ia},\psi_{ia})$, $(\lambda_{ib},\psi_{ib})$, $\cdots$. After integrating over these coordinates many points degenerate into the same on-shell values thanks to the bulk propagator. All end points of the bulk propagators stretching between superlines $i$ and $k$ collapse to points $(\lambda_{ik},\psi_{ik})$ and $(\lambda_{ki},\psi_{ki})$ respectively. In figure \ref{fig:onshellprops}, the off-shell coordinates in line $i$ and $k$ collapse to the same on-shell values: $\lambda_{ib}=\lambda_{ic}=\cdots \overset{\text{on-shell}}{=} \lambda_{ik}$ and $\lambda_{kb}=\lambda_{kc}=\cdots \overset{\text{on-shell}}{=}  \lambda_{ki}$. 
Because of this degeneration the superline propagators stretching between these points become trivially unity (the R-invariant (\ref{r123}) vanishes when two labels coincide). For instance, in the notation of figure \ref{fig:onshellprops}, we have:
\beq\label{eq:Deltaikk}
\Delta(\lambda_{ib},\lambda_{ic},\mu_{i})\quad\overset{\text{on-shell}}{\longrightarrow}\quad\Delta^{i}_{kk} \equiv \Delta(\lambda_{ik},\lambda_{ik},\mu_{i}) = 1
\eeq
while the non-trivial on-shell propagator $\Delta^{i}_{jk}$ stretches in superline $i$ between the bundles of bulk propagators $i$-$j$ and $i$-$k$ :
\beq\label{eq:DeltaOnShell}
\Delta(\lambda_{ia},\lambda_{ib},\mu_{i})\quad\overset{\text{on-shell}}{\longrightarrow}\quad\Delta^{i}_{jk} \equiv \Delta(\lambda_{ij},\lambda_{ik},\mu_{i}) \quad\text{with}\quad j\neq k~.
\eeq
 Up to now the Feynman rules are simplified to include the on-shell superline propagators in \eqref{eq:DeltaOnShell} and effective bulk propagators $d_{ij}$ coming within bundles. We define a bundle as a group of propagators order in a planar fashion. Then the genus of the full graph is determined by the way the group of bundles organizes.
\begin{figure}[t]
 \begin{minipage}[h]{1\textwidth}
 \centering 
  \includegraphics[width=1\textwidth]{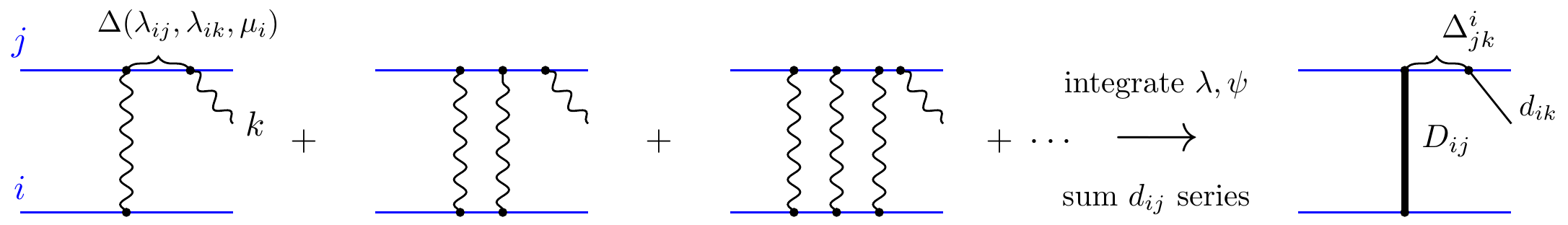}
 \end{minipage}
 \caption{Resummation of propagators within the same bundle and emergence of the effective propagator $D_{ij}$ with manifest ten-dimensional denominator. We include one extra propagator connecting to a third twistor line, such that the symmetry factors give the geometric series in \eqref{resummationD}. In the absence of this extra propagator we obtain the two-point function series in \eqref{eq:Otwopoint} instead.}
 \label{fig:resummationDprop}
 \end{figure}
 
We can further simplify these rules by noticing that the superline propagators $\Delta^{i}_{jk}$ are blind to the number of propagators in the bundles. Therefore it is possible to perform the resummation of all propagators within each bundle connecting a pair of operators. For three- and higher-point functions the symmetry factors are such that we obtain a geometric series for each bundle:
\beq\label{resummationD}
d_{ij} + (d_{ij})^2 + (d_{ij})^3+\cdots = \frac{d_{ij}}{1-d_{ij}}= \frac{-y_{ij}^2}{x_{ij}^2+y_{ij}^2} \equiv D_{ij} ~.
\eeq
In this way we obtain a new effective propagator with a ten-dimensional denominator.

\subsection{Summary of Feynman rules}\label{sec:Feynman}

For practical purposes, here we summarize and exemplify the effective Feynman rules obtained in the previous section. The new effective rules  use the propagator $D_{ij}$ for a bundle between operators $i$ and $j$, accounting for the resummation in \eqref{resummationD}. 
Besides, each operator $i$ comes with a weight $V^{i}_{j_{1}\cdots j_{m}}$ where $j_{m}$ labels the operators it connects with. This weight is given by the product of  on-shell propagators $\Delta^{i}_{j_{m}j_{m+1}}$ stretching between a pair of bundles $i$-$j_{m}$ and $i$-$j_{m+1}$:
 \bba\label{FeynmanVertex}
 V^{i}_{j_1,j_2,j_3,\cdots,j_{n}}\,=\,\raisebox{-.5\height}{\includegraphics[scale=1.25]{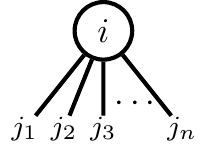}} \,=\, \Delta^{i}_{j_1j_2}\, \Delta^{i}_{j_2j_3}\, \Delta^{i}_{j_3j_4}\cdots \Delta^{i}_{j_{n}j_{1}}\quad \text{with }j_{m}\neq j_{m+1}
\end{align}
Notice consecutive bundles need to connect with different operators, but non-consecutive bundles can connecy to the same pair of operators. This already happens in the planar limit and we present examples when computing the four- and five-point function in the section below. 

Furthermore we can use properties in \eqref{generaln} to show that these vertices are $\mu_{i}$-independent:
\beq\label{eq:Vijkl}
V^{i}_{j k l}=\Delta^{i}_{j k l}\quad\text{and}\quad
V^{i}_{j_1\,j_2\,j_3\,\cdots\,j_{n}}= \prod_{m=2}^{n-1}\Delta^{i}_{j_1j_{m}j_{m+1}}~.
\eeq
Under these definitions the vertices with one or two bundles are trivial:
\beq\label{eq:V12trivial}
V^{i}_{j}=1 \qquad \text{and}\qquad V^{i}_{jk}=\Delta^{i}_{jk}\Delta^{i}_{kj}=1~.
\eeq
Finally we present an example on how to use our effective rules to read off the contribution of a graph to a five-point correlator:
\beq\label{eq:exampleVrules}
\raisebox{-.4\height}{\includegraphics[scale=.7]{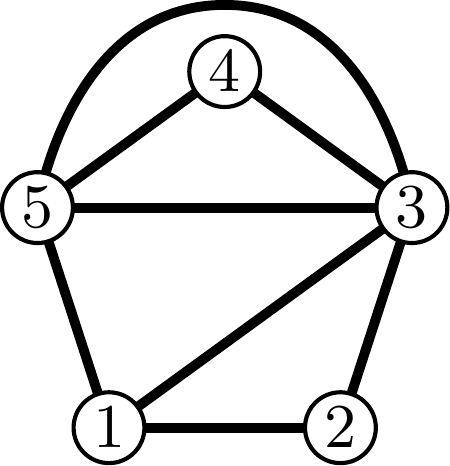}} \, = \,V^{1}_{235}\,V^{2}_{13}\,V^{3}_{12545}\,V^{4}_{35}\,V^{5}_{1343}\times D_{12}D_{23}D_{34}D_{45}D_{51}D_{13}D_{35}^2 ~. 
\eeq
In this example the fermionic part can be further simplified and written in terms of the superline propagator as:
\beq
V^{1}_{235}\,V^{2}_{13}\,V^{3}_{12545}\,V^{4}_{35}\,V^{5}_{1343}=\Delta^{1}_{235}\times 1 \times \Delta^{3}_{125}\times 1 \times 1~.
\eeq
In section \ref{sec:5NMHV} we reproduce this same example with the corresponding dual graph using the $\rho$-matrix integral derived in section \ref{section5}.

\section{Single-trace correlators in twistor space}\label{section4}

In this section we compute the correlation functions of the single-trace operator $\mathbb{O}$, whose construction in twistor space is shown in  \eqref{eq:logDetoutAlphaBeta}. We do this in the large $N_c$ limit by using the effective Feynman rules found in the previous section.

We will focus on connected correlators and organize them in components according to their Grassman degree:
\beq\label{eq:Gcorre}
G_{n}\equiv\langle\mathbb{O}_1 \cdots \mathbb{O}_{n}\rangle_c = G_{n,0}+G_{n,1}+\cdots+G_{n,n-4}
\eeq
where the subscript $c$ denotes the connected piece and $G_{n,k}$ is the part of the result of order $4k$ in the Grassman coordinate $\theta$ of superspace. We refer to it as the N$^{k}$MHV component. By superconformal symmetry this series truncates at $k=n-4$. However this is not obvious in the twistor Feynman rules and it requires some special identities satified by the R-invariants as we show in the  $n=4$ example in section \ref{sec:CF4}.

In the large $N_c$ limit the connected $n$-point correlator scales as:
\begin{equation}\label{eq:largeNcScale}
    \langle \mathbb{O}_1\cdots \mathbb{O}_n\rangle_c \sim  N_c^{2-n} \qquad \mbox{(genus 0)}~. 
\end{equation}
In order to compute this correlator at order $N_c^{2-n-2g}$ using the rules obtained above,
we first list all the graphs with $n$ vertices and genus $g$ including their inequivalent permutations.
The contribution of each graph is then found using the effective Feynman rules in section \ref{sec:Feynman}:  we weight the graph with propagators $D_{ij}$ for each bundle and the weight $V^{i}_{j_{1},j_{2}\cdots}$ in \eqref{eq:Vijkl} for each vertex. Combining all these contributions we obtain the final result for the correlator.

The task of listing graphs is straightforward for low $n$ but rapidly becomes delicate for larger numbers of operators, even at genus zero, on which we will focus in this section.
To partly alleviate this, for four and higher points, we consider only correlators in the
so-called single-particle basis \cite{Aprile:2020uxk}.
This basis is obtained by adding multi-traces to the single-trace operators, schematically:
\beq
\mathbb{O}^{\text{sp}} \,=\, \mathbb{O} + \text{multi-traces}~.
\eeq
The precise relation will be described below.  The upshot is that,
in the planar limit, the list of graphs which contribute in the single-particle basis is a strict
subset of the general graphs. This subset is obtained by omitting all the graphs that have at least one vertex with degree one. See \eqref{eq:G3Feynman} for an example on these degree-one graphs used for computing the three-point function in the single-trace basis.

In what follows we compute the two- and three-point functions in the single-trace basis in subsection \ref{sec:G23}. We then introduce the generating function of single-particle operators  in \ref{sec:singleParticle} and compute their four- and five-point functions in \ref{sec:CF4} and \ref{sec:CF5}, by evaluating the graphs in figures \ref{fig:fourskeletons} and \ref{fig:fiveskeletons}.

\subsection{Two- and three-point correlators}\label{sec:G23}

For the two-point correlator the resummation described around \eqref{resummationD} is modified by an extra symmetry factor of the single-trace operators. For higher-point functions this extra symmetry of the trace is broken due to the presence of extra propagators connecting to a third operator as shown in figure \ref{fig:resummationDprop}. However, in the absence of such extra connections the resummation for the two-point function contains an extra $1/k$ factor when we have $k$ propagators. This series is resummed to a logarithm:
\beq\label{eq:Otwopoint}
G_{2}=\langle \mathbb{O}_{1}\mathbb{O}_{2} \rangle_c =  d_{12} +\frac{d_{12}^2}{2}+\frac{d_{12}^3}{3}+\cdots = -\log(1-d_{12}) = \log(1+D_{12})~ + \mathcal{O}(N_c^{-2})~,
\eeq
where the absence of Grassmann dependence is due to the trivial superline propagators $\Delta^{1}_{22}=1$ and $\Delta^{2}_{11}=1$, as explained around \eqref{eq:Deltaikk}.

To confirm this result we can make a projection of the series to compare with known results for individual half-BPS operators. By taking the term of degree $(y_1y_2)^k$ on both the left- and right-hand side of \eqref{eq:Otwopoint}
for example we find
\beq
 \left\langle \frac{\tr[(y_1{\cdot}\phi(x_1))^k]}{k} \frac{\tr[(y_2{\cdot}\phi(x_2))^k]}{k}\right\rangle_c
= \frac{1}{k} d_{12}^k + \mathcal{O}(N_c^{-2})
\eeq
which is in precise agreement with a direct calculation
of Wick contractions using the spacetime propagator \eqref{normalization}.
The $1/k$ factors on the left come from expanding the logarithm in the definition of $\mathbb{O}$ (see \eqref{eq:introsingletrace}), and one of them is cancelled by the number of contractions.

Starting at three- and higher-point functions the effective Feynman rules of section \ref{sec:Feynman} apply without modifications. The three-point function only receives contributions from two distinct topologies: triangle and line.
\bba\label{eq:G3Feynman}
N_c\,G_{3} = N_c\,\langle \mathbb{O}_{1} \mathbb{O}_{2} \mathbb{O}_{3} \rangle  &= \,\raisebox{-.5\height}{\includegraphics[scale=1.25]{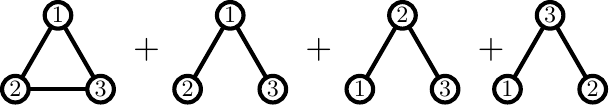}} \\
&=D_{12}D_{23}D_{13} + D_{12}D_{13} +D_{12}D_{23}+D_{13}D_{23}~.
\end{align}
Since all vertices have degree one or two then the Grassman dependence is trivial.  Non-trivial Grassmann dependence starts when graphs include operators with at least three bundles or more. This is the case for the four-point and five-point function, which we address in more detail in the following sections. Before that, we introduce the single-particle basis which allow us to drop all graphs with degree-one vertices (for instance the last three in \eqref{eq:G3Feynman}).

\subsection{Conversion to single-particle basis}\label{sec:singleParticle}

The ``single trace'' operators $\mathbb{O}$ are not orthogonal to multi-traces. Many formulas will be greatly simplified by adding multiples of double-traces, in such a way
as to make the operators orthogonal to multi-traces.  A general recipe to construct these \emph{single-particle} operators $\mathbb{O}^{\rm sp}$
is explained in \cite{Aprile:2020uxk}.
From eq.~(33) there, we obtain the first-order recipe
\begin{equation}
   {\rm Tr}[(y{\cdot}\phi)^k]^{\rm sp} = {\rm Tr}[(y{\cdot}\phi)^k] - \frac{1}{2N_c}\sum_{\ell=1}^{k-1}
  {\rm Tr}[(y{\cdot}\phi)^\ell]\  {\rm Tr}[(y{\cdot}\phi)^{k-\ell}]+ \mbox{higher traces or $1/N_c$}~.
\end{equation}
Multiplying by $1/k$ and summing over $k$, we can express a single-particle generating function $\mathbb{O}^{\rm sp}$
in terms of the single-trace one in \eqref{eq:introsingletrace}:
\begin{equation} \label{single-particle}
    \mathbb{O}^{\rm sp} \equiv \mathbb{O} -\frac{1}{2N_c}\left( y{\cdot}\partial_y \mathbb{O}\right)^2  + \mbox{higher traces or $1/N_c$}~.
\end{equation}
The derivatives simply pick up the homogeneity degree in $y$,
cancelling the factor $1/[\ell(k-\ell)]$ from the square of
\eqref{eq:introsingletrace}.

Importantly, the definition \eqref{single-particle}
is compatible with the planar scaling \eqref{eq:largeNcScale}.
In a $k$-point correlator, for example,
the second term in \eqref{single-particle} will generate
in the planar limit a product of two $a$- and $b$-point
connected correlators with $a+b=n+1$, thus contributing 
at the same order as the first term:
\begin{equation}
\frac{1}{N_c} N_c^{2-a} N_c^{2-b} = N_c^{2-n} \,.
\end{equation}
This means that to relate the genus-zero $n$-point correlators in the two basis, we only need
to know the genus-zero connected correlators with $m\leq n$ points.
Let us denote as $G_{n}^{\rm sp}$ the connected correlator of $n$ single-particle operators.
The relation between the $n=3$ correlators in the two basis, for example, is:
\begin{equation}\begin{aligned}
    G_{3}^{\text{sp}}-G_{3}
&= -\frac{1}{N_c}
    \left[y_1{\cdot}\frac{\partial}{\partial y_1}\langle \mathbb{O}_1\mathbb{O}_2\rangle\right]
    \left[y_1{\cdot}\frac{\partial}{\partial y_1}\langle \mathbb{O}_1\mathbb{O}_3\rangle\right]+\mbox{2 cyclic}
\\ &= -\frac{1}{N_c} (D_{12}D_{13}+D_{12}D_{23}+D_{13}D_{23}) +O(N_c^{-3})~, \label{shift 3pt}
\end{aligned}\end{equation}
where we have used the two-point function \eqref{eq:Otwopoint}.
Using the three-point function \eqref{eq:G3Feynman}, this gives the simple result
\begin{equation}
N_c\,
G_{3}^{\text{sp}}=
D_{12}D_{13}D_{23} + \mathcal{O}(N_c^{-2})~.
\end{equation}
Comparing with \eqref{eq:G3Feynman}, we see that this has simply removed the graphs with a vertex of valency one. Taking the coincidence limit of this result $(x_3,y_3)\to (x_2,y_2)$,
we also confirm immediately the orthogonality property $\< \mathbb{O}_1^{\rm sp}\mathbb{O}_2^2\>=0$.

To go to higher-points, we need terms with more traces
in \eqref{single-particle}, albeit we only need
the leading large-$N_c$ limit of the coefficient of each multi-trace.
By working out more examples of eq.~(33) of \cite{Aprile:2020uxk} we observe a simple pattern, giving:
\begin{equation}\begin{aligned} \label{single particle general}
    \mathbb{O}^{\rm sp}(x,y,\theta) &\equiv
    \mathbb{O}(x,y,\theta) +
    \sum_{k=1}^\infty \frac{(-1)^k(y{\cdot}\partial_y+1)_{k-1}}{N_c^k(k+1)!}
    \left( y{\cdot}\partial_y \mathbb{O}(x,y,\theta)\right)^{k+1} + \mbox{non-planar}
\\ &=    
    \mathbb{O} -\frac{1}{2N_c}\left( y{\cdot}\partial_y \mathbb{O}\right)^2 + \frac{(y{\cdot}\partial y+1)}{3!N_c^2}
    \left( y{\cdot}\partial_y \mathbb{O}\right)^3-\frac{(y{\cdot}\partial y+1)(y{\cdot}\partial y+2)}{4!N_c^3}\left( y{\cdot}\partial_y \mathbb{O}\right)^4+\ldots
\end{aligned}\end{equation}
We stress that the construction of \cite{Aprile:2020uxk} is exact in $N_c$ but here we have only worked
out the coefficients to the accuracy needed to compute genus-zero correlators of arbitrary multiplicity.

\begin{table}
\begin{minipage}[t]{1\textwidth}
 \centering 
\def\arraystretch{1}
\resizebox{6\totalheight}{!}{
\begin{tabular}{c|c|c|c|c|c}
$n$ & $\quad 3\quad$  &  $\quad 4\quad$ & $\quad 5\quad$ & $\quad 6\quad$ & $\quad 7\quad$ \\   \hline 
\#\;\text{primitive seeds} &1  &  3 & 10 & 49 & 332 \\   \hline 
\#\;\text{graphs} & 1  &  4 & 21 & 216 & 3278 
\end{tabular}
}
 \end{minipage}
 \caption{Counting of graphs whose vertices have valency greater or equal to two on each vertex.  Our graphs are obtained after incorporating decorations to each primitive seed due to the inner structure of vertices (single-traces) and edges (bundles of propagators), as discussed in the text.}
 \label{tab:graphcounting}
 \end{table}

Before we compute the four- and five-point correlators in the single-particle basis, it is helpful to describe the generation of relevant graphs. Diagrammatically, the switch to the single-particle basis removes any graph containing a vertex of valency one. 
Therefore, for single-particle correlators, we only need to list graphs that have degree-two and higher. This is still a challenging task even in the planar limit, since this number of graphs still grows exponentially with the number of operators.

Our strategy is to start by enumerating $n$-node connected planar graphs with minimum degree at least two, which we refer to as primitive seeds.
This is precisely the description of the OEIS sequence \href{https://oeis.org/search?q=A054381}{A054381} \cite{oeis:2023}  and their counting is shown in the first row of table \ref{tab:graphcounting} for different number of points. The primitive seeds for $n=4$ and $n=5$ correspond respectively to the first 3 graphs in figure \ref{fig:fourskeletons} and the first 10 graphs in figure \ref{fig:fiveskeletons}. 
 In order to go from the primitive seed to our desired list of graphs we need to include two types of modifications, as a consequence of the inner structure that we associate to vertices and edges. First, in our  context vertices represent single-traces. Hence, we should distinguish different orderings of the edges around a vertex, modulo cyclic identifications. Second, our edges represent bundles containing an infinite number of propagators, so we should allow for splitting of edges. These extra features require two decorations on the primitive seed. 
 The first is that we must choose a cyclic ordering for the edges around each vertex, compatible with the graph being drawn on the sphere. Thus, for example, the diagonal in the second graph of figure \ref{fig:fourskeletons} can be placed either along the front or back face of the square. The second decoration is that certain edges (which represent planar bundles of propagators) can be split into two or more non-adjacent edges while maintaining planarity, as in the fourth graph of figure \ref{fig:fourskeletons}.
By adding these decorations to the primitive seeds and keeping a single representative of each permutation orbit, we find the four graphs in figure \ref{fig:fourskeletons}. Similarly, we find the last eleven graphs in figure \ref{fig:fiveskeletons}, making a total of 21 graphs for $n=5$.
The numbers of sphere graphs modulo permutations are shown in the second row of table \ref{tab:graphcounting}. Once these graphs are obtained, we compute the correlator by summing over all inequivalent permutations of each decorated graph.

\subsection{Four-points}\label{sec:CF4}

\begin{figure}[t]
 \begin{minipage}[h]{1\textwidth}
 \centering 
  \includegraphics[width=1\textwidth]{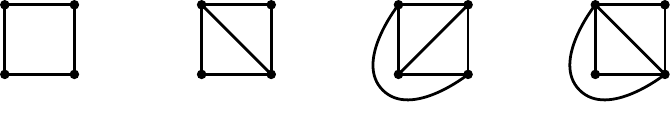}
 \end{minipage}
 \caption{Graphs with sphere topology for the planar four-point correlator in single-particle basis. The first three graphs contain a single bundle (or none) for each pair of points. The fourth graph is obtained from the second graph by splitting the diagonal into two bundles which are topologically inequivalent. One bundle goes on the front and the second on the back of the sphere.}
 \label{fig:fourskeletons}
 \end{figure}

In order to compute the four-point correlator in the single-particle basis we need to label the vertices in the four topologies of figure \ref{fig:fourskeletons} and consider all inequivalent permutations.
 \noindent
These permutations come in numbers: 3, 6, 2 and 12 respectively.  Using the effective Feynman rules in \eqref{FeynmanVertex} we define the following functions for each topology in figure \ref{fig:fourskeletons}:
\bba
I^{(1)}_{1234} \,&=\, \raisebox{-.4\height}{\includegraphics[scale=1.6]{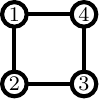}} \, = \, D_{12}D_{23}D_{34}D_{41} \label{eq:dress41} \\ 
I^{(2)}_{13;24} \,&=\, \raisebox{-.6\height}{\includegraphics[scale=1.6]{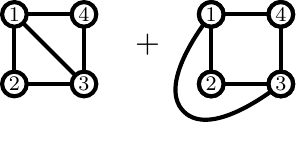}}  \,=\, D_{12}D_{23}D_{34}D_{41} D_{13}\,\left( \Delta^{1}_{234}\,\Delta^{3}_{124}\,+\,\Delta^{1}_{324}\,\Delta^{3}_{214}\right) \nonumber\\
\,&=\,2\prod\limits_{i<j}D_{ij}\times \frac{1+R^{1}_{234}R^{3}_{124}}{D_{24}} \label{eq:dress42}\\
I^{(3)}_{1234} \,&=\, \raisebox{-.6\height}{\includegraphics[scale=1.6]{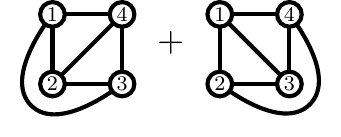}} =  \prod\limits_{i<j}D_{ij}\,\left(\Delta^{1}_{243}\,\Delta^{2}_{134}\,\Delta^{3}_{142}\,\Delta^{4}_{123}
\,+\,\Delta^{1}_{234}\,\Delta^{2}_{143}\,\Delta^{3}_{124}\,\Delta^{4}_{132}
\right)\nonumber\\
&=   \,2 \prod\limits_{i<j}D_{ij}\, \left(1-R^{1}_{234}R^{2}_{134}+R^{1}_{234}R^{3}_{124}-R^{2}_{134}R^{3}_{124}-R^{1}_{234}R^{4}_{123}+R^{2}_{134}R^{4}_{123}-R^{3}_{124}R^{4}_{123}\,\right.\nonumber\\
&\qquad\qquad\qquad\quad\left.+\,R^{1}_{234}R^{2}_{134}R^{3}_{124}R^{4}_{123}\right) \label{eq:dress43}\\
I^{(4)}_{13;24} \,&=\, \raisebox{-.6\height}{\includegraphics[scale=1.6]{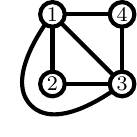}} \, = \, D_{12}D_{23}D_{34}D_{41} D_{13}^{2}\times\,\underbrace{V^{1}_{3234}}_{1}\,\underbrace{V^{3}_{1214}}_{1}\,~. \label{eq:dress44}
\end{align}
where we made direct use of the triviality of the valence two vertex, see \eqref{eq:V12trivial} and replaced the valence three vertices by $V^{i}_{jkl} = \Delta^{1}_{jkl}$. The only valence four vertices appear in the fourth topology and their Grassmann part reduces to one because it has repeated bundles: $V^{1}_{ijik}= \Delta^1_{iji}\Delta^1_{iki} =1$. Notice also that we already added one permutation in the definitions of $I^{(2)}$ and $I^{(3)}$, so we only need to consider six and one permutations of these functions respectively.  These groupings allow us to cancel the terms of Grassman degree two and six thanks to the antisymmetry of the $R$-invariant:  $R^l_{ijk}= -R^l_{jik}$. 
Summing over all inequivalent permutations of \eqref{eq:dress41} to \eqref{eq:dress44}  we obtain the single-particle correlator as:
\bba\label{eq:G4sp}
N_c^2\,G^{\text{sp}}_{4} \,&=\,\left(  I^{(1)}_{1234} + I^{(1)}_{1324} +I^{(1)}_{1243}  \right)+ \left(I^{(2)}_{12;34}+I^{(2)}_{13;24}+I^{(2)}_{14;23}+I^{(2)}_{23;14}+I^{(2)}_{24;13}+I^{(2)}_{34;12}\right)\nonumber\\
&\qquad\, +\,I^{(3)}_{1234}\,+\,\left(I^{(4)}_{12;34}+\text{5 permutations}\right) ~.
\end{align}
Considering the Grassmann degree two of the $R$-invariant we can organize the supercorrelator \eqref{eq:G4sp}  in components: 
\beq
G^{\text{sp}}_{4} = G^{\text{sp}}_{4,0} + G^{\text{sp}}_{4,1} + G^{\text{sp}}_{4,2}~,
\eeq
where $G_{4,1}$ and $G_{4,2}$ are the NMHV and N$^{2}$MHV components with Grassmann degree four and eight respectively. These components contain the loop-integrands for the two and three-point correlators and they should vanish according to the partial non-renormalization theorem. This holds in both the single-particle and single-trace basis. In order to verify this statement we will need some identities which we present below.

First, we compute the bosonic MHV component by setting $R\to0$:
\begin{equation}\begin{aligned}\label{eq:G4MHVsp}
N_c^{2}\,G_{4,0}^{\text{sp}}&= \,D_{12}D_{23}D_{34}D_{14}(1 + 2 D_{13}+ D_{13}^{2} + 2 D_{24} + D_{24}^2) + (1\leftrightarrow 2) + (1\leftrightarrow 4)\\
&\qquad\quad+ 2 D_{12}D_{13}D_{14}D_{23}D_{24}D_{34}  + \mathcal{O}(N_c^{-2})~.
\end{aligned}\end{equation}
The NMHV component receives contributions only from the second and third topologies:
\bba
N_c^2\,G_{4,1}^{\text{sp}} &= \left(I^{(2)}_{12;34}+\text{5 permutations}\right)\big{|}_{R^{2}\text{ part}}  \,+\,  I^{(3)}_{1234}\big{|}_{R^{2}\text{ part}}  \nonumber\\
&=   \,2 \prod\limits_{i<j}D_{ij}\, \left(\left(1+\frac{1}{D_{24}}\right)R^{1}_{234}R^{3}_{124}+ \text{5 more terms}\right).
\end{align}
The parenthesis turns out to identically vanish:
\beq\label{sixterms}
G_{4,1}^{\text{sp}}\propto
-\frac{R^{1}_{234}R^{2}_{134}}{d_{34}}+\frac{R^{1}_{234}R^{3}_{124}}{d_{24}}-\frac{R^{2}_{134}R^{3}_{124}}{d_{14}}-\frac{R^{1}_{234}R^{4}_{123}}{d_{23}}+\frac{R^{2}_{134}R^{4}_{123}}{d_{13}}-\frac{R^{3}_{124}R^{4}_{123}}{d_{12}}\,=\,0~.
\eeq
Finally the N$^2$MHV comes entirely from the third topology and it also vanishes:
\beq\label{eq:R4zero}
G_{4,2}^{\text{sp}} 
\propto R^{1}_{234}R^{2}_{134}R^{3}_{124}R^{4}_{123} = 0~.
\eeq
The vanishing of $G_{4,1}$ and $G_{4,2}$ was expected on grounds of superconformal symmetry, see \eqref{eq:Gcorre}. We verified the identities \eqref{sixterms}  and \eqref{eq:R4zero} numerically for random  sets of twistor kinematics.
The ``six-term identity'' \eqref{sixterms} was discussed in earlier twistor-space calculations of stress-tensor multiplet correlators, see for example eq.~4.12 of \cite{Chicherin:2014uca}.

In conclusion, the four-point correlator in the single-particle basis is given by its MHV component in \eqref{eq:G4MHVsp}. Finally, a calculation analogous to \eqref{shift 3pt} gives us the difference between the correlators in both basis:
\bba
N_c^2\times\left(G_{4}-G_{4}^{\text{sp}}\right) &=D_{12}D_{13}D_{14}(2+D_{12}+D_{13}+D_{14})+ (3\, \text{permutations}) \cr 
    &\quad+D_{12}D_{34}D_{14}D_{13}(2+D_{14}+D_{13} )+(11\, \text{permutations}) \cr 
    &\quad +D_{12}D_{34}D_{14}(1+D_{14})+ (11\, \text{permutations})+ \mathcal{O}(N_c^{-2})~.
\end{align}
This difference was also given in eq.~3.17 of \cite{Caron-Huot:2021usw} and it corresponds diagrammatically to the contribution of graphs with minimum degree one.

\subsection{Five-points}\label{sec:CF5}

\begin{figure}[t]
 \begin{minipage}[h]{1\textwidth}
 \centering 
  \includegraphics[width=1\textwidth]{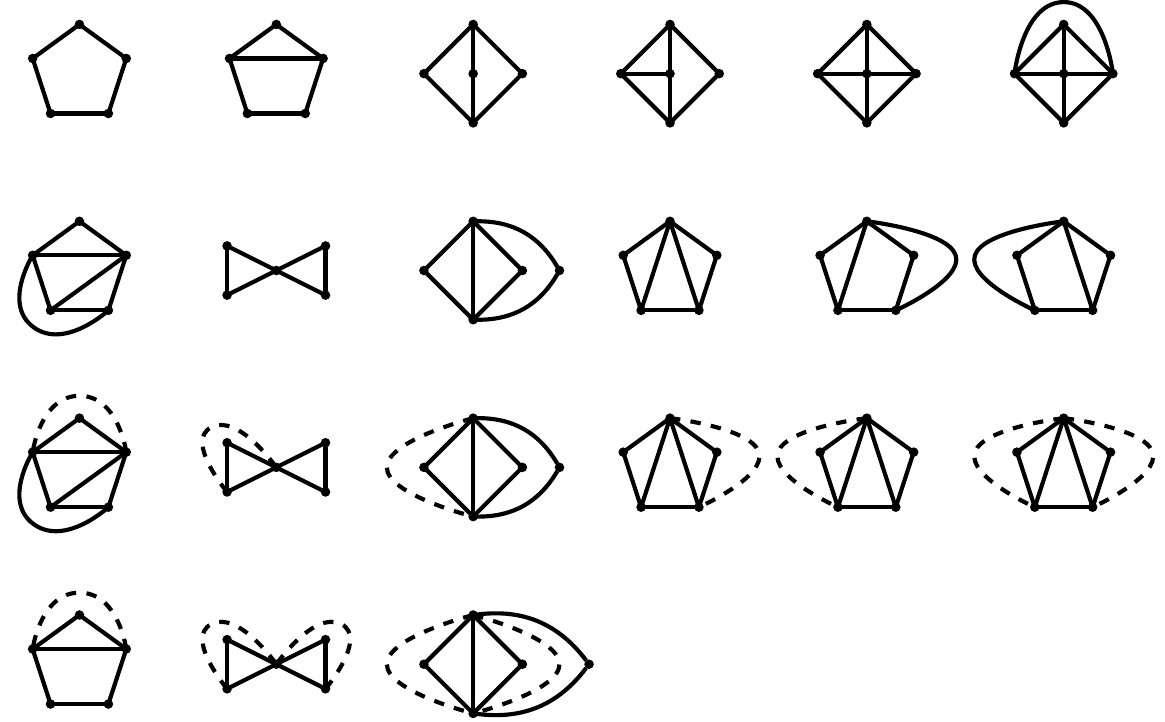}
 \end{minipage}
 \caption{Graphs for planar five-point correlator in the single-particle basis. The dashed lines denote repeated bundles which are topologically inequivalent.}
 \label{fig:fiveskeletons}
 \end{figure}

\begin{figure}[t]
 \begin{minipage}[h]{1\textwidth}
 \centering 
  \includegraphics[width=.9\textwidth]{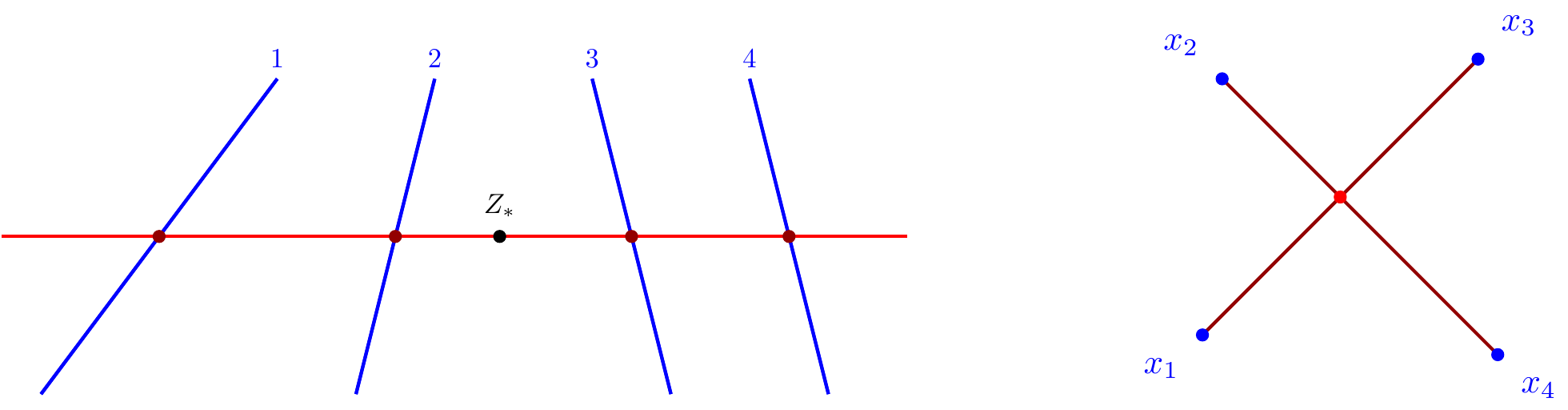}
 \end{minipage}
 \caption{Illustration of special choice of twistor $Z_{*}$ to simplify our five-point computation. On the left, in twistor space, $Z_{*}$ lies on the line that intersects the four twistor lines corresponding to $x_{i}$. On the right, in spacetime, the dual spacetime point is  null-separated from $x_1, x_2, x_3$ and $x_4$. The fifth point $x_5$ is in generic position with respect to the other points and to $Z_{*}$.}
 \label{fig:specialtwistor}
 \end{figure}

In order to compute the five-point function we could follow the same procedure as in the previous section: label all topologies in figure \ref{fig:fiveskeletons}, use effective Feynman rules and sum over inequivalent permutations. Instead here we take a short-cut by taking advantage of the reference twistor $\mathcal{Z}_{*}$. Since the final result should not depend on it, we can make a choice that simplifies intermediate steps. 
We set the fermionic part to zero and choose the bosonic part $Z_{*}$ to lie on the line that intersects the four twistor lines associated to spacetime points $x_{1}, x_2, x_3$ and $x_4$. This special twistor line is  dual to the spacetime point which is null-separated from the first four spacetime points as shown in figure \ref{fig:specialtwistor}.
In this special kinematics all $R$-invariants vanish except for the ones living on the twistor line of $x_5$: $R^{i}_{jkl}=0$, if $i\neq 5$. This implies the vanishing of all N$^k$MHV components for $k$ equal or higher than two. While the NMHV component only receives contributions from eight topologies in figure \ref{fig:fiveskeletons}: the six in the first row and the last two in the second row. Furthermore, we only need to consider inequivalent permutations where the point $5$ lies on the vertices with highest degree. The final result, in this special gauge, can be reduced to:
\beq \label{G51 from RR}
N_c^3\,G^{\text{sp}}_{5,1}=  \frac{2\, \prod\limits_{1\leq i<j\leq 4}x_{i,j}^2\prod\limits_{1\leq i \leq4}d_{i5}}{\prod\limits_{1\leq i<j \leq5}(1-d_{ij})}\,\left(d_{12}d_{34}\,R^{5}_{123}R^{5}_{124}-d_{13}d_{24}\,R^{5}_{123}R^{5}_{134}+d_{14}d_{23}\,R^{5}_{124}R^{5}_{134}\right)~.
\eeq
In order to find this compact representation we made use of the identity (implied by \eqref{four term identity} and \eqref{eq:fusionRule}):
\beq
R^{5}_{123}\,R^{5}_{124}\,-\,R^{5}_{123}\,R^{5}_{134}+R^{5}_{124}\,R^{5}_{134}=0~.
\eeq
We have then checked numerically for various components
that our result is proportional to the unique 5-point superconformal invariant $\mathcal{I}_{5,1}$ given in eq.~5.22 of \cite{Chicherin:2014uca} (see also appendix B of \cite{Eden:2011we}):
\beq \label{G51}
N_c^3\,G_{5,1}\,=\, N_c^3\,G^{\text{sp}}_{5,1}\,=\, \frac{2\,\mathcal{I}_{5,1}}{\prod\limits_{1\leq i<j\leq 5}X_{i,j}^2} 
\eeq
where the denominator is given by the product of ten-dimensional distances $X_{i,j}^2$ that combine spacetime and R-charge space distances, see \eqref{eq:introX10d}.  We also highlight that the NMHV component is the same in both the single-trace and single-particle basis.

In particular, the part of the NMHV component that has maximum
Grassmann degree on the fifth point is given by:
\beq
N_c^3\,\langle \mathsf{O}_1\,\mathsf{O}_{2}\,\mathsf{O}_{3}\,\mathsf{O}_{4}\,L_{\rm int}(x_5,y_{5})\rangle
=N_c^3\, G_{5,1}\big{|}_{(\theta_{5})^4}=
\frac{2\,\mathcal{I}_{5,1}\big{|}_{(\theta_{5})^4}}{\prod\limits_{1\leq i<j\leq 5}X_{i,j}^2}  =\frac{2\,\mathcal{R}_{1234}  \prod\limits_{1\leq i<j\leq 4}x_{i,j}^2}{\prod\limits_{1\leq i<j\leq 5}X_{i,j}^2}
\label{G51 L}
\eeq
with 
\beq
\mathcal{R}_{1234}= d_{13}^{2}d_{24}^{2}x_{13}^{2}x_{24}^2\,+\, d_{12}d_{23}d_{34}d_{41}(x_{13}^2\,x_{24}^2-x_{12}^{2}x_{34}^2-x_{14}^2 x_{23}^2) \,+\, (1\leftrightarrow 2)\,+\, (1\leftrightarrow 4)~.
\eeq
We also use $\mathsf{O}_i\equiv \mathsf{O}(x_i,y_i)$  to denote the bottom component of the single-trace generating function (called ``master operator" in \cite{Caron-Huot:2021usw}).
The fifth operator $L_{\rm int}(x_5,y_5)$
combines the chiral Lagrangian $L_{\rm int}(x_5)$
with a tower of its R-charged counterparts.
By setting $y_5=0$ we reproduce the result in eq.~3.19 from that reference for the four-point one-loop integrand of arbitrary half-BPS operators.
To our knowledge, this is the first time that the $y$ dependence of $L_{\rm int}(x_5,y_5)$
is explicitly calculated. We find it very pleasing that \eqref{G51 L} confirms the ten-dimensional structure expected from \cite{Caron-Huot:2021usw}.

For each $n$ there exists a unique superconformal invariant of maximal Grassmann degree $4(n{-}4)$ \cite{Eden:2011we}. Denoting it as $\mathcal{I}_{n,n-4}$, this implies for example that the N$^{2}$MHV component of our six-point supercorrelator must be proportional to this unique invariant:
\beq
G_{6,2} \propto \mathcal{I}_{6,2}~.
\eeq
A simple guess for the proportionality factor can be made by writing the two-loop integrand in terms of ten-dimensional invariant, explicitly:
\bba
N_c^4\,G_{6,2}\big{|}_{(\theta_{5})^4(\theta_{6})^4}\;&=\;N_c^4\,\langle \mathsf{O}_1\,\mathsf{O}_{2}\,\mathsf{O}_{3}\,\mathsf{O}_{4}\,L_{\rm int}(x_5,y_{5})\,\,L_{\rm int}(x_6,y_6)\rangle \nonumber\\ &=2\,\mathcal{R}_{1234}  \prod\limits_{1\leq i<j\leq 4}x_{i,j}^2\times\frac{X_{1,2}^2\,X_{3,4}^2\,X_{5,6}^2+14\text{ permutations}}{\prod\limits_{1\leq i<j\leq 6}X_{i,j}^2}~.
\end{align}
We have compared this formula numerically with the twistor space Feynman rules outlined above and found perfect agreement.
This confirms and extends the ten-dimensional structure found in \cite{Caron-Huot:2021usw} to the case where $y_5, y_6\neq 0$.  In section \ref{section6} below, we will discuss NMHV 6-point correlators for which multiple invariants exist.

\section{Matrix duality: determinant correlators as a matrix integral}\label{section5}

Matrix duality relates the expectation values of determinant operators in Gaussian matrix ensembles in which the number of determinants and the rank of the matrices are exchanged: the $n$-point function for $N_c\times N_c$ matrices is equal to a $N_c$-point function for $n\times n$ matrices. Diagrammatically, it amounts to  a graph duality which trades the faces and vertices of Feynman diagrams, as will be described below.
It was proposed to be related to open-closed dualities between string models \cite{GopakumarTalk}, considering as a prime example the FZZT case \cite{Fateev:2000ik, Teschner:2000md}.


In the context of $\mathcal{N}=4$ SYM, this duality has been used in \cite{Jiang:2019xdz, Budzik:2021fyh, Chen:2019gsb} to study the correlation functions of operators dual to (maximal)  giant gravitons in AdS$_5\times S^5$, see also \cite{Brown_2011} for an earlier use in a special kinematics without spacetime dependence and \cite{Bargheer:2019kxb} for a large-charge case. These authors consider determinants of the scalar matrix of the form $\det(y{\cdot}\phi(x))$ or  $\det(1-y{\cdot}\phi(x))$, restricting to the scalar sector of the free theory.  Their $n$-point correlator was recasted as a Gaussian integral over a $n\times n$ matrix $\rho$ with an insertion of $\det(\rho)^{N_c}$ or $\det(\mathbb{I}-\rho)^{N_c}$. In this reformulation, which we denote as the $\rho$-matrix integral, the number of colors $N_c$ appears as a coupling and it becomes more amenable to study large-$N_c$ expansions around various possible saddle points.

In this section we go beyond the scalar sector of \cite{Jiang:2019xdz,Budzik:2021fyh,Chen:2019gsb} and apply matrix duality to correlators of supersymmetrized determinants in self-dual SYM. We make use of the Gaussian twistor reformulation of SDYM in axial gauge and follow the steps in figure \ref{fig:Graphduality}, integrating in and out auxiliary fields. As a result we obtain the $\rho$-matrix integral, but with a modified version of the determinant insertion which now includes superspace coordinates $\theta$ appearing from the on-shell $\mathbb{CP}^{1|2}$ propagators \eqref{eq:DeltaOnShell}. When expanding the determinant into vertices the modification will amount to: $\cdots\rho_{ij}\,\rho_{jk}\cdots\mapsto \cdots\rho_{ij}\,\Delta^{j}_{ik}\rho_{jk}\cdots$. We also highlight that the $\rho$-matrix integral manifests the appearance of ten-dimensional denominators in the large-$N_c$ expansion around the trivial saddle.

After deriving the $\Delta\rho$-matrix integral dual to the correlator of supersymmetrized determinants in SDYM in subsection \ref{sec:DualitySteps}, we exemplify some of its properties in subsection \ref{sec:ExamplesRho} and discuss its large-$N_c$ expansion in \ref{ssec:largeN}. 
Finally, in subsections \ref{sec:replica} and  \ref{sec:singletrace}, we present two methods to extract single-trace correlators from those of determinants: by applying the replica trick to the $\Delta\rho$-matrix integral; and by using combinatorial relations
between the two types of correlators.

\subsection{The $\Delta\rho$-matrix dual}\label{sec:DualitySteps}

In this section we rewrite the correlator of $n$ determinants in SDYM in terms of an integral over an $n\times n$ matrix. Our starting point is the twistor action of self-dual YM \eqref{SDYM} in CSW gauge coupled to $n$ copies
of the $\mathbb{CP}^{1|2}$ Gaussian model \eqref{mainConjecture} representing each determinant:
\begin{align}\label{combinedCP}
&\hspace{-8mm}\left\langle \mathbb{D}(x_1,y_1,\theta_1) \mathbb{D}(x_2,y_2,\theta_2) \ldots \mathbb{D}(x_n,y_n,\theta_n)   \right\rangle_{\scriptscriptstyle\text{SDYM}} \cr 
&\hspace{-8mm}= \!\!\int\! \mathcal{D}\mathcal{A}\mathcal{D}\alpha\mathcal{D}\beta \exp\bigg[N_c\!\int\Omega^{3|4}{\rm tr}\left( \frac{1}{2}\mathcal{A} \bar{\partial}\mathcal{A}+\mbox{g.f.}\right) 
+ \sum_{i=1}^n {\int \Omega_i^{1|2} \,\alpha_i\left(\bar{\partial}+\mathcal{A}+{\color{black}\bar{\delta}^{1|{\color{black}2}}_{\mu,\lambda}}\right)\!\Big{|}_{\mathbb{CP}^{1{\color{black}|2}}}\,\beta_i}\bigg]\!.
\end{align}
Whereas in section \ref{section3} we started by integrating out the superfields $\alpha$-$\beta$, the main idea here will be to first integrate out $\mathcal{A}$.
In the CSW gauge, see \eqref{twistor propagator}, the action for the gauge field is Gaussian and so this can be done exactly. 
The final result will be a finite-dimensional matrix integral because
the interactions between different $\mathbb{CP}^{1|2}$ lines localize to discrete points (see \eqref{tilde_lambda}). 
\begin{figure}[t]
 \begin{minipage}[h]{1\textwidth}
 \centering 
  \includegraphics[width=1\textwidth]{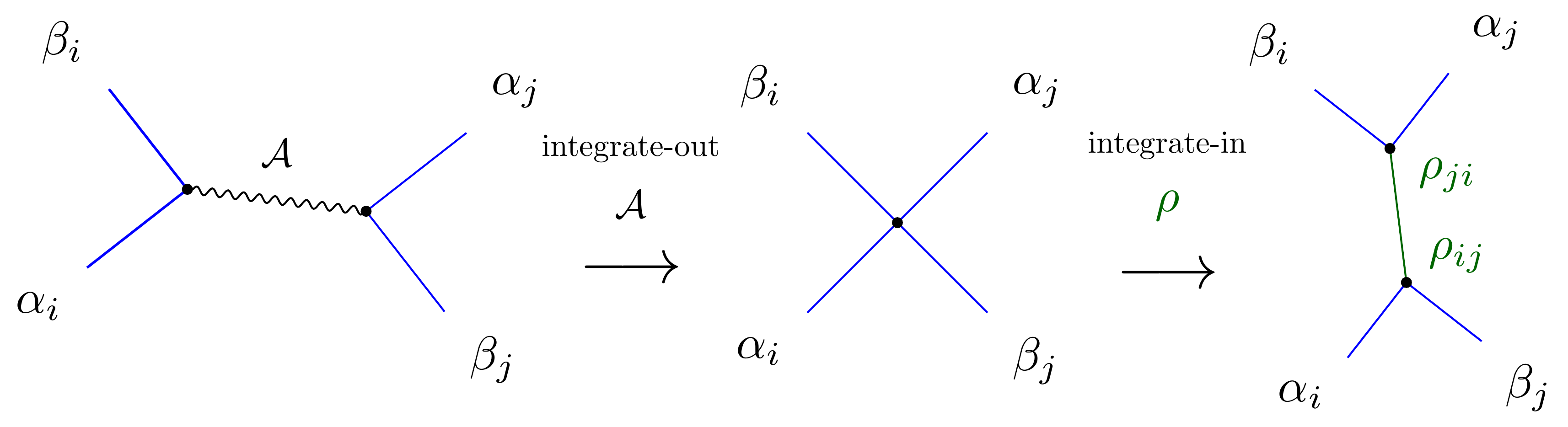}
 \end{minipage}
 \caption{Schematic illustration of matrix duality. First the gauge field is integrated-out, collapsing the first diagram into the four-valence vertex of the middle diagram. Then we integrate-in the $\rho$-matrix to open up the four-valence vertex into new bi-local three-valence vertices on the right diagram. 
 This interchanges the vertices and faces of a
 diagram (see figure \ref{fig:Graphduality 2} below). The old $\mathcal{A}$-edge and the new $\rho$-edge are dual in the sense of graph duality.
 }
 \label{fig:Graphduality}
\end{figure}

First, to remove the term linear in the gauge field in (\ref{combinedCP}),
we shift $\mathcal{A}$ by a term proportional to the $\mathbb{CP}^{3|4}$ propagator $\Delta_{*}(\mathcal{Z},\mathcal{Z}_i)$ (\ref{eq:delta44})
\begin{equation}
    \mathcal{A} \rightarrow \mathcal{A} + \frac{1}{N_c} \sum_{i=1}^n \int_{\mathbb{CP}^{1|2}} \Omega_i^{1|2} \Delta_*(\mathcal{Z},\mathcal{Z}_i)\beta_i \alpha_i~.
\end{equation}
Notice that $\beta_i \alpha_i$ is a $(N_c\times N_c)$-dimensional matrix.
We can then integrate out $\mathcal{A}$ and obtain an action solely in terms of the $\mathbb{CP}^{1|2}$ variables
\begin{equation}
S[\alpha,\beta] =\,\sum_{i=1}^{n}\,\int  \Omega_i^{1|2}\,\alpha_i\left(\bar{\partial}+{\color{black}\bar{\delta}^{1|{\color{black}2}}_{\mu,\lambda}}\right)\!\Big{|}_{\mathbb{CP}^{1{\color{black}|2}}}\beta_i
-\frac{1}{N_c}\sum_{i,j =1}^n\int\Omega_i^{1|2} \Omega_j^{1|2}\,\Delta_*(\mathcal{Z}_i ,\mathcal{Z}_j) \,\tr\left(\frac{1}{2}\,\alpha_i \beta_j \alpha_j \beta_i\right)~.
\end{equation}
Using the $4|4$ representation of $\Delta_{*}$ \eqref{eq:delta44}, its factorization \eqref{eq:deltaBosonicCP34} and integrating over the $\mathbb{CP}^{1|2}$ parameters we obtain the four-valent interaction
\beq\label{final_CP12}
{ S[\alpha,\beta] =\,\sum_{i=1}^{n}\,\int_{\mathbb{CP}^{1|2}} \Omega_i^{1|2}\,\alpha_{i}\,\left(\bar{\partial}+{\color{black}\bar{\delta}^{1|{\color{black}2}}_{\mu,\lambda}}\right)\!\Big{|}_{\mathbb{CP}^{1{\color{black}|2}}}\beta_{i}
\,+\,\sum_{i\neq j=1}^{n}\,\frac{d_{ij}}{2N_c}(\tilde{\alpha}\tilde{\beta})_{ij}\,(\tilde{\alpha}\tilde{\beta})_{ji} }~,
\eeq
where $d_{ij}\equiv -{y_{ij}^2}/{x_{ij}^2}$ (\ref{dij_propagator}). We use $\tilde{\alpha}$ and $\tilde{\beta}$ to indicate that we evaluate them ``on-shell" using (\ref{tilde_lambda}) and introduce the notation
\begin{equation}
 ( \tilde{\alpha}\tilde{\beta})_{ij}  \equiv \alpha_{i}(\tilde{\lambda}_{ij},\tilde{\psi}_{ij})\beta_j(\tilde{\lambda}_{ji},\tilde{\psi}_{ji})~.
\end{equation}
Diagrammatically, the four-valent interaction comes from collapsing the $\mathcal{A}$-propagators in the original Feynman graphs. This is depicted in figure \ref{fig:Graphduality}, on going from the left panel to the middle one.

\begin{figure}[t]
 \begin{minipage}[h]{1\textwidth}
 \centering 
  \includegraphics[width=1\textwidth]{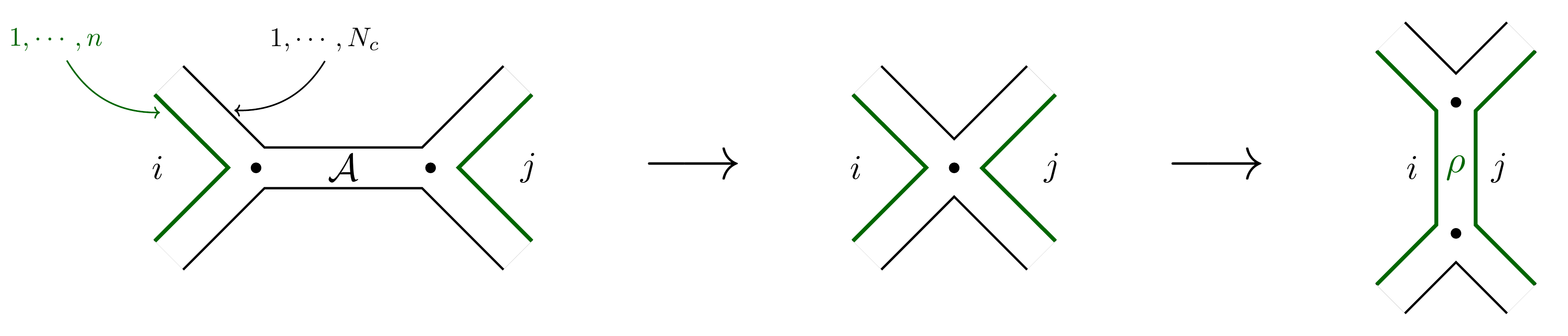}
 \end{minipage}
 \caption{Schematic illustration of matrix duality using the fat-line notation to highlight the matrix structure. The black lines run on color indexes ($1,\cdots,N_c$) and the green lines run on the number of operators ($1,\cdots,n$). The gauge field  $\mathcal{A}$ is a  $N_c\times N_c$ matrix, the fields $\alpha$ and $\beta$ are $n\times N_c$ and $N_c\times n$ matrices respectively, and the $\rho$-matrix has dimension $n\times n$.
 }
 \label{fig:FatGraphduality}
\end{figure}
	
Next, we perform a Hubbard-Stratonovich transformation to \eqref{final_CP12}. We integrate-in a color-neutral and spacetime-independent
$n\times n$  bosonic matrix $\rho$ in order to split the four-valent vertices into new three-valent vertices:
\beq
 S[\rho] =\sum_{i,j=1}^{n}\left(\frac{-N_c}{2d_{ij}}\rho_{ij}\rho_{ji}+\rho_{ij}(\tilde{\alpha}\tilde{\beta})_{ji}\right)+\sum_{i=1}^{n}\,\int_{\mathbb{CP}^{1|2}} \Omega_i^{1|2}\,\alpha_{i}\left(\bar{\partial}+{\color{black}\bar{\delta}^{1|{\color{black}2}}_{\mu,\lambda}}\right)\!\Big{|}_{\mathbb{CP}^{1{\color{black}|2}}}\beta_{i}~.
\eeq
This splitting is depicted in figure \ref{fig:Graphduality}, on going from the middle panel to the right. We also represent these steps in figure \ref{fig:FatGraphduality}, using the fat-graph notation to highlight the matrix structure of each field. 

From the spacetime viewpoint, the three-valent vertices $\rho_{ij}(\tilde{\alpha}\tilde{\beta})_{ji}$ are bi-local since they depend simultaneously on the kinematics of operators $i$ and $j$. The elements $\rho_{ij}$ are in one-to-one correspondence with the old gauge propagators connecting superlines $i$ and $j$, they are edge duals in the diagrammatic sense of figures \ref{fig:Graphduality},\ref{fig:FatGraphduality} and \ref{fig:Graphduality 2}. Hence $\rho_{ii}=0$, since we do not consider self-contractions.

Finally, the path integral on the fields $\alpha$ and $\beta$ is Gaussian and can be performed exactly, resulting on $N_c$ insertions of determinants in the $\rho$-matrix integral:
\begin{equation}\label{MM}
 \int [\mathcal{D}\rho] e^{\sum\limits_{i<j}\frac{N_c}{d_{ij}}\rho_{ij}\rho_{ji}}
 \det\left(\mathbb{I}_{n(n-1)}-\bar{\Delta}(\lambda,\psi,\mu,\mathcal{Z}_*) \bar{\rho}\right)^{N_c}~,
\end{equation}
where we defined the measure 
\begin{equation}
[\mathcal{D}\rho]  \equiv \prod_{i<j} \dd \rho_{ij}\dd\rho_{ji}~,\quad  \rho_{ij}^* = \rho_{ji}~,
\end{equation}
and the matrix $\bar{\Delta}$, defined below, depends on the on-shell values of $\lambda$ (\ref{tilde_lambda}) and $\psi$ (\ref{onshellpsi}), as well as the reference twistors in the superlines ($\mu_{i}$) and in the bulk ($\mathcal{Z}_{*}$).
The determinant in \eqref{MM} involves matrices of size $n(n-1)\times n(n-1)$,
defined as:
\beq
\NiceMatrixOptions{code-for-first-row=\footnotesize,code-for-last-row=\footnotesize} 
   \bar{\Delta}\equiv \begin{pNiceMatrix}[first-row,last-col=7]
      [12] & & \quad\cdots &  \quad[ij] & \cdots &\quad \quad \quad &\\
      0 & & &  &  &  & [12] \\
       & & \ddots &  &  &  & \;\;\vdots \\
       &  & &   \delta_{jk} {\Delta}^j_{il\mu}  &  &  & [k l]\\
         &  & &    &  \ddots & & \;\;\vdots\\
          &  & &    & &  & \\
          \end{pNiceMatrix}~\quad \text{and}\quad  \bar{\rho} \,\equiv\, \left(\begin{matrix}
\rho_{12} & 0  &\qquad\qquad &\\
0  & \rho_{21} & \qquad\qquad &\\
  &  & \ddots &\\
  &  & \qquad & \rho_{ij} & 0\\ 
  & & \qquad & 0 & \rho_{ji}\\
  &  & \qquad & & & \ddots\\
\end{matrix}\right)_{1 \leq i < j \leq n} \hspace{-3mm}
\label{big matrices}
\eeq
where ${\Delta}^j_{il\mu} = 1+ R({\lambda}_{ji},{\lambda}_{jl},\mu_j)$ is the ``on-shell" $\alpha$-$\beta$  propagator   (see \ref{CP12_prop} and \ref{eq:DeltaOnShell}) on the $\mathbb{CP}^{1|2}$ superline $j$,  between special points connecting to superlines $i$ and $l$; $\mu_j$ is the reference position on superline $j$. 

Intuitively, the $n(n-1)$ rows and columns of the matrices \eqref{big matrices} are in one-to-one correspondence
with the special points on the various lines, and each nonzero entry of the matrix $\bar\Delta$ can be interpreted as a hopping term: along the line $j=k$, the $\alpha$-$\beta$ propagator can hop between the special points that connect to lines $i$ and $l$, thus connecting $[ij]$ to $[kl]$.

As an example we present the matrix $\bar{\Delta}$ for  $n=3$:
\begin{equation}\label{Delta3}
    \bar{\Delta} = \begin{pmatrix}
    0 & 1 & 0 & 0 & \Delta^{2}_{13\mu_2} & 0 \\ 
    1 & 0 & \Delta^{1}_{23\mu_1} & 0 & 0 & 0 \\
    0 & 0 & 0 & 1 & 0 & \Delta^{3}_{12\mu_3} \\
    \Delta^{1}_{23\mu_1} & 0 & 0 & 0 & 0 & 0 \\ 
    0 & 0 & 0 & \Delta^{3}_{12\mu_3} & 0 & 1 \\
    0 & \Delta^{2}_{13\mu_2} & 0 & 0 & 1 & 0
    \end{pmatrix}~,
\end{equation}
where we used the simplification $\Delta_{jj\mu_i}^i=1$ for $R$-invariants at coincident points.

We highlight again the result of this section. Applying matrix duality to (\ref{combinedCP}) we rewrote the $n$-point supercorrelator of determinants $\mathbb{D}(x,y,\theta)$ \eqref{mainConjecture} in terms of an integral over a color-neutral and spacetime-independent matrix $\rho$:
\begin{equation}\label{matrix_duality1}
 \left\langle \prod_{i=1}^n\mathbb{D}(x_i,y_i,\theta_i) \right\rangle_{\scriptscriptstyle\text{SDYM}}
= \frac{1}{\mathcal{M}}\times\int [\mathcal{D}\rho] e^{N_c\sum\limits_{i<j}\frac{\rho_{ij}\rho_{ji}}{d_{ij}}}\det\left(\mathbb{I}_{n(n-1)}-\bar{\Delta}(R) \bar{\rho}\right)^{N_c}~,
\end{equation}
where the normalization is the Gaussian matrix integral
\begin{equation}\label{eq:normalisationMM}
\mathcal{M} = \int [\mathcal{D}\rho] e^{N_c\sum\limits_{i<j}\frac{\rho_{ij}\rho_{ji}}{d_{ij}}} = \prod_{i<j} \left(\frac{-d_{ij}\pi}{N_c}\right)~,
\end{equation}
and the $R$-invariant (\ref{r123}) captures the dependency of $(\lambda,\psi,\mu,\mathcal{Z}_*)$. 
Note that the integrals converge for $d_{ij}<0$.  All results are rational functions of $d_{ij}$'s which can be readily continued.

\subsection{Examples and properties}\label{sec:ExamplesRho}

The formula in \eqref{matrix_duality1} will be further simplified in the next section (see \eqref{matrix_duality simplified}), but let us start by illustrating some of its features.

In the special case where we drop out the Grassmann dependency in the $\alpha$-$\beta$ propagator, so that $\Delta_{il\mu}^j\to1$, the $n(n-1)\times n(n-1)$ matrix in (\ref{matrix_duality1}) can be transformed into a block matrix. Using properties of block matrix determinants we then obtain the simplified determinant
\begin{equation}\label{deMello}
\det\left(\mathbb{I}_{n(n-1)}-\bar{\Delta}(\lambda,\psi,\mu,\mathcal{Z}_*) \bar{\rho}\right) \big{|}_{\theta=0}= \det \left(\mathbb{I}_n - \rho\right) \qquad \mbox{(no fermions)}~.
\end{equation}
In particular the size of the matrix on the right hand side is significantly smaller. Here the $n\times n$ matrix $\rho$ is given by
\beq\label{rhoMatrix}
\rho \equiv \left(\begin{matrix}
0  &\,\rho_{12}\,& \,\rho_{13}\, & \cdots  \\
\,\rho_{21}\,& 0 &\,\rho_{23}\, &  \\
\,\rho_{31}\,&\,\rho_{32}\,& 0 &  \\
\vdots & & & \ddots  \\
\end{matrix}\right)~,\quad \rho_{ij}^* = \rho_{ji}~.
\eeq
This latter matrix integral was discussed previously in refs.~\cite{Jiang:2019xdz,Budzik:2021fyh,Chen:2019gsb}
and is known to generate Wick contractions between scalar operators in the free theory. We will refer to it as the ``bosonic $\rho$-integral''.
\newline

We stress that due to the presence of $d_{ij}$ in (\ref{matrix_duality1}) we are not 
dealing with a standard solvable one-matrix integral.
An exception is $n=2$, which corresponds to the two-point function of determinants, where we have a single matrix integral with
\begin{equation}
\rho = \begin{pmatrix} 0  & \rho_{12} \\ \rho_{21} & 0 \end{pmatrix}~, \quad  \bar{\rho} = \begin{pmatrix} \rho_{12} & 0 \\ 0 & \rho_{21} \end{pmatrix}~,\quad \bar{\Delta} = \begin{pmatrix} 0 & {\Delta}_{11\mu_2}^2 \\  {\Delta}^1_{22\mu_1}& 0  \end{pmatrix}~.
\end{equation}
Since the $R$-invariant (\ref{r123}) vanishes if two points coincide we have ${\Delta}_{11\mu_2}^2= {\Delta}_{22\mu_1}^2=1$.
Thus 
\begin{equation}\label{n2a}
\det\left(\mathbb{I}_{2}-\bar{\Delta}(\lambda,\psi,\mu,\mathcal{Z}_*) \bar{\rho}\right) = 1- \rho_{12}\rho_{21}~,
\end{equation}
which is also in agreement with \eqref{deMello} since the fermions drop out.
The absence of Grassmann dependence was of course as expected since
there is no superconformal invariant at two points.
The correlator of two determinants is then equal to
\begin{align}
    \left\langle\,\mathbb{D}_1\mathbb{D}_2\,\right\rangle_{\scriptscriptstyle\text{SDYM}} 
 &= \frac{1}{\mathcal{M}} 
 \int  [\mathcal{D}\rho]  e^{N_c\frac{\rho_{12}\rho_{21}}{d_{12}}} (1-\rho_{12}\rho_{21})^{N_c}
= \sum_{k=0}^{N_c} \frac{N_c!}{(N_c-k)!} \left(\frac{d_{12}}{N_c}\right)^k
\cr &\to \frac{1}{1-d_{12}} - \frac{1}{N_c} \frac{d_{12}^2}{(1-d_{12})^3} + \mathcal{O}(N_c^{-2})
 \label{n2 sum}~,
\end{align}
where $\mathbb{D}_i\equiv \mathbb{D}(x_i,y_i,\theta_i)$.  The sum on the first line could be expressed exactly as an incomplete Gamma function; in the second line we have given its large-$N_c$ expansion.
Notice that $\frac{1}{1-d_{12}}=\frac{x_{12}^2}{x_{12}^2+y_{12}^2}$, revealing the emergence of
ten-dimensional denominators at large-$N_c$. The above result is compatible with the single-trace planar correlator $\langle \mathbb{O}_1  \mathbb{O}_2\rangle = -\log(1-d_{12})$ obtained in \eqref{eq:Otwopoint}, as will be further discussed in section \ref{sec:singletrace}.
\newline
The first example with non-trivial Grassmann dependence within $\bar\Delta$ is $n=3$. 
For the correlator of three determinants we have the integral
\begin{align}\label{n3}
    \left\langle\,\mathbb{D}_1\mathbb{D}_2\mathbb{D}_3\,\right\rangle_{\scriptscriptstyle\text{SDYM}} =
\int  [\mathcal{D}\rho] \, e^{N_c\left(\frac{\rho_{12}\rho_{21}}{d_{12}}+\frac{\rho_{23}\rho_{32}}{d_{32}}+\frac{\rho_{13}\rho_{31}}{d_{13}}\right)}\det\left(\mathbb{I}_{6}-\bar{\Delta}(\lambda,\psi,\mu,\mathcal{Z}_*) \bar{\rho}\right)^{N_c}~,
\end{align}
where $\bar{\Delta}$ is given in (\ref{Delta3}).
Due to the presence of $R$-invariants we should worry that the result might depend on the arbitrary references 
$\mu_k$ that appears on each $\mathbb{CP}^{1|2}$ as well as the reference twistor $\mathcal{Z}_*$
We find however that both cancel out upon integration.
A simple way to see this is to effect the following rescaling:
\begin{equation}
\rho_{ij}\rightarrow \rho_{ij}\times {\Delta^k_{ij\mu_k}}~,
\quad  i, j \neq k \in \{1,2,3\}~.
\end{equation}
This transformation
has unit Jacobian (thanks to the identity $\Delta_{ij\mu_k}^{k}\Delta_{ji\mu_k}^{k}=1$, discussed around \eqref{eq:DeltaIdentities}), and completely cancels the fermions in the determinant.
We are thus reduced to the bosonic model determinant,
\begin{equation}
    \det(\mathbb{I}_3-\rho_{3\times 3})= 1- \rho_{12}\rho_{21} - \rho_{13}\rho_{31} - \rho_{23}\rho_{32} - \rho_{12}\rho_{23}\rho_{31} -\rho_{21}\rho_{13}\rho_{32}~.
\end{equation}
Again it makes sense that there is no Grassmann dependence for $n=3$ since
there exists no superconformal invariant function with $n\leq 4$ points (see \cite{Eden:2011we}).

\paragraph{$\mu$-independence for general $n$.}
The general formula \eqref{matrix_duality1} superficially depends on gauge-fixing choices
inherent to the twistor formalism: the points $\mu_k$ on each $\mathbb{CP}^{1|2}$, and the
gauge-fixing $\mathcal{Z}_*\in \mathbb{CP}^{3|4}$.
Dependence on these quantities must cancel out of the final expressions for correlation functions.

Independence on the $\mu$'s can be demonstrated for any $n$
using a variant of the $\rho$-rescaling trick that we have just used.
Let us compare the matrix integral computed using a given $\mu_1$ to
that using a different choice $\nu_1$.  We find that a  
similarity transformation $C$ and rescaling $\rho'$ exist such that:
\begin{equation}\label{symmetryTransfo}
C(\mu_1,\nu_1) \bar\Delta(\mu_1,\ldots) \bar{\rho} \,C(\mu_1,\nu_1)^{-1}=
\bar\Delta(\nu_1,\ldots) \bar{\rho}',
\end{equation}
where 
\begin{equation} \label{rho rescaling}
\rho'_{ij} = \rho_{ij} \times \frac{\Delta^1_{ij\nu_1}}{\Delta^1_{ij\mu_1}}\,.
\end{equation}
The elements $\rho_{1k}$ and $\rho_{k1}$, for $k=2,\ldots n$,
are left untransformed. The equivalence in  (\ref{symmetryTransfo}) can be easily verified by raising both sides to an arbitrary power and taking traces.
This change of variable leaves invariant both the measure and exponent in \eqref{matrix_duality1}, and so we conclude that the correlator is independent of $\mu_1$, as it should.  A similar argument works for the other $\mu_k$, giving that:
\begin{equation}
\int [\mathcal{D}\rho] e^{N_c\sum\limits_{i<j}\frac{\rho_{ij}\rho_{ji}}{d_{ij}}}\det\left(\mathbb{I}_{n(n-1)}-\bar{\Delta}(\lambda,\psi,\mu,\mathcal{Z}_*) \bar{\rho}\right)^{N_c}
=\mbox{independent on $\mu$'s}
\end{equation}
consistent with the arguments in section \ref{section3}. 
One could use this for example
to simplify the computation of the matrix integral by
``gauge-fixing'' each $\mu_k$ to some special point on line $k$ so as to set some $R$-invariants to zero.

\subsection{Determinants at large $N_c$}\label{ssec:largeN}

\begin{figure}[t]
 \begin{minipage}[h]{1\textwidth}
 \centering 
  \includegraphics[width=.5\textwidth]{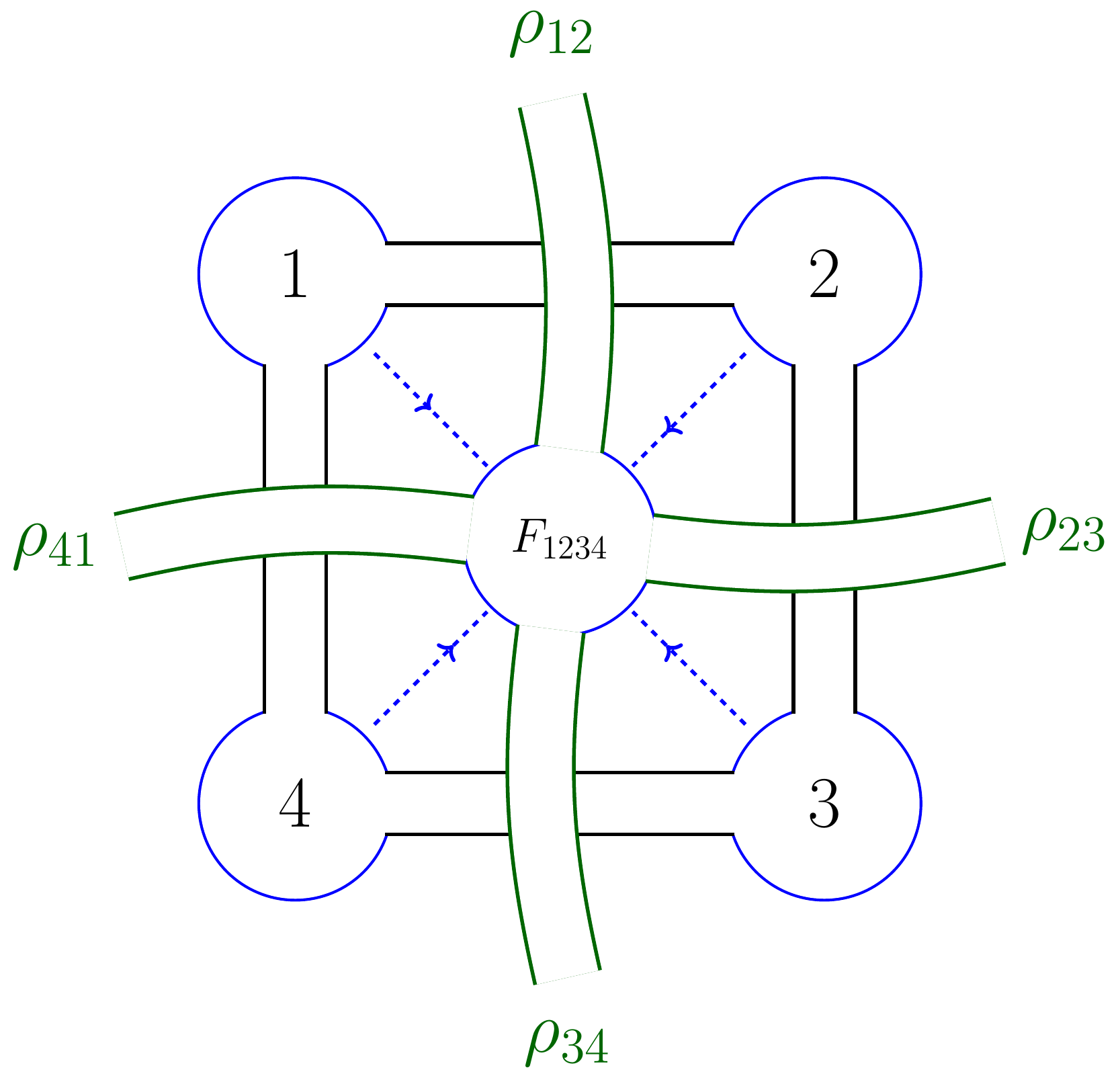}
 \end{minipage}
 \caption{Illustration of graph duality, which interchanges faces and vertices. The dashed blue lines indicate how the face $[1234]$ is deformed to become the four-valent vertex $F_{1234}$. The curved green legs are associated to the $\rho$-matrix and the blue solid lines in the vertex can be associated to the on-shell $\mathbb{CP}^{1|2}$ propagators: $\Delta^{1}_{42},\,\Delta^{2}_{13},\,\Delta^{3}_{24},\,\Delta^{4}_{31}$.
 In the replica method, used in section \ref{sec:replica} to obtain single-trace correlators, the elements $\rho_{ij}$ will be generalized from c-numbers to $m_i\times m_j$ matrices.
 }
 \label{fig:Graphduality 2}
 \end{figure}

An important feature of the $\rho$-matrix integral is that the rank $N_c$ of the gauge group now appears explicitly as a coupling in the action. This means that in the large $N_c$ limit we can expand the determinant in single-trace vertices and only keep a finite number of them when interested in a fixed genus correction. Explicitly, upon taking the logarithm, the determinant can be expanded as
\beq\label{eq:logdetexpansion}
-\log\det(\mathbb{I}_{n(n-1)}-\bar{\Delta}\bar{\rho}) \,=\, \sum_{k=1}^{\infty} \frac{1}{k}\text{tr}[(\bar{\Delta}\,\bar{\rho})^{k}]=\frac{1}{2}\text{tr}[(\bar{\Delta}\,\bar{\rho})^{2}]+ \frac{1}{2}\text{tr}[(\bar{\Delta}\,\bar{\rho})^{3}]+ \ldots~.
\eeq
The $k=1$ term on the right hand side vanishes because $\rho_{ii}=0$.
As can be seen from the form of the matrix, the generic term from the trace is an alternating product of $\rho$'s and $\Delta$'s:
\begin{equation} \label{schematic delta rho}
   \Delta^{i_1}_{i_n i_2}\,\rho_{i_1 i_2}\, \Delta^{i_2}_{i_1 i_3}\,
    \rho_{i_2 i_3}\, \cdots\,
  \rho_{i_{n-1}i_n}\Delta^{i_n}_{i_{n-1} i_1}  \rho_{i_n i_1} \,.
\end{equation}
This alternating structure can also be seen to arise diagrammatically, see figure \ref{fig:Graphduality 2}, when collapsing the faces of the original Feynman graphs to become the vertices of the $\rho$-matrix integral.

The $k=2$ term in \eqref{eq:logdetexpansion} is purely bosonic thanks to the property $\Delta^{i}_{jj}=1$:
\beq\label{squareDeltarho}
\frac{1}{2}\text{tr}(\bar{\Delta}\,\bar{\rho})^{2} =\frac{1}{2}\sum_{i,j}\Delta^{i}_{jj}\,\rho_{ij}\,\Delta^{j}_{ii}\,\rho_{ji} = \sum_{i<j}\rho_{ij}\rho_{ji}~.
\eeq
This constitutes a correction to the Gaussian term in \eqref{matrix_duality1} which can be absorbed to redefine the matrix integral as: 
\begin{equation}\label{MM2}
\left\langle\,\mathbb{D}(x_1,y_1,\theta_1))\ldots \mathbb{D}(x_n,y_n,\theta_n)\,\right\rangle_{\scriptscriptstyle \text{SDYM}} =\frac{1}{\mathcal{M}}\times 
 \int [\mathcal{D}\rho]\, e^{\sum\limits_{i<j}\frac{N_c}{D_{ij}}\rho_{ij}\rho_{ji} - \sum\limits_{k=3}^{\infty} \frac{N_c}{k}\text{tr}(\bar{\Delta}\,\bar{\rho})^{k}}~,
\end{equation}
where, using that $d_{ij}\equiv -y_{ij}^2/x_{ij}^2$,
the new propagator exhibits a ten-dimensional denominator \cite{Caron-Huot:2021usw} 
\begin{equation}\label{propagatorrho_10D}
\langle \rho_{ij} \rho_{kl}\rangle  = -\frac{D_{ij}}{N_c}\delta_{il}\delta_{kj}~,
\qquad D_{ij}=\frac{d_{ij}}{1-d_{ij}} = \frac{-y_{ij}^2}{x_{ij}^2+y_{ij}^2}~,
\end{equation}
which we encountered previously in (\ref{resummationD}). 

The higher-valence vertices are sums of products of the form \eqref{schematic delta rho}
\beq\label{eq:Fs_largeN}
\hspace{-2mm}\frac{1}{k}\text{tr}(\bar{\Delta}\,\bar{\rho})^{k}=\!\sum_{1\leq i_l \leq n}\,c^{\{i\}}_F\,F_{i_1 i_2\cdots i_k}~,\quad F_{i_1 i_2\cdots i_k}
\equiv \prod_{p=1}^{k}\,\Delta^{i_p}_{i_{p-1}\,i_{p+1}}\times f_{i_1 i_2\cdots i_k} 
~,
\eeq
where the $c_{F}$'s are symmetry factors and only terms with $i_l\neq i_{l+1}$ contribute. The $F$-vertices can be split into a $\mathbb{CP}^{1|2}$ component, carrying the fermionic part, and the purely bosonic $f$-vertices given by the cyclic products:
\beq\label{eq:fvertices}
f_{i_1 i_2\cdots i_k}\,\equiv\,\rho_{i_1i_2}\,\rho_{i_2i_3}\,\cdots \rho_{i_{k-1}i_k}\, \rho_{i_ki_1}~.
\eeq
These are the vertices in the original bosonic $\rho$-matrix integral of \cite{Jiang:2019xdz}.

Since we are interested in the genus expansion of the matrix integral, we can use Euler's formula $(F+V-E=2-2g)$ to find a truncation on the number of vertices we need to use at a given genus. In particular at genus $g$ the valence of a vertex can be at most $2n-2+4g$. More generally, denoting as $\mathsf{f}_{k}$ the number of $k$-valent $\rho$-vertices, the terms which contribute at genus $g$ satisfy
\beq\label{EuelrFormula}
2n-4+4g = \mathsf{f}_{3}+2 \mathsf{f}_{4}+3 \mathsf{f}_{5} +\cdots + (2n-4+4g)\,\mathsf{f}_{2n-2+4g}\,.
\eeq

We conclude this section with the large $N_c$ expansion of the $n=3$ example (\ref{n3}). From (\ref{EuelrFormula}) we infer that for $n=3$ and $g=0$ we can either have two three-valent vertices or one four-valent vertex leading to the matrix integral (\ref{MM2}) at 
\begin{equation}
   \frac{ \langle \mathbb{D}_1\mathbb{D}_2\mathbb{D}_3\rangle_{\scriptscriptstyle \text{SDYM}}}{\prod_{i<j}(1+D_{ij})} = \frac{1}{\mathcal{M}_{\scriptscriptstyle\text{10D}}}\times 
 \int [\mathcal{D}\rho]\, e^{\sum\limits_{i<j}\frac{N_c}{D_{ij}}\rho_{ij}\rho_{ji} - N_c(F_{123}+ \frac{1}{2}F_{1212}+F_{1213}+ \text{perm}+\text{non-planar})}~,
\end{equation}
with $F_{i_1\ldots i_k}$ defined in (\ref{eq:Fs_largeN}); $\mathcal{M}_{\scriptscriptstyle \text{10D}}$ denotes the Gaussian matrix integral with 10D propagator (\ref{MM2}) and we divide the left hand side by the ratio $\mathcal{M}/\mathcal{M}_{\scriptscriptstyle\text{10D}}$ (\ref{eq:normalisationMM}). For example we have
\begin{equation}
F_{123}= \Delta_{32\mu}^1\Delta_{13\mu}^2\Delta_{21\mu}^3\times \left(\rho_{12}\rho_{23}\rho_{31}\right)~,\quad F_{1213}= \rho_{12}\rho_{21}\rho_{13}\rho_{31}~.
\end{equation}
Performing Gaussian integrals and using the 10D propagator (\ref{propagatorrho_10D}) as well as the relations (\ref{eq:DeltaIdentities}) we obtain
\begin{multline}\label{eq:3DetMM}
  \hspace{-2mm} \frac{ \langle \mathbb{D}_1\mathbb{D}_2\mathbb{D}_3\rangle_{\scriptscriptstyle \text{SDYM}}}{\prod_{i<j}(1+D_{ij})}
   = 1-\frac{1}{N_c}(D_{12}D_{23}D_{31}+ D_{12}D_{23}+D_{23}D_{31}+ D_{31}D_{12}+ D_{12}^2+ D_{13}^2+D_{23}^2)
   \cr +\mathcal{O}(N_c^{-2})~.
\end{multline}
The term of order $N_c^{-1}$ is the connected genus-zero contribution.

In order to compare this result (and others to come) with single-trace correlators,
it is important to note that we have two distinct types of operators: determinants, and single traces:
\begin{equation}
    \mathbb{D}(x,y,\theta) = \det(1-y{\cdot}\phi(x)) + {\cal O}(\theta)~, \qquad
    \mathbb{O}(x,y,\theta) = \sum_{k=1}^\infty \frac{1}{k} {\rm Tr} [(y{\cdot}\phi(x))^k] +{\cal O}(\theta)
\label{det vs O}
\end{equation}
where the $\theta$-dependent terms are determined by supersymmetry.
We now explain two distinct methods to extract single-trace correlators: first using a replica method,
then using the direct large-$N_c$ expansion of the relation $\mathbb{D}=e^{-\mathbb{O}}$.

\begin{figure}[t]
\begin{center}
\begin{tikzpicture}[scale=1,thick,black!60!green]   
\draw[] (0,0) --(0,.3);
\draw[] (.25,0) --(.25,.3);
\draw[] (0,0) --(-.6,-.4);
\draw[] (.25,0) --(.75,-.35);
\draw[] (.125,-.25)-- (-.25,-.5);
\draw[] (.125,-.25)-- (.7,-.65);

\draw[] (2,0) --(2,.3);
\draw[] (2.25,0) --(2.25,.3);
\draw[] (2,0) --(1.5,-.35);
\draw[] (2.25,0) --(2.85,-.4);
\draw[] (2.125,-.25)-- (1.6,-.6);
\draw[] (2.125,-.25)-- (2.5,-.5);

\draw (.75,-.35) to[out=-30,in=-150] (1.5,-.35);
\draw (.7,-.65) to[out=-30,in=-150] (1.6,-.6);

\draw (0,.3) to[out=90,in=90] (2.25,.3);
\draw (0.25,.3) to[out=60,in=120] (2.,.3);

\draw (-.6,-.4) to[out=-115,in=-65] (2.85,-.4);
\draw (-.25,-.5) to[out=-135,in=-45] (2.5,-.5);

\node[scale=.5] at (.4,.1)   {$1$ ~ };
\node[scale=.5] at (1.96,.1)   {$1$ ~ };
\node[scale=.5] at (-0.05,.1)   {$2$ ~ };
\node[scale=.5] at (2.42,.1)   {$2$ ~ };
\node[scale=.5] at (.2,-.4)   {$3$ ~ };
\node[scale=.5] at (2.2,-.4)   {$3$ ~ };
\node[scale=.75,black] at (1.2,-1.7)   {$\langle F_{123}\,F_{321}\rangle={D}_{12}{D}_{23}{D}_{31}$ ~ };

\draw[] (6,0) -- (6,.7);
\draw[] (6.3,0) -- (6.3,.4);
\draw[] (6.3,0) -- (6.7,0);
\draw[] (6.3,-.3) -- (7,-.3);
\draw[] (6.3,-.3) -- (6.3,-1);
\draw[] (6.,-.3) -- (6.,-.7);
\draw[] (6.,-.3) -- (5.6,-.3);
\draw[] (6.,0) -- (5.25,0);

\draw (6,.7) to[out=70,in=30] (7,-.3);
\draw (6.3,.4) to[out=70,in=30] (6.7,0);

\draw (5.25,0) to[out=-160,in=-90] (6.3,-1);
\draw (5.6,-.3) to[out=-160,in=-90] (6.,-.7);

\node[scale=.5] at (5.95,.15)   {$2$ ~ };
\node[scale=.5] at (6.47,-.42)   {$2$ ~ };
\node[scale=.5] at (5.95,-.45)   {$1$ ~ };
\node[scale=.5] at (6.47,.15)   {$1$ ~ };

\node[scale=.75,black] at (6.2,-1.7)   {$\langle F_{1212}\rangle={D}_{12}^2$ ~ };

\draw[] (10,0) -- (10,.7);
\draw[] (10.3,0) -- (10.3,.4);
\draw[] (10.3,0) -- (10.7,0);
\draw[] (10.3,-.3) -- (11,-.3);
\draw[] (10.3,-.3) -- (10.3,-1);
\draw[] (10.,-.3) -- (10.,-.7);
\draw[] (10.,-.3) -- (9.6,-.3);
\draw[] (10.,0) -- (9.25,0);

\draw (10,.7) to[out=70,in=30] (11,-.3);
\draw (10.3,.4) to[out=70,in=30] (10.7,0);

\draw (9.25,0) to[out=-160,in=-90] (10.3,-1);
\draw (9.6,-.3) to[out=-160,in=-90] (10.,-.7);

\node[scale=.5] at (9.95,.15)   {$1$ ~ };
\node[scale=.5] at (10.47,-.42)   {$1$ ~ };
\node[scale=.5] at (9.95,-.45)   {$2$ ~ };
\node[scale=.5] at (10.47,.15)   {$3$ ~ };

\node[scale=.75,black] at (10.2,-1.7)   {$\langle F_{1213}\rangle={D}_{12}{D}_{13}$ ~ };

\end{tikzpicture}
\end{center}
\caption{Genus zero contributions for three operators.}
\label{figureMO3}
\end{figure}
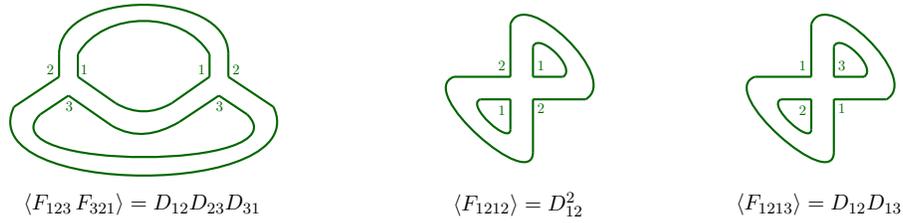

\subsection{From determinants to single-traces: replica method}\label{sec:replica}

\begin{figure}[t]
 \begin{minipage}[h]{1\textwidth}
 \centering 
  \includegraphics[width=1\textwidth]{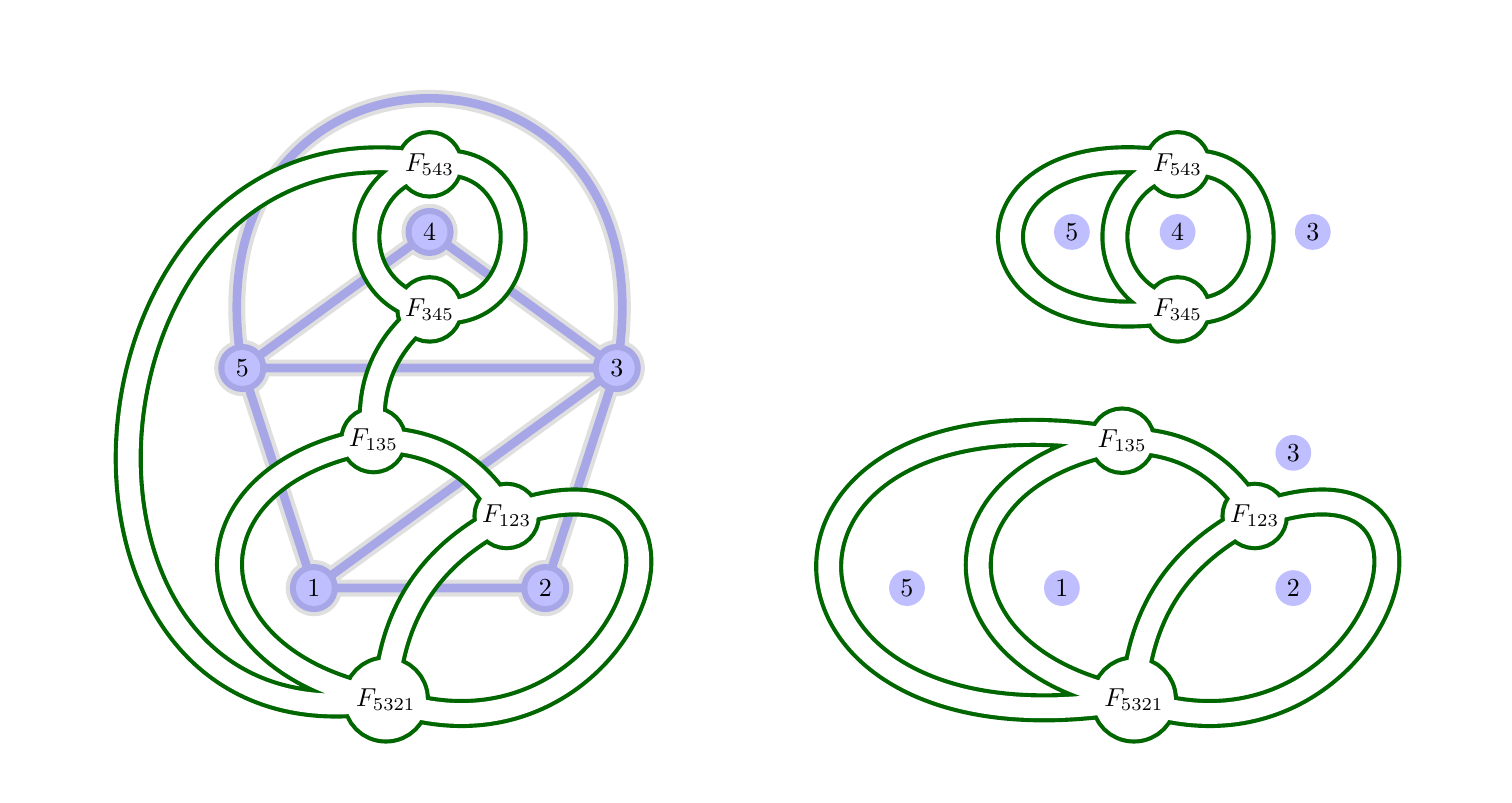}
 \end{minipage}
 \caption{Wick contractions of the five F-vertices in \eqref{eq:exampleFreplica}. On the left: the fully connected piece in green. The F-vertices are the faces of the dual graph in shaded blue. On the right: the disconnected piece with two extra faces $3$ and $5$. This latter does not survive the replica limit.}
 \label{fig:dualgraphreplica}
 \end{figure}

Here we explain how to use the replica method to extract correlators of single-trace operators  $\mathbb{O} = -\log \mathbb{D}$ from the matrix integral. This method relies on the limit $\lim\limits_{m\to0} \frac{x^m-1}{m}=\log x$ to produce the logarithm.  As a first step we consider a correlation function with  $m_{i}$ copies of the operator at position $x_{i}$. Then we take the limit $m_{i}\to 0$ to extract the single-traces:
\beq\label{eq:replicalimit}
\left\langle \prod_{i=1}^{n}\mathbb{O}(x_{i},y_i,\theta_i)\right\rangle \,=\, \lim_{m_i\to0}\,\frac{1}{\prod_{i=1}^{n}\,m_{i}}\,\left\langle \prod_{i=1}^{n}(1-\mathbb{D}(x_{i},y_i,\theta_i)^{m_{i}})\right\rangle \,.
\eeq
The idea is that, for integer $m_i$, the correlator of $\prod_i \mathbb{D}^{m_{i}}$
is equivalent to a correlator of $M=\sum_i m_i$ determinants. We can thus
adapt the matrix integral \eqref{matrix_duality1}
to get the replicated correlator, which will give us formulas that can then be analytically continued in $m_i$: 
\beq\label{eq:MMreplica}
\left\langle \prod_{i=1}^{n}\mathbb{D}(x_{i},y_i,\theta_i)^{m_{i}}\right\rangle_{\!\!\!\scriptscriptstyle\text{SDYM}} \, = \,   \frac{1}{\mathcal{M}}\times\int \mathcal{D}\rho\,e^{N_c\sum\limits_{i<j}\frac{\tr\left(\rho_{ij}\rho_{ji}\right)}{d_{ij}}}\det\left(\mathbb{I}-\bar{\Delta}\bar{\rho}\right)^{N_c}~.
\eeq
The integral on the right hand side depends on the same kinematics as described around \eqref{MM}. The main difference is that now the exponents $m_i$ on the left hand side control the dimension of the matrix on the right hand side. The elements of $\bar{\rho}$, which were just c-numbers in \eqref{matrix_duality1}, now become rectangular block matrices of dimension $m_{i}\times m_{j}$ and also the matrix $\bar{\Delta}$ changes accordingly:
\beq
\rho_{ij} \,\to\, (\rho_{ij})_{m_{i}\times m_{j}}\qquad \text{and}\qquad \Delta^{i}_{jk} \to \Delta^{i}_{jk}\,\mathbb{I}_{m_{i}\times m_{i}}~.
\eeq
Besides, the Gaussian term in \eqref{eq:MMreplica} now carries a trace taken over this new inner structure of the rectangular matrices. This also happens when writing $\det(1-\bar{\Delta}\bar{\rho})$ in terms of  single-trace vertices:
\beq\label{eq:logdetreplica}
\log\det(\mathbb{I}-\bar{\Delta}\bar{\rho}) \,=\, \sum_{k=1}^{\infty} \frac{1}{k}\text{tr}(\bar{\Delta}\,\bar{\rho})^{k}\,=\,\sum_{k=1}^{\infty} \sum_{1\leq i_l \leq n}\,c^{\{i\}}_F\,F_{i_1 i_2\cdots i_k}
\eeq
where now the vertices of type $F$ carry a trace over the $m_{i}$ structure:
\beq
F_{i_1 i_2\cdots i_k}
\,=\,\prod_{p=1}^{k}\,\Delta^{i_p}_{i_{p-1}\,i_{p+1}}\times\text{tr}(\rho_{i_1i_2}\,\rho_{i_2i_3}\,\cdots \rho_{i_{k-1}i_k}\, \rho_{i_ki_1}\,)~.
\eeq
We can perform a brute-force and systematic computation of the replicated correlator \eqref{eq:MMreplica} by performing Wick contractions on the $F$-vertices. For this purpose we use the propagator:
\beq
\langle \left(\rho_{ij}\right)_{ab}\,\left(\rho_{kl}\right)_{cd} \rangle\,=\,-  \frac{D_{ij}}{N_c}\,\delta_{il}\delta_{jk}\, \delta_{ad}\delta_{bc}\,.
\eeq
Similarly to the discussion around \eqref{MM2}, the latter propagator carries the ten-dimensional structure $D_{ij}$ instead of the four-dimensional propagator $d_{ij}$. This is due to the correction to the Gaussian in \eqref{eq:MMreplica} at large-$N_c$,
coming from the two-valence vertex in the expansion of the determinant. Hereafter all $F$-vertices have valency three or higher.

After performing the Wick contractions the result for the replicated correlator is a polynomial on $m_{i}$ and the limit in \eqref{eq:replicalimit} is straightforward to take.  As an example we present a contribution to the five-point replicated correlator in the planar limit $N_c\to\infty$:
\bba\label{eq:exampleFreplica}
\langle F_{123}F_{135}F_{345}F_{543}F_{5321}\rangle \, &= \, \langle F_{123}F_{135}F_{345}F_{543}F_{5321}\rangle_{c} \,+\,\langle F_{345}F_{543}\rangle_c\times \langle F_{123}F_{135}F_{5321}\rangle_c \nonumber\\
&= m_1 m_2 m_3 m_4 m_5\,\left(1+m_3\,m_5\right)\Delta^{1}_{253}\Delta^{3}_{152}\,D_{12}D_{13}D_{15}D_{23}D_{34}D_{35}^2D_{45}~.
\end{align}
As shown in figure \ref{fig:dualgraphreplica}, the two pieces in the first line of \eqref{eq:exampleFreplica} correspond to different dual graphs. The connected piece is dual to the graph in example \eqref{eq:exampleVrules}.
While the disconnected piece is proportional to $(m_3 m_5)^2$ and would contribute to a correlator with double-traces $\mathbb{O}(x_3)^2$ and $\mathbb{O}(x_5)^2$.  By applying the replica limit we isolate the contribution to the 5-point single-trace correlator:
\beq
 \lim_{m_i\to0}\,\frac{\langle F_{123}F_{135}F_{345}F_{543}F_{5321}\rangle}{m_1 m_2 m_3 m_4 m_5}= \Delta^{1}_{253}\Delta^{3}_{152}\,D_{12}D_{13}D_{15}D_{23}D_{34}D_{35}^2D_{45}~.
\eeq
This result reproduces the contribution of the graph given as example in \eqref{eq:exampleVrules}, where we used Feynman rules in twistor space.

Using this replica method we have calculated the planar correlation functions of the single-trace operator up to seven points. These results will be displayed in section \ref{section6} and have been used to cross-check the alternative method described there.

\subsection{From determinants to single-traces: algebraic relations}\label{sec:singletrace}

At large-$N_c$, there is also a more direct way to relate determinants to single-trace correlators,
simply by expanding the relation between them:
\begin{equation}
    \mathbb{D}(x,y,\theta) = e^{-\mathbb{O}(x,y,\theta)}. \label{det from log det}
\end{equation}
The important step is to understand the scaling of the ingredients at large-$N_c$.
The genus expansion applies most straightforwardly to single-traces, for which the leading (genus-zero) contribution scales like:
\begin{equation}
    \langle \mathbb{O}_1\cdots \mathbb{O}_n\rangle_c \sim  N_c^{2-n} \qquad \mbox{(genus 0)}~, 
    \label{largeNc counting}
\end{equation}
where the subscript $c$ denotes the connected part.
This result, together with the exponential form in \eqref{det from log det},
allows to understand correlators of determinants.
For example the logarithm of the correlator of $n=2$ determinants can be expanded as
\begin{equation} \label{log 2pt}
\log \langle \mathbb{D}_1\mathbb{D}_2\rangle =
\underbrace{\langle \mathbb{O}_1\mathbb{O}_2\rangle_c}_{\sim 1} - \tfrac12
\underbrace{
\left(\langle \mathbb{O}_1^2\mathbb{O}_2\rangle_c+
\langle \mathbb{O}_1\mathbb{O}_2^2\rangle_c\right) }_{\sim N_c^{-1}}
+ \ldots~.
\end{equation}
Higher terms involve more connected correlators of multi-point functions (in coincidence limits) and so are further suppressed
in the large-$N_c$ expansion.
Starting with the result in \eqref{n2 sum}, we thus deduce that
\begin{equation}
    \langle \mathbb{O}_1\mathbb{O}_2\rangle_c = \log(1+D_{12}) + \mathcal{O}(1/N_c^2) \label{two point O}
\end{equation}
which is in perfect agreement with the single-trace calculation in \eqref{eq:Otwopoint}.
The size of the error will be verified shortly.
Note that while the standard large-$N_c$ rules in a theory with adjoint fields imply that single-trace correlators admit expansions in integer powers of $1/N_c^2$, correlators of determinants contain both odd and even powers of $1/N_c$.

The fact that two-point functions are of order $N_c^0$
has an important implication for correlators of $n$ determinants.  The logarithm of such correlators will always contain a sum of $n(n-1)$ pairwise copies of \eqref{two point O}, plus terms that are suppressed at large-$N_c$. Exponentiating back, this gives for example for $n=3$:
\begin{equation} \label{3pt dets from log dets}
 \frac{\langle \mathbb{D}_1\mathbb{D}_2\mathbb{D}_3\rangle}{\prod_{i<j}^3 (1+D_{ij})}
 = 1 -\underbrace{\langle \mathbb{O}_1\mathbb{O}_2\mathbb{O}_3\rangle_c}_{\sim N_c^{-1}}
-\tfrac{1}{2}\sum_{6\,\rm perm.} \underbrace{\langle \mathbb{O}_1^2 \mathbb{O}_2\rangle}_{\sim N_c^{-1}} + \mathcal{O}(1/N_c^2)~.
\end{equation}
For any $n$, the ``divided correlator'' $\<\cdots\>/\prod_{i<j}(1+D_{ij})$ evaluates
to $1+\mathcal{O}(N_c^{-1})$. This observation can be used to extract correlators of $\mathbb{O}$'s starting from those of determinants computed by the non-replicated matrix integral \eqref{matrix_duality1}.
The difficulty is that several unknowns appear on the right.
The trick is to consider appropriate combinations:
define a ``connected divided correlator'' by combining divided correlators using the familiar combinatorics of connected parts:
\begin{align}
\left[\frac{\langle \mathbb{D}_1\mathbb{D}_2\mathbb{D}_3\rangle}{\prod_{i<j}^3 (1+D_{ij})}\right]_c
&\equiv
\frac{\langle \mathbb{D}_1\mathbb{D}_2\mathbb{D}_3\rangle}{\prod_{i<j}^3 (1+D_{ij})}
-\left(\frac{\langle \mathbb{D}_1\mathbb{D}_2\rangle\langle \mathbb{D}_3\rangle}{1+D_{12}}+\mbox{2 perm.}\right)
+2\langle \mathbb{D}_1\rangle\langle \mathbb{D}_2\rangle\langle \mathbb{D}_3\rangle
\nonumber\\ &= -\langle \mathbb{O}_1\mathbb{O}_2\mathbb{O}_3\rangle_c + \mbox{subleading}~.
\end{align}
Inserting the two- and three-point functions of determinants computed in 
\eqref{n2 sum} and \eqref{eq:3DetMM} (and $\langle \mathbb{D}_i\rangle=1$)
we thus find
\begin{equation} \label{three point O}
    \langle \mathbb{O}_1\mathbb{O}_2\mathbb{O}_3\rangle_c =
    \frac{1}{N_c}\left(D_{12}D_{23}D_{13} + D_{12}D_{13}+D_{12}D_{23}+D_{13}D_{23}\right) + \mathcal{O}(N_c^{-3})~,
\end{equation}
again in perfect agreement with the diagrammatic calculation in \eqref{eq:G3Feynman}.

There is an interesting cross-check with \eqref{log 2pt}: by taking the coincidence limit $y_3\to y_2$, $x_3\to x_2$ of \eqref{three point O},
we deduce that
\begin{equation}
    \langle \mathbb{O}_1\mathbb{O}^2_2\rangle_c = \frac{D_{12}^2}{N_c} +\mathcal{O}(N_c^{-3})~.
\end{equation}
Note that it is important in this limit
to take $y_3\to y_2$ faster than $x_3\to x_2$, so that self-contractions $D_{23}$ are set to zero. This is the correct prescription because the standard calculation of BPS correlators does not include self-contractions.  Substituting this and \eqref{two point O} into \eqref{log 2pt} and exponentiating, we then get
\begin{equation}
    \frac{\langle \mathbb{D}_1\mathbb{D}_2\rangle}{1+D_{12}} = 1-\frac{D_{12}^2}{N_c} +\mathcal{O}(N_c^{-2})~.
\end{equation}
This is in perfect agreement with 
\eqref{n2 sum} including now the $1/N_c$ correction.

The above examples confirm that the matrix integral \eqref{matrix_duality1} correctly predicts correlators of determinants as well as of single traces, which are related exactly as they should.  Since the intermediate steps in the two cases are somewhat distinct, this will provide useful cross-checks on the new calculations presented in the next section.

\section{NMHV $n$-point correlators}\label{section6}

In this section we discuss the
NMHV component of $n$-point correlation functions, that is, the component of Grassmann degree four for any $n$.
We will first obtain a simplified matrix-integral representation of the relevant correlator of determinants, valid for any finite $N_c$.
We will then investigate the gauge-independence of this expression and show how it can be exploited to obtain more concise integral representations.

We will denote averages in the bosonic $n\times n$ $\rho$-integral as:
\begin{equation} \label{avgbosonic}
\avgbosonic{\cdots} \equiv
\frac{1}{\mathcal{M}}\int [\mathcal{D}\rho]e^{N_c\sum\limits_{i<j}\frac{\rho_{ij}\rho_{ji}}{d_{ij}}}\det \left(\mathbb{I}_n - \rho\right)^{N_c}\left(\cdots\right)~,
\end{equation}
where the dots stand for an arbitrary function of $\rho$'s and the normalization
$\mathcal{M}$ is the Gaussian integral \emph{without} the determinant.
\newline

This section is organised as follows.
In subsection \ref{sec:NMHVmatrix} we introduce a change of variables to simplify the matrix integral \eqref{MM2},
which pushes the fermions outside the determinant
to give an average of the form \eqref{avgbosonic}.
We work it out up to order $R^2$ which gives a formula for any $n$-point NMHV correlator at finite $N_c$, see \eqref{Gn1}.
We then highlight in \ref{sec:Dyson} the virtues of this reformulation in manifesting the cancellation of $Z_*$ and $\mu_i$-dependent spurious poles through Schwinger-Dyson equations. In subsections \ref{sec:5NMHV} and \ref{sec:6NMHV} we obtain new finite-$N_c$ forms for the 5-point and 6-point NMHV correlators, and we highlight their planar limit,
which are expressed in terms of superconformal invariants times polynomials in $D_{ij}$ with manifestly ten-dimensional denominators.
Finally in subsection \ref{sec:countinginvariants} we count the $n$-point NMHV superconformal invariants which can appear with this method.

\subsection{Simplification of determinants including fermions}
\label{sec:NMHVmatrix}
For generic $n$, 
the $n(n-1)$-determinant in \eqref{matrix_duality1}
is a rather complex object.
We will now demonstrate a major simplification of it, proceeding empirically order-by-order in fermions but uniformly in $n$. At zeroth order, we have the identity \eqref{deMello} which reduces it to the much simpler $n\times n$ determinant in \eqref{avgbosonic}.
Here we will find that the fermions can be re-introduced by simply shifting the entries of that matrix.

For conciseness, we will omit the subscript on $\mu_k$ and denote the $R$-invariant \eqref{r123} on the $k^{\text{th}}$ line as
\begin{equation}
    \basicR{i}{j}{k} \equiv R(\lambda_{ki},\lambda_{kj},\mu_k)~.
\end{equation}

Let us first collect some data on the determinant in \eqref{matrix_duality1}.
At linear order in $R$'s (quadratic order in fermions), for $n=3$ points, the coefficient of $\basicR{1}{2}{3}$ is simply (other $R$-invariants can be obtained by symmetry):
\begin{equation}
    \det\left(\mathbb{I}_{6}-\bar{\Delta}(R) \bar{\rho}\right)\Big|_{\basicR{1}{2}{3}} = f_{123}-f_{132}~,
\end{equation}
where we write $\bar{\Delta}(R)$ to highlight that this matrix depends on the external
variables $(x,y,\theta)$ and references $\mu,\mathcal{Z}_*$ through the $R$-invariants \eqref{r123}.
For $n=4$, the analogous coefficient is
\begin{equation}
    f_{1423}-f_{1324} +f_{13}(f_{234}-f_{243}) + f_{23}(f_{143}-f_{134})
    +(1-f_{34})(f_{123}-f_{132})~,
\end{equation}
where as above $f_{i_1i_2\ldots i_k}\equiv \rho_{i_1i_2}\,\rho_{i_2i_3}\,\cdots \rho_{i_{k-1}i_k}\, \rho_{i_ki_1}$.  These display no clear pattern and the expressions for higher $n$ grow rapidly in complexity.

However, after some trial and error, we observe that the coefficients can be written uniformly
as derivatives of the bosonic determinant:
\begin{equation}
        \det\left(\mathbb{I}_{n(n-1)}-\bar{\Delta}(R) \bar{\rho}\right)\Big|_{\basicR{1}{2}{3}} =
        \left(\rho_{13}\rho_{32}\frac{\partial}{\partial \rho_{12}}
        -\rho_{23}\rho_{31}\frac{\partial}{\partial \rho_{21}}\right)\det \left(\mathbb{I}_n - \rho_{n\times n}\right)~. \label{linear shift determinant}
\end{equation}
This simple formula works for all $n$.

This can be interpreted as a small shift of the matrix entries and suggests trying to obtain the contributions with higher powers of $R$'s through a finite substitution:
\begin{equation}
    \tilde{\rho}_{ij}(\rho,R) = \rho_{ij} + \tilde{\rho}_{ij}^{(1)}(\rho,R) + \tilde{\rho}_{ij}^{(2)}(\rho,R) + \ldots \label{rho shift}~,
\end{equation}
where the $k^{\text{th}}$ shift has degree $k$ in $R$-invariants. The derivative
\eqref{linear shift determinant} is equivalent to the linear shift
\begin{equation}
    \tilde{\rho}_{ij}^{(1)} = \sum_{k\neq i,j} \rho_{ik}\rho_{kj} \basicR{i}{j}{k}~.
\end{equation}
The idea is to look for a shift $\tilde\rho$ which generates the entire Grassmann dependence of the $n(n-1)$-determinant: 
\begin{equation}
    \det\left(\mathbb{I}_{n(n-1)}-\bar{\Delta}(R) \bar{\rho}(\rho)\right)
    \stackrel{?}{=} \det\left(\mathbb{I}_{n} - \tilde{\rho}(\rho,R)\right)~,
    \label{nnm1 det from n}
\end{equation}
which is satisfied to linear order in $R$'s thanks to \eqref{linear shift determinant}.

To look for a second-order shift $\tilde{\rho}_{ij}^{(2)}$, we used the following strategy.
After applying the linear shift, the quadratic-in-$R$ error
in \eqref{nnm1 det from n} is a polynomial in $\rho$ variables and
the question is whether it can be written as a linear combination of derivatives
$\frac{\partial}{\partial \rho_{ab}}\det \left(\mathbb{I}_n - \rho\right)$.
This polynomial division problem can be answered
algorithmically using a Groebner basis for the ideal generated by the derivatives.
Quite to our surprise, we found empirically that a solution always exists.
The solution is not unique (the determinant is annihilated by various differential operators), and the solutions that come out of the Groebner method are unfortunately not very concise and do not generalize well.  However, by analyzing the solution space at low $n$, we were able to find a concise general solution:
\begin{equation}
    \tilde{\rho}_{ij}^{(2)} = \sum_{k\neq i,j} \basicR{i}{j}{k}\basicR{j}{k}{i} \rho_{ik}(\rho_{kj}-\rho_{ki}\rho_{ij})+\sum_{k,l\neq i,j} \basicR{j}{k}{l}(\basicR{j}{l}{k}-\basicR{i}{l}{k}) \rho_{ik}\rho_{kl}\rho_{lj}~.
    \label{rho shift quadratic}
\end{equation}
Eq.~\eqref{nnm1 det from n} is then satisfied up to an error which is cubic in $R$-invariants
(ie. degree six in Grassmann variables).

The identity \eqref{nnm1 det from n} greatly simplifies the evaluation of correlation functions.
The idea is to substitute the inverse change of variables $\rho\mapsto \rho(\tilde{\rho})$,
which can be computed in an iterative fashion order by order in fermions,
into the matrix integral \eqref{matrix_duality1}.
This eliminates the $n(n-1)$-dimensional determinant altogether and pushes all fermions into the exponent.  The Jacobian of the change of variable turns out to be exactly unity (at least up to quadratic order in $R$'s) and so
\eqref{matrix_duality1} becomes:
\begin{equation} \label{matrix_duality simplified}
\left\langle\,\mathbb{D}_1\ldots \mathbb{D}_n\,\right\rangle_{\scriptscriptstyle\text{SDYM}} 
=
\frac{1}{\mathcal{M}} \int [\mathcal{D}\rho]
e^{N_c\left(\sum\limits_{i<j}\frac{\rho_{ij}\rho_{ji}}{d_{ij}}+
S_{\rm eff}^{(1)}(\rho,R)+ S_{\rm eff}^{(2)}(\rho,R)+\ldots\right)} \det(\mathbb{I}_n-\rho)^{N_c}.
\end{equation}
Here $S_{\rm eff}^{(k)}$ is of order $R^k$ and
the determinant is now the purely bosonic $n\times n$ one that controls scalar correlators in \eqref{deMello}.

The effective action, obtained by inserting the inverse of the shift \eqref{rho shift}, \eqref{rho shift quadratic} into the Gaussian part of the action (and then relabelling $\tilde\rho \mapsto \rho$), reads:
\begin{equation}\begin{aligned} \label{Seff}
S_{\rm eff}^{(1)} &= \sum_{i<j<k} (f_{ijk}-f_{ikj})
\left(
\frac{\basicR{j}{k}{i}}{d_{jk}}+\frac{\basicR{k}{i}{j}}{d_{ki}}+\frac{\basicR{i}{j}{k}}{d_{ij}}\right),
\cr
S_{\rm eff}^{(2)} &= \sum_{i\neq j\neq k}
(f_{ijk}+f_{ij}f_{ik})\frac{\basicR{j}{k}{i}\basicR{i}{k}{j}}{d_{ik}}
+\sum_{i\neq j\neq k\neq l} f_{ijkl}
\left[
\frac{\basicR{j}{l}{i}\basicR{k}{l}{j} - R_{jkl}^i\basicR{i}{k}{j}}{d_{kl}}
-\frac{\basicR{j}{l}{i}\basicR{j}{l}{k}}{2d_{jl}}\right]\,.
\end{aligned}\end{equation}
Eq.~\eqref{matrix_duality simplified} will be the main tool of this section.  Its main features are that it is exact in $N_c$
and that $S^{(k)}_{\rm eff}$ is a finite-degree polynomial in $\rho$'s.
Note however that we have only established it to quadratic order in $R$'s (ie. we have not established that a shift ensuring \eqref{nnm1 det from n} always exists). Nonetheless, we conjecture that the form \eqref{Seff} exists to all orders.\footnote{
We sketch a derivation which may generalize to all orders. Start from the expansion 
of the logarithm of determinants in
\eqref{eq:Fs_largeN} and expand $\Delta^i_{jk}=1+R^i_{jk\mu}$ in each term.
In each monomial, factors where the ``1'' is used will be connected by strings of $R$'s.  Arranging these strings into a matrix:
\begin{equation}
  M_{ij} \equiv \rho_{ij} + \sum_{k\neq i,j} \rho_{ik}R^k_{ij\mu}\rho_{kj} +
  \sum_{k\neq i} \sum_{l\neq k,j}
   \rho_{ik}R^k_{il\mu}\rho_{kl}R^l_{kj\mu}\rho_{lj} + \ldots,
\end{equation}
we then have the following combinatorial identity:
\begin{equation}
    \det\left(\mathbb{I}_{n(n-1)}-\bar{\Delta} \bar{\rho}\right)
  = \det\left(\mathbb{I}_{n(n-1)}-\bar{R} \bar{\rho}\right) \times \det\left(\mathbb{I}_n - M\right)\,.
\end{equation}
The first term involves the matrix $\bar\Delta$ in \eqref{big matrices} with $\Delta\mapsto R$ and accounts for monomials with the maximal number of $R$'s, ie. where the ``1'' is not used.
We believe that \eqref{matrix_duality simplified} can be derived from this identity, essentially by making a change of variables $\rho_{ij}\mapsto M_{ij}$.
A subtlety is that since $M_{ii}\neq 0$, one first needs to shuffle the rows of $(\mathbb{I}_n - M)$ so as to eliminate the diagonal entries of $M$; we believe that this accounts, for example, for the terms quadratic-in-$\rho$ in \eqref{rho shift quadratic}.
This approach suggests that adding auxiliary degrees of freedom corresponding to diagonal entries of $\rho$, and possibly gauge redundancies related to the row shuffling, could
further simplify the matrix representation.
}
In any case, it gives us a relatively explicit formula for the $n$-point NMHV correlators in terms of averages in the bosonic measure \eqref{avgbosonic}:
\newline
\begin{align} \label{Gn1}
&\left\langle\,\mathbb{D}(x_1,y_1,\theta_1)\ldots \mathbb{D}(x_n,y_n,\theta_n)\,\right\rangle_{\scriptscriptstyle\text{SDYM}}^{\scriptscriptstyle\rm NMHV}
\cr
 \quad =&\phantom{+}
    \frac{N_c^2}{2}\!\sum_{(ijk),(lmn)}\avgbosonic{\! (f_{ijk}{-}f_{ikj})(f_{lmn}{-}f_{lnm})\!}
\left(\frac{\basicR{j}{k}{i}}{d_{jk}}+\mbox{2 cyclic}\right)
\left(\frac{\basicR{m}{n}{l}}{d_{mn}}+\mbox{2 cyclic}\right)
\cr&
+N_c\!\sum_{i\neq j\neq k\neq l} \avgbosonic{f_{ijkl}}
\left[
\frac{\basicR{j}{l}{i}\basicR{k}{l}{j} - R_{jkl}^i\basicR{i}{k}{j}}{d_{kl}}
-\frac{\basicR{j}{l}{i}\basicR{j}{l}{k}}{2d_{jl}}\right]
\cr&
+N_c
 \sum_{i\neq j\neq k} \avgbosonic{f_{ijk}{+}f_{ij}f_{ik}}\frac{\basicR{j}{k}{i}\basicR{i}{k}{j}}{d_{ik}}~.
\end{align}
The first sum runs over possibly overlapping triplets $(ijk)$ and $(lmn)$ and originates
from squaring the first line of \eqref{Seff}.
Note that since $R$-invariants square to zero,
each nonvanishing term occurs exactly twice in the sum, cancelling the factor $1/2$.

Eq.~\eqref{Gn1} is an exact formula for supersymmetrized correlators at finite $N_c$.  The averages produce polynomials in $1/d_{ij}$'s. In practice, we are able to compute them either in the large-$N_c$ limit, or for specific low values of $N_c$.
We find that it is highly redundant: individual terms depend on gauge-fixing
parameters $\mu_i$ and $\mathcal{Z}_*$,
but this dependence has to cancel out in the sum. In the rest of this section we analyze the cancellation mechanism and use it to infer more concise expressions.

\subsection{Cancellation of spurious poles and Schwinger-Dyson equations}\label{sec:Dyson}

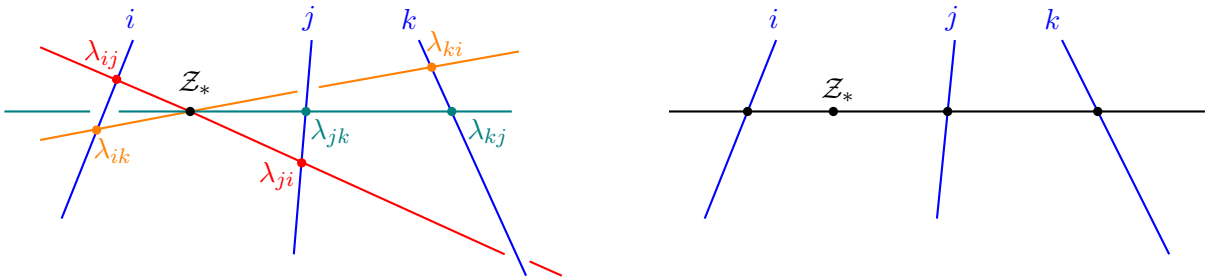
\begin{figure}[t]
\begin{center}
\begin{tikzpicture}[thick, scale=.95]

\node[scale=1,blue] at (.1,.3)   {$i$ ~ };
\node[scale=1,blue] at (2.6,.3)   {$j$ ~ };
\node[scale=1,blue] at (4.,.3)   {$k$ ~ };
\node[scale=1,red] at (-.3,-.2)   {$\lambda_{ij}$ ~ };

\node[scale=1,orange] at (-.15,-1.55)   {$\lambda_{ik}$ ~ };
\node[scale=1,red] at (2.15,-1.9)   {$\lambda_{ji}$ ~ };
\node[scale=1,teal] at (2.9,-1.3)   {$\lambda_{jk}$ ~ };
\node[scale=1,teal] at (5.1,-1.3)   {$\lambda_{kj}$ ~ };
\node[scale=1,orange] at (4.5,-.05)   {$\lambda_{ki}$ ~ };
\node[scale=1,black] at (1,-.6)   {$\mathcal{Z}_*$ ~ };
\draw[blue] (0,0) --(-1,-2.5);
\draw[blue] (2.5,0) -- (2.25,-3);
\draw[blue] (4,0) -- (5.5,-3.3);
\draw[orange] (.8,-1)  -- (2.3,-.72);
\draw[orange] (2.6,-.66)  -- (5.4,-.16);
\draw[orange] (.8,-1)  -- (-1.3,-1.4);
\draw[teal] (-.2,-1.)  -- (5.3,-1.);
\draw[teal] (-.6,-1.)  -- (-1.8,-1.);
\draw[red] (-1.3,-.1)  -- (.8,-1);
\draw[red] (.8,-1)  -- (5.2,-3);
\draw[red] (5.55,-3.1)  -- (6,-3.3);
\draw [fill,black] (.8,-1) circle [radius=0.05];

\node[scale=1,blue] at (9.1,.3)   {$i$ ~ };
\node[scale=1,blue] at (11.6,.3)   {$j$ ~ };
\node[scale=1,blue] at (13.,.3)   {$k$ ~ };
\node[scale=1,black] at (10,-.7)   {$\mathcal{Z}_*$ ~ };
\draw[blue] (9,0) --(8,-2.5);
\draw[blue] (11.5,0) -- (11.25,-2.5);
\draw[blue] (13,0) -- (14.5,-3);
\draw[black] (7.5,-1)  -- (15,-1.);
\draw [fill,black] (9.8,-1) circle [radius=0.05];

\draw [fill,red] (-.23,-.55) circle [radius=0.05];

\draw [fill,orange] (-.51,-1.26) circle [radius=0.05];

\draw [fill,red] (2.36,-1.72) circle [radius=0.05];

\draw [fill,teal] (2.42,-1.0) circle [radius=0.05];

\draw [fill,teal] (4.46,-1.0) circle [radius=0.05];

\draw [fill,orange] (4.18,-.38) circle [radius=0.05];

\draw [fill,black] (8.6,-1.0) circle [radius=0.05];
\draw [fill,black] (11.4,-1.0) circle [radius=0.05];
\draw [fill,black] (13.5,-1.0) circle [radius=0.05];

\end{tikzpicture}
\caption{Geometry of spurious poles.
Left: given a twistor $\mathcal{Z}_*$ and three generic (non-intersecting) $\mathbb{CP}^{1}$ lines $i, j, k$, there exists a unique line which passes
through $\mathcal{Z}_*$ and any two of the $\mathbb{CP}^{1}$'s.  This defines six special points $\lambda_{ab}$.
Right: When two special points collide on one line, a simultaneous collision occurs on all three lines. This creates the $\mathcal{Z}_*$-dependent spurious poles.
\label{fig:spurious poles}
}
\end{center}
\end{figure}

The invariant $R^i_{jkl}$ on line $i$, defined in \eqref{r123},
contains three factors of the form $1/\langle {\lambda_{i j}\ \lambda_{i k}}\rangle$ in its denominator. These diverge whenever two special points on line $i$ collide. These singularities are spurious since the position of these points depends on the axial gauge reference $\mathcal{Z}_*\in \mathbb{CP}^{3|4}$.
As depicted in figure \ref{fig:spurious poles}, this happens
when the line passing through $\mathcal{Z}_*$, $X_i$ and $X_j$ also passes through $X_k$.  This condition is symmetrical, so the same collision simultaneously occurs
on three lines:
\begin{equation}
    \frac{1}{\langle \lambda_{ij}\ \lambda_{ik}\rangle},\
    \frac{1}{\langle \lambda_{ji}\ \lambda_{jk}\rangle}\ \mbox{and}\ 
    \frac{1}{\langle \lambda_{ki}\ \lambda_{kj}\rangle}\ 
    \mbox{share the same spurious singularity}\,.
\label{spurious}
\end{equation}
This means that each spurious singularity is shared between $R$-invariants on three different lines and their cancellation will require nontrivial identities between them.
At linear order in $R$'s we find that the unique linear combination of $R$'s which is free of this spurious pole is the three-term sum:
\begin{equation} \label{spurious free}
    \frac{R^i_{jk\mu_i}}{d_{jk}}+\frac{R^j_{ki\mu_j}}{d_{ik}}+\frac{R^k_{ij\mu_k}}{d_{ij}}
\quad  \mbox{is free of the spurious pole \eqref{spurious}}~,
\end{equation}
where the $\mu_a$ are arbitrary (the residue of $R^i_{jk\mu}$ on $1/\langle \lambda_{ij}\ \lambda_{ik}\rangle$ is independent of $\mu$). The same combination appeared
previously in eq.~4.35 of \cite{Chicherin:2014uca}.

We conclude that the first line of \eqref{Gn1} is already manifestly free of such spurious poles.  In fact, the same happens in each line. For example, on the last line, $\basicR{j}{k}{i}$ and $\basicR{i}{k}{j}$ share the same $1/\langle \lambda_{ij}\ \lambda_{ik}\rangle$ pole but it cancels in their product because of a double zero from the fermionic numerator.  This is easy to verify numerically.  The second line is safe thanks to a combination of these mechanisms.

Following standard arguments, since
the correlator is a rational function in $\mathcal{Z}_*$ of homogeneity degree zero,
the cancellation of all $\mathcal{Z}_*$-dependent spurious poles in \eqref{Gn1} implies its complete $\mathcal{Z}_*$-independence.

Note that it is not quite true that \eqref{Gn1} is $\mathcal{Z}_*$-independent termwise, because $\mathcal{Z}_*$ also enters through poles of the form
$1/\langle \lambda_{ij}\ \mu_i\rangle$, where $\mu_i$ is the arbitrary point
on line $i$ introduced to remove the zero-mode (see \eqref{mainConjecture}).
However, we have already established in \eqref{symmetryTransfo} that these poles cancel out after integrating over $\rho$. Altogether, we have thus
established that \eqref{Gn1} is independent of both $\mu$'s and $\mathcal{Z}_*$.
The question now is how to exploit this in practice.
While
$\mu$-cancellation could be proved in \eqref{symmetryTransfo} by simply rescaling $\rho$'s, it has become less evident after the change of variable \eqref{rho shift} and it is nontrivial to simultaneously manifest the independence on $\mu$ and $\mathcal{Z}_*$.

To better understand the cancellation of $\mu_i$, consider for example the coefficient of $\basicR{j}{k}{i}/\langle \lambda_{lm}\ \mu_l \rangle$ where $l,m\neq i,j,k$;
only the first line in \eqref{Gn1} contributes and its cancellation requires the identity:
\begin{equation} \label{desired IBP identity}
    \sum_{n\neq l,m} \frac{1}{d_{mn}} \avgbosonic{(f_{ijk}-f_{ikj})(f_{lmn}-f_{lmn})} \stackrel{?}{=}0 \qquad (l,m\neq i,j,k)~.
\end{equation}

How should we understand this relation from the viewpoint of the matrix integral?
The main tool at our disposal is its set of Schwinger-Dyson equations, ie. integration-by-parts in $\rho$.
We will now construct a differential operator which annihilates the determinant in \eqref{avgbosonic} and demonstrates the above identity.

A good starting point is the following identity:
\begin{equation}\label{dgl_det}
\mathcal{D}^{k,l}_{\rho} \det(\mathbb{I}_n- \rho_{n\times n}) = \rho_{lk}\rho_{kl}\det(\mathbb{I}_{n-1}- \rho_{n-1\times n-1})|_{l,l}~,
    \end{equation}
where we indicate by the bar $|_{i,j}$ the minor with the $i^{\text{th}}$ row and $j^{\text{th}}$ column removed, and $\mathcal{D}^{k,l}_{\rho}$ is the following differential operator
\begin{equation}
\mathcal{D}^{k,l}_{\rho}\equiv -\rho_{kl}\frac{\partial}{\partial\rho_{kl}} +\sum_{j\neq k,l}\rho_{jk}\rho_{kl}\frac{\partial}{\partial\rho_{jl}}~.
\end{equation}
The form of this operator is inspired by the change of variables in \eqref{rho shift} 
and the relation \eqref{dgl_det} follows from the Laplace expansion of the determinant with respect to the $l^{\text{th}}$ column.  We sketch a proof of (\ref{dgl_det}). Expanding $\det(\mathbb{I}_n- \rho_{n\times n})$ with respect to the $l^{\text{th}}$ column we obtain
\begin{equation}\label{Lexpansion}
\det(\mathbb{I}_n- \rho_{n\times n}) =-  \sum_{l \neq i=1}^{n}(-1)^{i+l}\rho_{il}\det(\mathbb{I}_n- \rho_{n\times n})|_{i,l}+\det(\mathbb{I}_n- \rho_{n\times n})|_{l,l}~.
\end{equation}
After acting with $\mathcal{D}^{k,l}_{\rho}$ on (\ref{Lexpansion}) it remains to show that 
\begin{equation}
(-1)^{l+k}\det(\mathbb{I}_n- \rho_{n\times n})|_{k,l} - \sum_{j\neq k,l}(-1)^{l+j}\rho_{jk}\det(\mathbb{I}_n- \rho_{n\times n})|_{j,l}- \rho_{lk}\det(\mathbb{I}_n- \rho_{n\times n})|_{l,l}=0~.
\end{equation}
However this is nothing than the Laplace expansion of the original matrix $\mathbb{I}_n- \rho_{n\times n}$ where the $l^{\text{th}}$ column is replaced by the $k^{\text{th}}$ column. Since however that matrix has two equal columns its determinant must vanish. 

We now study the implications of (\ref{dgl_det}).
Since the determinant is invariant under taking the transpose, we could equally well expand with respect to the $l^{\text{th}}$ row and obtain \eqref{dgl_det} with a transpose differential equation $(\mathcal{D}^{k,l}_{\rho})^*$, where ${}^*$ replaces $\rho_{ij}$ by $\rho_{ji}$.
In particular, subtracting these two relations gives an identity
\begin{equation}
0=\int [\mathcal{D}\rho]e^{N_c\sum\limits_{i<j}\frac{\rho_{ij}\rho_{ji}}{d_{ij}}}
F(\rho)
\left(\mathcal{D}_\rho^{l,k}-(\mathcal{D}^{l,k}_\rho)^{*}\right)\det \left(\mathbb{I}_{n} - \rho_{n\times n}\right)^{N_c}~,
\end{equation}
which is valid for any polynomial $F(\rho)$.
After an integration by parts it leads to
\begin{equation}
\avgbosonic{ F(\rho) \sum_{j\neq k,l}\frac{f_{kjl}-f_{klj}}{d_{jl}}}=\frac{1}{N_c}\avgbosonic{
\mathcal{D}_\rho^{l,k}F(\rho)-(\mathcal{D}^{l,k}_\rho)^{*}F(\rho))}
\end{equation}
which is valid for any pair with $k\neq l$.
We can now generate identities like \eqref{desired IBP identity} by choosing $F(\rho)=f_{ijk}-f_{ikj}$ and using $\mathcal{D}_\rho^{l,m}$.
Depending on whether the index sets overlap, the derivative on the right-hand-side produces different 
terms:
\begin{subequations}\label{IBP relations}\begin{align}
 \label{IBP relations 1}
\hspace{-6mm}    \sum_{n\neq l,m} \frac{1}{d_{mn}} \avgbosonic{(f_{ijk}{-}f_{ikj})(f_{lmn}{-}f_{lnm})} & = 0 & (m\neq i,j,k)~, \\
    \sum_{n\neq l,i} \frac{1}{d_{in}} \avgbosonic{(f_{ijk}{-}f_{ikj})(f_{lin}{-}f_{lni})} & =
    \frac{1}{N_c}\avgbosonic{ f_{ikjl}{+}f_{iljk}{-}f_{ijkl}{-}f_{ilkj}} &(l\neq i,j,k)~, \\
    \sum_{n\neq i,j} \frac{1}{d_{jn}} \avgbosonic{(f_{ijk}{-}f_{ikj})(f_{ijn}{-}f_{inj})} & = \frac{1}{N_c} \avgbosonic{2f_{ij}f_{ik}{+}f_{ijk}{+}f_{ikj}}~. &
\end{align} \end{subequations}
These relate all the averages entering \eqref{Gn1} and we expect that these ensure its gauge invariance.
Let us see how this works for $n=3$ and $n=4$, where the NMHV component of the correlator \eqref{Gn1} should actually vanish due to the absence of corresponding superconformal invariants.
For $n=3$, five averages enter \eqref{Gn1} and we find that the relations \eqref{IBP relations} specialize to
\begin{equation}\begin{aligned}
\tfrac{N_c}{2}\avgbosonic{(f_{123}{-}f_{132})^2} &= d_{23}\avgbosonic{f_{12}f_{13}}+\avgbosonic{f_{123}}(d_{23})\,, \\
d_{13}\avgbosonic{f_{12}f_{23} } &= d_{23}\avgbosonic{f_{12}f_{13}}+\avgbosonic{f_{123}}(d_{23}-d_{13})\,,
\\
d_{12}\avgbosonic{ f_{13}f_{23} } &= d_{23}\avgbosonic{f_{12}f_{13}}+\avgbosonic{f_{123}}(d_{23}-d_{12})\,. \label{3pt relations}
\end{aligned}\end{equation}
We stress that this specialization is only valid for $n=3$.
Inserting these relations into \eqref{Gn1} and collecting terms,
we indeed find that the coefficient of each independent average vanishes identically:
\begin{equation}
    \langle\mathbb{D}_1\mathbb{D}_2\mathbb{D}_3 \rangle_\SDYM^{\scriptscriptstyle \rm NMHV} =
    \avgbosonic{f_{12}f_{13}} \times 0 + \avgbosonic{f_{123}} \times 0\,.
\end{equation}
In other words, after using the relations \eqref{3pt relations}, we do not need to compute any average to see that the Grassmann terms in the three-point correlator vanish.

For $n=4$, seventeen independent averages appear and
we find that after using all the identities \eqref{IBP relations}
the correlator \eqref{Gn1} becomes proportional to a single combination:
\begin{equation} \label{six term B}
\langle\mathbb{D}_1\cdots \mathbb{D}_4 \rangle_\SDYM^{\scriptscriptstyle \rm NMHV} =    
\left[N_c \avgbosonic{ \frac{N_c d_{34}}{d_{13}d_{24}} (f_{123}{-}f_{132})(f_{124}{-}f_{142})
  - f_{1234}-f_{1432}}\right] \times \Big( \mbox{eq.~\eqref{sixterms}}  \Big) = 0\,.
\end{equation}
Again, no matter what the average in the square bracket evaluates to, this correlator vanishes because it is proportional to the six-term identity in \eqref{sixterms}.
The only input were the relations \eqref{IBP relations}, which hold at any $N_c$.

These examples confirm that the relations \eqref{IBP relations}
ensure gauge invariance of our NMHV formula \eqref{Gn1}. At higher points, they will enable us to write the supercorrelators in terms of manifestly gauge-invariant building blocks.

\subsection{NMHV five-points}\label{sec:5NMHV}

For the correlator of $n=5$ determinants, the number of distinct averages
entering \eqref{Gn1} is very large, however we find that the above identities reduce them to
just six independent ones.  This number can be understood as follows.
First, it is useful to consider only identities resulting from 
\eqref{IBP relations 1} and linear combinations thereof, which involve only quadratic averages in $f$'s. They imply relations like
\begin{multline} \label{IBP relation 5pt}
\avgbosonic{ (f_{123}{-}f_{132})(f_{245}{-}f_{254}) }
\left(\frac{1}{d_{12}d_{35}}- \frac{1}{d_{15}d_{23}}\right) \cr
=
\avgbosonic{ (f_{134}{-}f_{143})(f_{245}{-}f_{254}) }
\left(\frac{1}{d_{14}d_{35}}- \frac{1}{d_{15}d_{34}}\right).
\end{multline}
Together with its permutations, this shows that all
averages involving two $f$'s with five distinct labels are proportional to each other.
This means that there is a single degree of freedom which multiplies
a truly five-index invariant.

The remaining five degrees of freedom multiply four-index invariants that are simply
permutations of the six-term identity \eqref{six term B} which appeared at four-points.
These can thus be discarded.  We conclude that the five-point NMHV correlator is proportional to a single, rigid linear combination of products of $R$'s.

This invariant is symmetrical under $S_5$ permutations
so we find it convenient to organize its terms in symmetry orbits.
In an explicitly $\mu$-independent form it is:
\begin{equation} \label{I5}
    \tilde{\mathcal{I}}^{(5)}_{12345} =
        \sum_{10}\frac{R_{345}^1 R_{345}^2}{d_{12}d_{34}d_{45}d_{35}}+
        \sum_{60}\frac{R_{234}^1 R_{135}^2}{d_{15}d_{24}d_{34}d_{35}}+
        \sum_{15}{}^{\prime} \frac{R_{234}^1 R_{345}^1}{d_{23}d_{34}d_{45}d_{52}}~,
\end{equation}
where each sum runs over the permutations of $\{1,2,3,4,5\}$ which act nontrivially on its summand, and the subscripts display
the resulting number of terms.
Recall that $R$'s (eq.~\eqref{r123}) are antisymmetric in their three subscripts, and $d$'s are symmetrical.
The prime in the last sum emphasizes the identity $R^1_{234}R^1_{345}=R^1_{245}R^1_{345}$ (see \eqref{eq:fusionRule}),
which reduces the number of distinct permutations from 30 to 15. The total
number of terms in \eqref{I5} is thus 85.

The function $\tilde{\mathcal{I}}^{(5)}_{12345}$ is superconformal invariant since it is homogeneous in twistors and independent of references $\mathcal{Z}_*$ and $\mu_i$. It should thus be proportional to the unique solution $\mathcal{I}_{5;1}$ to superconformal Ward identities at five points from \cite{Eden:2011we},
discussed already around \eqref{G51 L}.  We find, indeed,
\begin{equation} \label{I51 versus Itilde}
   \mathcal{I}_{5;1} = -\tilde{\mathcal{I}}^{(5)}_{12345} \times \prod_{1\leq i\leq j\leq 5}  y_{ij}^2\,.
\end{equation}

The coefficient of \eqref{I5} in \eqref{Gn1} can be found by matching with any single term.
To take the coefficient of $R^1_{34\mu}R^{2}_{45\mu}$ in \eqref{I5}, for example, we simply
expand out the $R$-invariants using \eqref{four term identity} in reverse, giving: 
\begin{equation}
    \mathcal{I}^{(5)}_{12345} \big|_{R^1_{34\mu}R^{2}_{45\mu}} = 
\frac{1}{d_{34}d_{45}} \left(\frac{1}{d_{12}d_{35}} - \frac{1}{d_{15}d_{23}}\right)\,.
\end{equation}
This procedure is unambiguous because there are no linear relations between $R_{ij\mu}^k$ (each has exactly one $\mu$-independent pole), nor quadratic relations (except those on the same line implied by \eqref{eq:fusionRule}). Comparing with \eqref{Gn1} we thus find
\begin{equation}
  \langle\mathbb{D}_1\cdots \mathbb{D}_5 \rangle_\SDYM^\NMHV = \mathcal{I}^{(5)}_{12345} \times N_c^2\left(\frac{1}{d_{12}d_{35}} - \frac{1}{d_{15}d_{23}}\right)^{-1}
 \avgbosonic{(f_{134}{-}f_{143})(f_{245}{-}f_{254})}\,.
\end{equation}
This gives the full Grassmann and $N_c$ dependence of the 5-point correlator of half-BPS determinants.
It is nontrivial that the right-hand-side is permutation-invariant, but this is ensured by the identities \eqref{IBP relation 5pt}.

In the planar limit, the integral can be expanded using the method from section \ref{ssec:largeN}.
We find in this way a result which is indeed permutation-invariant:
\begin{equation} \label{G51 planar}
\frac{\langle\mathbb{D}_1\cdots \mathbb{D}_5 \rangle_\SDYM^\NMHV}{
\prod_{i<j}^5 (1+D_{ij})}
= \frac{2 \mathcal{I}^{(5)}_{12345}}{N_c^3} \prod_{i<j}^5 D_{ij}  + \mathcal{O}(N_c^{-4})~.
\end{equation}
The single-trace and single-particle correlators can then
be deduced using the formulas from sections \ref{sec:singleParticle} and \ref{sec:singletrace}, which are rather trivial here since lower-point correlators do not depend on Grassmann variables.
Thus
\begin{equation} \label{G51 planar from dets}
    G_{5,1} = G_{5,1}^{\rm sp} = \frac{-2 \mathcal{I}^{(5)}_{12345}}{N_c^3} \prod_{i<j}^5 D_{ij} + \mathcal{O}(N_c^{-5})\,,
\end{equation}
again in perfect agreement with the diagrammatic calculation in \eqref{G51 from RR} and in \eqref{G51}.

\subsection{NMHV six-points}\label{sec:6NMHV}

For higher-points, we follow a similar strategy:
we enumerate all quadratic products of $f$'s and use the relations \eqref{IBP relations 1} to find the truly independent ones.  
For $n=6$, there are 190 such averages
and we find exactly $23=15+8$ independent ones. 15 are uninteresting as they simply correspond to permutations of the identity \eqref{sixterms} which appeared at four points.  Six of the remaining ones multiply permutations of $\tilde{\mathcal{I}}^{(5)}_{12345}$, and the remaining two are genuinely six-point NMHV invariants.  

To understand these two invariants, we take inspiration by the planar limit.  We find the following form for a six-label average:
\begin{multline} \label{six-label average}
    N_c^4\frac{\avgbosonic{(f_{123}{-}f_{132})(f_{456}{-}f_{465})}}{\prod_{i<j}^6 D_{ij}(1+D_{ij})} \cr= \frac{2}{d_{14}d_{25}d_{36}} + 
\frac{2}{d_{14}d_{15}d_{24}d_{26}d_{35}d_{36}}
\pm \mbox{(perm.~of $(123)$)} + \mathcal{O}(N_c^{-5})~,
\end{multline}
where the $\pm$ sign denotes antisymmetrization in $\{1,2,3\}$. Other averages involving six labels are related to it by permutations.
The right-hand-side can be written as a sum of two $3\times 3$ determinants. Because the relations are homogeneous in $d$,
we conclude that each of them can be used to seed a distinct invariant, which is then fully determined by 
the relations \eqref{IBP relations} up to a multiple of the five-term identities. Requiring full permutation symmetry, we obtain the following two explicit representatives:
\begin{align} \label{I6a}
    \mathcal{I}^{(6a)}_{123456} =& \phantom{+}\sum_{90} \frac{R^1_{234}R^4_{561}}{d_{23}d_{56}}
   \ \det \left[d_{ij}^{-1}\right]^{i=1,2,3}_{j=4,5,6}
  \nonumber\\&+ \sum_{360}\frac{R^1_{234}R^2_{135}}{d_{34}d_{35}}
        \left[\frac{1}{d_{16}d_{26}d_{45}}-\frac{1}{d_{15}d_{26}d_{46}}-\frac{1}{d_{16}d_{24}d_{56}}+\frac{1}{d_{12}d_{46}d_{56}}\right]
         \nonumber\\
   & -\sum_{90} \frac{R^1_{234}R^2_{134}}{d_{34}}
        \left[\frac{1}{d_{16}d_{26}d_{35}d_{45}}+\frac{1}{d_{12} d_{36} d_{45} d_{56}}+\frac{1}{d_{12}d_{35}d_{46}d_{56}}+\frac{1}{d_{15}d_{25} d_{36}d_{46}}\right]
         \nonumber\\
    & -\sum_{72}{}^\prime
    \frac{R^1_{234}R^1_{245}+R^1_{234}R^1_{256}+R^1_{245}R^1_{256}}{d_{23}d_{34}d_{45}d_{56}d_{62}}~,
\end{align}
and\newline
\begin{align} \label{I6b}
    \mathcal{I}^{(6b)}_{123456} =& \phantom{+}\sum_{90} \frac{R^1_{234}R^4_{561}}{d_{23}d_{56}}
  \  \frac{\det \left[d_{ij}\right]^{i=1,2,3}_{j=4,5,6}}{\prod_{\substack{i=1,2,3\\ j=4,5,6}} d_{ij}}
  \nonumber\\&+
  \sum_{360} \frac{R^1_{234} R^2_{135}}{d_{34}d_{35}d_{36} d_{12}d_{24}d_{45}d_{51}}
\left[\frac{d_{12}}{d_{16}d_{26}} -\frac{d_{15}}{d_{16}d_{56}}-\frac{d_{24}}{d_{26}d_{46}} + \frac{d_{45}}{d_{46}d_{56}}\right]
  \nonumber\\&
 +\sum_{90} \frac{R^1_{234}R^2_{134}}{d_{12}d_{34}d_{35}d_{36}d_{45}d_{46}}\left[\frac{1}{d_{15}d_{26}}+\frac{1}{d_{16}d_{25}}\right]
-\sum_{180} \frac{R^1_{234}R^1_{235}}{d_{16}d_{23}d_{24}d_{25}d_{34}d_{35}d_{46}d_{56}}\,.
\end{align}
The prime in the last sum of \eqref{I6a} accounts for the fact that the summand involves the dihedrally-invariant 5-point amplitude on the line $1$ (see the second line of \eqref{5points}), which reduces the number of its independent permutations to 72.

We conclude that the NMHV correlator of six determinants at any $N_c$ must be a sum of the above
eight invariants.  Absorbing convenient factors of $D_{ij}'s$, we can write:
\begin{equation} \label{expansion 6pt}
\frac{\langle\mathbb{D}_1\cdots \mathbb{D}_6 \rangle_\SDYM^\NMHV}{\prod_{i<j}^6 (1+D_{ij})} =
\prod_{i<j}^6 D_{ij}\times\left(C^{(6a)}_{6,1} \mathcal{I}^{(6a)}_{123456} +
C^{(6b)}_{6,1} \mathcal{I}^{(6b)}_{123456}\right) +
\left[\prod_{i<j}^5D_{ij}\times C^{(5)}_{6,1} \tilde{\mathcal{I}}^{(5)}_{12345}+ \mbox{5 perm}\right].
\end{equation}
This is the main result of this analysis.
The $C$'s are rational functions of $d_{ij}$'s that can be written at finite $N_c$
as $\rho$ averages by matching any 8 independent coefficients against \eqref{Gn1}. Here we record their values in the planar limit. For correlators of determinants we find that 
\begin{equation}\begin{aligned}
     N_c^4C^{(6a)\rm det}_{6,1} &= 2 +\mathcal{O}(N_c^{-1}) = -N_c^4C^{(6b)\rm det}_{6,1}~,\qquad
     \\
     N_c^4C^{(5)\rm det}_{6,1}&=
2N_c+
4D_{16}D_{26}D_{36}D_{46}D_{56}+2\sum_5 D_{16}D_{26}D_{36}D_{46} -2\sum_5 D_{16}(D_{16}+2)
\\ &\phantom{=} +\sum_{10}\left(8-8(D_{12}{+}1)(D_{13}{+}1)(D_{23}{+}1)
-4\left(D_{16}D_{26}+ D_{12}(D_{16}D_{26}+D_{16}+D_{26})\right)\right),
\end{aligned}\end{equation}
where the first two terms simply follows from \eqref{six-label average}.
Each sum runs over inequivalent $\{1,2,3,4,5\}$-permutations of its summand, with the resulting number of terms counted by the subscript.

As before, things simplify somewhat when we extract single-trace correlators: notably,
the disconnected contributions at order $N_c^{-3}$ and $N_c^{-4}$ cancel out and some coefficients change slightly, due to products like $\< \mathbb{O}_i^2\mathbb{O}_j\>$ similar to \eqref{3pt dets from log dets}.
Expanding the single-trace correlator $G_{6,1}$ as on the right-hand-side of \eqref{expansion 6pt}
with $C^{\rm det}\mapsto C$, we find the following coefficients:
\begin{equation}\begin{aligned} \label{c65 st}
N_c^4C_{6,1}^{(6a)} &= 2 +\mathcal{O}(N_c^{-1})
=-C_{6,1}^{(6b)}~, \\
N_c^4C_{6,1}^{(5)}
 &= 4D_{16}D_{26}D_{36}D_{46}D_{56}+2\sum_5 D_{16}D_{26}D_{36}D_{46} -4\sum_5 D_{16}
\\ &\phantom{=}-2\sum_{10} \left(D_{16}D_{26}+ D_{12}(D_{16}D_{26}+D_{16}+D_{26})\right) + \mathcal{O}(N_c^{-1})~.
\end{aligned}\end{equation}
Note that the coefficients of $\mathcal{I}^{(6a)}$ and $\mathcal{I}^{(6b)}$ are unchanged.
Finally,  converting to the single-particle basis removes further terms in the $\tilde{\mathcal{I}}^{(5)}$ coefficient, due to the double-trace term in \eqref{single particle general}, which gives a derivative
of the $n=5$ correlator times a derivative of two-point functions.
The former can be computed using \eqref{G51 planar from dets} that the superconformal invariant $\tilde{\mathcal{I}}^{(5)}$ is homogeneous of weight $-2$ in each $y$:
\begin{equation}
    y_i{\cdot}\frac{\partial}{\partial y_i} \tilde{\mathcal{I}}^{(5)}_{12345} = -2 \tilde{\mathcal{I}}^{(5)}_{12345}
    \quad\mbox{for $i\in 1\ldots5$}~.
\end{equation}
We find:
\begin{equation}\begin{aligned} \label{c65 sp}
N_c^4C_{6,1}^{(6a)\rm sp} &= 2 +\mathcal{O}(N_c^{-1}) = -N_c^4C_{6,1}^{(6b)\rm sp}~,
\\
N_c^4C_{6,1}^{(5)\rm sp} &=
    4 D_{16}D_{26}D_{36}D_{46}D_{56}+2\sum_{5} D_{16}D_{26}D_{36}D_{46}
    -2\sum_{10} D_{16}D_{26}(1+D_{12})+\mathcal{O}(N_c^{-1})\,.
\end{aligned}\end{equation}
This is our main result regarding the planar six-point NMHV correlator. It can be
inserted into the right-hand-side of \eqref{expansion 6pt} to give the correlator $G^{\rm sp}_{6,1}$ of six single-particle operators.

An important question is whether the coefficients $C_{6,1}^{\rm sp}$ display any kind of ten-dimensional symmetry.  Clearly, the coefficients take particularly
concise forms when written using the ten-dimensional building block $D_{ij}=\frac{-y_{ij}^2}{x_{ij}^2+y_{ij}^2}$,
and all denominators are ten-dimensional, but the numerators do not exhibit the full ten-dimensional rotational symmetry, let alone conformal symmetry.
The same observation applies to the MHV correlators for any $n$ (such as $G^{\rm sp}_{4,0}$ in \eqref{eq:G4MHVsp}). In the present case there is an additional ambiguity since the invariants
$\mathcal{I}^{(6a)}$, $\mathcal{I}^{(6b)}$
are only canonically defined modulo multiples of $\mathcal{I}^{(5)}$, which can shift $C_{6,1}^{(5)\rm sp}$ somewhat. The symmetry seems in some sense only mildly broken in \eqref{c65 sp} but we do not quite know how to quantify this.

It is also interesting to consider the $y\to 0$ limit,
where the coefficient of $y_i^2$ should match with the
stress-tensor multiplet correlators obtained in eq.~3.20 of \cite{Chicherin:2015bza}.
These authors considered the SU$(N_c)$ theory but we can use the fact that
single-particle operators in the with SU(4) weight $k\geq 2$ are the same in the U$(N_c)$ and SU$(N_c)$ theories.
By considering the homogeneity degree of the various invariants in \eqref{expansion 6pt} we find that only the
$\mathcal{I}^{(6b)}$ term contributes:
\begin{equation}
    N_c^4 G_{6,1}^{\rm sp}\Big|_{y_i^2} = 
    -2 \mathcal{I}^{(6b)}_{123456} \prod_{i<j}^6 \frac{y_{ij}^2}{x_{ij}^2}  + \mathcal{O}(N_c^{-1})
=\frac{1}{480}\frac{A_2-2A_1-8 B_2}{\prod_{i<j}^6 x_{ij}^2}~,
\end{equation}
where $A_i$ and $B_i$ are defined in \cite{Chicherin:2015bza}.
We found numerically agreement (up to an overall sign)
for the sub-components $\theta_{6}^4$ and $\theta_{5}^2\theta_{6}^2$ given in the ancillary of that reference.

\subsection{Counting superconformal invariants}\label{sec:countinginvariants}

In twistor space, any $n$-point half-BPS correlator has the following structure: it is a linear combination of products of R-invariants \eqref{r123} with coefficients that are polynomials in $d_{ij}$'s. The combination must be gauge-invariant: free of both $\mathcal{Z}_*$ and $\mu_i$ dependence.
It is interesting to try to classify such invariants generally.

As a warm up, consider gauge-invariants that are linear in $R$'s. To be free
of $\mathcal{Z}_*$ spurious poles it must be built out of the three-term combinations \eqref{spurious free}.
The antisymmetric coefficients $c_{ijk}$ must satisfy further relations to ensure $\mu$-independence:
\begin{equation}
    \sum_{(ijk)} c_{ijk} \left(\frac{R^i_{jk\mu_i}}{d_{jk}}+\frac{R^j_{ki\mu_j}}{d_{ik}}+\frac{R^k_{ij\mu_k}}{d_{ij}}\right) \quad \mbox{is invariant iff}\quad E^i_j \equiv \sum_{k\neq i,j} \frac{c_{ijk}}{d_{jk}}=0 \quad \forall\, i\neq j\,.
\label{linear gauge independence}
\end{equation}
The constraint $E^i_j$ ensures cancellation of the $1/\<\lambda_{ji}\,\mu_i\>$ pole on line $i$.
The solutions to \eqref{linear gauge independence} can be readily counted.
Naively there are $n(n-1)$ constraints, but they turn out to not be all independent due to the following redundancies:
\begin{equation}
    \sum_{i\neq j} E^i_j =0, \qquad \sum_{j\neq i} \frac{1}{d_{ij}}E^i_j=0~,
\end{equation}
which are identically satisfied for any choice of $c$'s.
There are $2(n{-}1)$ such redundancies. The total number of solutions of \eqref{linear gauge independence} at $n$ points is thus equal to the number of triplets, minus $n(n-3)$ constraints:
\begin{equation}
    N_1(n) = \tfrac{1}{6}n(n{-}1)(n{-}2) - n(n{-}3) = \tfrac{1}{6}n(n{-}4)(n{-}5)\,.
\end{equation}
We have verified numerically, for various random configurations of $x$ and $y$'s,
that this is indeed equal to the number of gauge-invariant combinations of \eqref{linear gauge independence}. (In this check, we do not impose that coefficients are polynomial in $d_{ij}$'s.)

The numbers $N_1$ are displayed in table \ref{tab:invariants}. Notice that the first nontrivial solutions occur for $n=6$ points, where there are 2 solutions.  Symbolic solutions can be found
but they are rather unwieldy (for example there is a unique solution
for which $c_{456}=0$, but one of its coefficient contains a factor with 312 terms).
In particular, we were not able to find solutions that transform simply under permutations of 6 labels,
and it is unclear what would constitute a ``good basis''.

\begin{table}
\centering\small
\renewcommand{\arraystretch}{1.2}
\begin{tabular}{c|c|c|c|c|c|c|c}
$n$ &$4$&5&6&7&8&9&$n$\\   \hline
\# linear & 0 & 0&  2 & 7 & 16 & 30 & $N_1=\tfrac{1}{6}n(n{-}4)(n{-}5)$ \\[2pt]  \hline
\# quadratic & 0  &  1 & 8 & 43 & 171 & 535 & $N_2 =\tfrac12 N_1(N_1{+}1)
+ \tfrac{1}{24}(n{-}1)(n{-}2)(n{-}3)(n{-}4)$
\end{tabular}
\caption{Number of linear and quadratic gauge-invariant linear combinations of twistor $R$-invariants
 for different number of points.}
\label{tab:invariants}
\end{table}

There is a clear parallel between the gauge-invariance constraints \eqref{linear gauge independence}
and the Schwinger-Dyson equations \eqref{IBP relations 1} of the matrix integral.
In fact, given any two solutions of the former, their product will define an ``average'' that satisfies the latter.  To better understand this connection, we set up a numerical problem
analogous to \eqref{linear gauge independence} to look for quadratic linear combinations of $R$'s that are gauge-invariant (accounting for the relations implied by \eqref{eq:fusionRule}).
We found, for all $n\leq 9$, that the number of quadratic gauge invariants is precisely equal to the number of solutions to \eqref{IBP relations 1}, which is also equal to the number of solutions to all of \eqref{IBP relations}.  For any $n$ we expect a one-to-one correspondence between gauge-invariant combinations of $R$'s and solutions to the Schwinger-Dyson equations.

The number of quadratic-in-$R$ gauge invariants for various $n\leq 9$ is displayed in table \ref{tab:invariants}. For each $n$ we have subtracted 
$\frac{1}{4!}n(n-1)(n-2)(n-3)$
permutations of the six-term identity \eqref{sixterms}, since it yields trivial solutions.
We have verified numerically that all remaining solutions are genuinely independent
by evaluating them for different Grassmann components for a given value of $x,y$'s.
The numbers were fit to the degree-6 polynomial in $n$ recorded in the table,
and its prediction $N_2=1401$ for $n=10$ was then
confirmed by counting solutions to \eqref{IBP relations 1}.

Currently, the principal obstacle for writing human-readable forms of NMHV correlators with $n\geq 7$
is the lack of a practical basis for superconformal invariants:
computing the averages in \eqref{Gn1} is otherwise possible
(we were able to do it explicitly for $n=7$).
Other perspectives, such as the correlahedron \cite{Eden:2017fow}, might offer additional insight on this question.  From the spacetime viewpoint, only three independent invariants were found for $n=6$  \cite{Chicherin:2015bza}.  This does not necessarily contradict the fact that we find eight, since we consider invariants that are potentially of higher degree in $y$ variables, but the relation should be better understood.

\section{Discussion}\label{section8}

In this paper we studied correlators of half-BPS operators with arbitrary R-charge in self-dual $\mathcal{N}$=4 super-Yang-Mills (SYM) theory.
Our main results are:  a new twistor-space representation  \eqref{mainConjecture} which unites all half-BPS single-trace operators,
and a new formula \eqref{matrix_duality1} which expresses the correlator of $n$ determinants as a $n\times n$ matrix integral. Since we retained the full dependence on (chiral) superspace coordinates $(x,y,\theta)$, these formulas give access to correlators involving the SYM Lagrangian, a crucial bridge between the self-dual and full SYM theory.

Using these methods we have confirmed that the correlators with the maximal Grassmann degree
enjoy a dependence on the 10-vector $X=(x,y)$ that is fully compatible with ten-dimensional symmetry,
as conjectured in \cite{Caron-Huot:2021usw}. We also find a new concrete formula \eqref{Gn1} for the $n$-point correlator of Grassmann degree four
(NMHV) for any $n$, which is of non-maximal degree for $n\geq 6$ operators.
A challenging feature of twistor space calculations is that it is often difficult to see how gauge dependence cancels from final expressions.  While we have not fully solved this problem,
we hope that the matrix integral representation will prove a useful step towards manifesting these cancellation, for example by adding
suitable gauge redundancies or auxiliary degrees of freedom to it.

There are many limits that one would like to analyze using the matrix integral,
many of which exhibit remarkable simplificity.
For example, the terms with maximal Grassmann degree should match the high-loop integrands computed in \cite{Bourjaily:2016evz}, which admit extremely rich combinatorial structure.
In the limit where subsets of points approach a null 10D polygon, $X_{ij}^2\to 0$,
the planar correlator was conjectured to give the loop integrand for planar scattering amplitudes on the Coulomb branch \cite{Caron-Huot:2021usw},
which is known to enjoy a 10-dimensional dual conformal symmetry \cite{Caron-Huot:2010nes}.
Finally, one may ask if there are further simplifications in the limit $y\to 0$, where one singles out the stress-tensor multiplet.
Currently, a main difficulty in analyzing all these limits from the viewpoint of the matrix
integral is the presence of the twistor-space gauge-fixing in intermediate expressions.
More speculatively, could the matrix integral help define correlators
in a more ten-dimensional friendly way, relaxing the constraint $y_i^2=0$?

While half-BPS operators naturally correspond to $\mathbb{CP}^{1|2}$ defects in twistor space,
it may be interesting to consider analogous models supported on $\mathbb{CP}^{1|k}$.
Could the $k=1$ case be related to quarter-BPS operators?
The action \eqref{combinedCP} shows how multiple $\mathbb{CP}^{1|2}$ models (each associated with an $m=2$ amplituhedron) can be coupled together to describe SYM correlators
that are known to be related to the $m=4$ amplituhedron.  One may wonder if coupling multiple $m=4$ models in a similar fashion could be related to other generalizations such as the $m\geq 6$ amplituhedrons.

It would be interesting to further connect with D-instantons in string theory.
Loosely speaking, to the extent that (following \cite{Witten:2003nn}) our twistor space setup can be interpreted as D1 branes coupled to a D5 brane in topological string theory,
and that the twistor transform is a two-dimensional Fourier transform,
one may naively expect that our correlators are related to D$(-1)$-instantons coupled to the familiar stack of $N_c$ D3 branes (see for example \cite{Lechtenfeld:2005xi}). 
Can the degrees of freedom in the matrix integral \eqref{matrix_duality1}
be interpreted as D$(-1)$-D$(-1)$ strings?
This is somewhat distinct from the perspective of \cite{Jiang:2019xdz, Budzik:2021fyh, Chen:2019gsb}, where
certain determinant operators called giant gravitons were related to D3 branes in (twisted) holography.
In any case, since the correlators we studied are defined within the self-dual Yang-Mills theory, it would be interesting to have an ${\rm AdS}_5\times S^5$ bulk dual description for them. The string dual of free $\mathcal{N}=4$ SYM using twistor space
could provide a natural starting point \cite{Gaberdiel:2021qbb}.

\section*{Acknowledgements}
It is a great pleasure to acknowledge Shota Komatsu and Matteo Parisi for useful discussions.
BM is supported in part by the Simons Foundation Grant No. 385602 and the Natural Sciences and Engineering Research Council of Canada (NSERC), funding reference number SAPIN/00047-2020. 
Work of SCH and FC are supported in parts by the Simons Collaboration on the Nonperturbative Bootstrap.
Work of SCH is also supported by the Simons Fellowships in Theoretical Physics and the Canada Research Chair program.  SCH is grateful for hospitality at the Institute for Advanced Study, where part of this work was carried out.

\bibliographystyle{JHEP}
\bibliography{references}

\end{document}